\documentclass[PAPER, atlasdraft=false, UKenglish, cernpreprint, texmf, orcidlogo]{atlasdoc}
\usepackage[subfigure]{atlaspackage}
\usepackage{atlasbiblatex}
\usepackage{pgffor}

\usepackage[journal,misc,particle,unit,xref,jetetmiss,snippets]{atlasphysics}
\graphicspath{{logos/}{figures/}}

\usepackage{ANA-FTAG-2023-06-PAPER-defs}

\AtlasTitle{%
A continuous calibration \\
of the ATLAS flavour-tagging classifiers \\
via optimal transportation maps%
}

\AtlasAbstract{%
A calibration of the ATLAS flavour-tagging algorithms using a new
calibration procedure based on optimal transportation maps is presented.
Simultaneous, continuous corrections to the $b$-jet, $c$-jet, and
light-flavour jet classification probabilities from jet-tagging algorithms in
simulation are derived for $b$-jets using $t\bar t \to e\mu\nu\nu bb$
data.
After application of the derived calibration maps, closure between simulation
and observation is achieved for jet flavour observables used in ATLAS analyses
of Large Hadron Collider (LHC) Run 2 proton--proton collision data.
This continuous calibration opens up new possibilities for the future use of
jet flavour information in LHC analyses and also serves as a guide for
deriving high-dimensional corrections to simulation via transportation maps,
an important development for a broad range of inference tasks.
}

\PreprintIdNumber{CERN-EP-2025-026}

\AtlasDate{\today}

\AtlasJournalRef{Eur. Phys. J. C 85 (2025) 1272}
\AtlasDOI{10.1140/epjc/s10052-025-14682-0}

\hypersetup{pdftitle={ATLAS document},pdfauthor={The ATLAS Collaboration}}

\begin{document}

\maketitle

\tableofcontents

%
%

% The next lines are included from the .//intro.tex input file
\section{Introduction}
\label{sec:intro}

The identification and classification of jets containing heavy-flavour hadrons
(flavour tagging) is a crucial component of many analyses of the Large Hadron
Collider~\cite{Evans:2008zzb} (LHC) data collected by the ATLAS
detector~\cite{PERF-2007-01}.
For example, searches and measurements targeting processes producing top quarks
($t$) or Higgs bosons ($H$) are particularly dependent on flavour tagging
because of the high branching fraction of these particles to heavy-flavour
quarks~\cite{HIGG-2021-23, HIGG-2020-03, HIGG-2021-12, HIGG-2018-51,
TOPQ-2018-26, TOPQ-2017-03, EXOT-2018-48, HDBS-2019-29, HDBS-2022-03,
HDBS-2018-40, HDBS-2018-34, HIGG-2020-20}.

Recent progress in flavour-tagging algorithms for classifying jets containing
\bhadrons (\bjets) and \chadrons (\cjets) has greatly increased the sensitivity
of ATLAS analyses to possible new phenomena~\cite{FTAG-2019-07,
ATL-PHYS-PUB-2022-027, ATL-PHYS-PUB-2020-014}.
These algorithms are trained on samples produced by generators of proton-proton
($pp$) collision events that are subsequently passed through a sophisticated
simulation of the ATLAS detector~\cite{Buckley:2011ms, Agostinelli:2002hh,
SOFT-2010-01}.
Contemporary algorithms take as input various observables related to a jet --
primarily calorimeter and tracking information -- and compute probabilities that
each jet contains a \bhadron (\pb), a \chadron (\pc), or neither of these (\pu).
Traditionally, ATLAS analyses using flavour tagging define a summary
observable based on these outputs and categorize jets by using this summary
observable to separate signal from the background jets.
The most common summary observable in use is the log-likelihood ratio of \pb,
\pc, and \pu assuming some mixture of jet populations as a background
hypothesis.
For example, for \bjet identification (\enquote{\btagging{}}), the ratio
\begin{equation}
\label{eq:Db}
D_b = \log \frac{\pb}{f_c \pc + (1 - f_c) \pu}
\end{equation}
is an optimal discriminant between \bjets and non-\bjets when the \cjet fraction
of non-\bjets is $f_c$.\footnote{$f_c$ should be interpreted as an effective
parameter that controls the \cjet fraction at inference time; during the
optimization of the classification algorithm, a different fraction might be
chosen.}
Alternatively, if the goal is to identify \cjets and reject \bjets and
light-flavour jets (\enquote{charm-tagging}), the discriminant used is
\begin{equation}
\label{eq:Dc}
D_c = \log \frac{\pc}{f_b \pb + (1 - f_b) \pu}.
\end{equation}
Typical values of $f_c$ used in previous ATLAS analyses range between 0.018 and 0.08 for
\btagging, while for charm-tagging, choices of 0.3 or 0.4 for $f_b$ are common.

From these discriminants, operating points are defined; e.g.\ the \enquote{70\%
\btagging operating point} is the value of $D_b$ for which 70\% of \bjets
with $\pt > 20$~GeV in a Standard Model (SM) \ttbar sample have a higher
discriminant score.

The probability of jets to pass or fail selections based on these operating
points is conventionally measured directly in data, and the flavour-tagging
information about simulated jets is corrected such that on average it reflects
the observed performance in data.
These corrections are calculated in bins\footnote{In this work, a set of
\enquote{bins}, such as those used to define a histogram, is a collection of
contiguous subsets that are used to build a piece-wise approximation of a
function. For example, the ranges (0, 0.5], (0.5, 0.75], (0.75, 1.0) define a
\emph{binning} of the open unit interval (0, 1) made of three \enquote{bins}.} of the
jet's transverse momentum (\pt) because of the known dependence of tagger
performance on this quantity~\cite{FTAG-2019-07}; however, the corrections are
averaged over other jet properties.
It is worth emphasizing that the calibration is derived purely based on the
jet flavour-tagging information:\ other aspects of the simulation may only be
altered through correlations with the flavour-tagging observables.

In ATLAS analyses of LHC \RunTwo data, the calibration is obtained by
measuring the tagging efficiencies (the probability that a signal jet passes
some selection) or mis-tag rates (the probability that a background jet passes
some selection) for \bjets, \cjets, and the remaining jet population (light
jets) in regions enriched in each jet flavour.
These tagging rates in data (\effdata) are compared with the expected rates in
simulation (\effmc), and correction scale factors ($\effdata/\effmc$)
are applied as weights to simulated jets~\cite{FTAG-2019-02, FTAG-2020-08,
FTAG-2018-01}.
For \bjets, these scale factors typically deviate from unity by less than 5\%,
but for $c$-jets and light-flavour jets, efficiency corrections may reach 20\%
or 50\%, respectively.
This enables extraction of physical parameters from ATLAS data that are unbiased
by mismodelling of flavour-tagging performance in the simulation.

The resulting calibrations have a limited applicability:\ they only correct
identification efficiencies for a particular choice of discriminant binning.
For example, if an analysis requires a different combination of $b$- and \cjet
tagging (e.g.\ a new \btagging operating point or change in $f_c$ or $f_b$), a
completely new set of corrections must be derived for the relevant combination;
this is especially pertinent in the search for $H \to cc$
decays~\cite{HIGG-2021-12,HIGG-2020-20}.
Similarly, while calibrations have been performed with up to five bins of the
tagging discriminant score (\enquote{pseudo-continuous calibrations}), they do
not allow a truly continuous use of the tagging algorithms' output in physics
analyses.
Other analyses that would benefit from non-standard operating points are those
that need very strong \cjet rejection, e.g.\ searches for rare top-quark
production processes~\cite{TOPQ-2018-06}; those that use a very loose \bjet
veto, such as measurements of $H \to W W^*$ decays~\cite{HIGG-2020-25,
HIGG-2018-49}; or those for which the optimal operating point is strongly
dependent on event kinematics:\ for example searches for high-mass resonances
decaying into \bjets~\cite{EXOT-2016-33}.

This paper describes an alternative calibration based on changing (or
\emph{transporting}) the three flavour-tagging classifier outputs (\pu, \pc,
\pb) such that the resulting probability density of these features in simulation
matches data.
The transportation map that calibrates the simulation is fully continuous
(i.e.\ it is not binned); it does not involve reweighting simulated events but
rather alters the output of the flavour-tagging algorithms for each jet.
It is moreover designed such that it \emph{minimally} alters the simulation
in a well-defined sense that is made more formal in Section~\ref{sec:methods}.
The derivation of such minimal transport maps between probability distributions
is the topic of optimal transportation theory~\cite{monge1781mémoire,
villani2008optimal}.

ATLAS uses optimal-transportation-based corrections for the energy calibrations
of muons~\cite{MUON-2022-01}, electrons~\cite{EGAM-2018-01}, and hadronic
jets~\cite{JETM-2018-05}, but these each rely on the shape of the
detector energy responses being nearly Gaussian.
In the Gaussian case, a transportation-based calibration amounts to correcting
the first two moments of a distribution; as such, these calibrations are not
usually formulated using the optimal-transportation nomenclature.
However, in the case of jet flavour tagging, the probability density of
discriminants is highly non-Gaussian and has no well-motivated simple
parametric form.
Nevertheless, recent advances in neural optimal-transport solvers allow the
derivation of optimal transportation maps between distributions without known
\emph{a priori} densities as long as samples can be drawn from
them~\cite{pmlr-v119-makkuva20a, Pollard:2021fqv, bunne2022supervised}.
Optimal transport has been proposed for several other purposes in particle
physics, including for background estimates~\cite{Manole:2022bmi}, measurements
of multijet event shapes~\cite{STDM-2020-20}, and for anomaly
detection~\cite{Craig:2024rlv}.
In this paper, the focus is on the use of optimal transport for the calibration
of the ATLAS flavour-tagging algorithms.

This calibration procedure has several advantages over more-traditional binned
efficiency scale factors.
Given a suitable choice of transport distance, there is a unique optimal
transport map that yields a simulation that reflects the data while
\emph{minimally} altering the simulation.
Transporting the observables of simulated jets is inherently unitary in the
sense that the number of jets in a sample is unaltered by applying the
calibration; this is not in general the case when weighting by the
data--simulation efficiency ratio, even after having applied corresponding scale
factors for jets that fail the selection (\enquote{inefficiency scale
factors})~\cite{FTAG-2018-01}.
Furthermore, instead of correcting the tagging probabilities of particular
operating points or jet categories, their continuous density is calibrated; this
allows much more flexibility in using said observables in data analysis.
Finally, while flavour-tagging calibrations in ATLAS have so far been limited to
a single one-dimensional discriminant with a fixed binning, in large part due to
the \enquote{curse of dimensionality} that makes multidimensional binned
distributions difficult to work with, calibrations based on neural optimal
transport have the ability to handle high-dimensional calibration problems with
relative ease and without any explicit binning choice.

The primary disadvantage of the transport-based calibration identified during
the course of this study was that this approach does not yet allow for
\emph{in situ} constraints on nuisance parameters, such as the flavour fractions
that are fitted in more-conventional flavour-tagging calibrations as in
Ref.~\cite{FTAG-2018-01}.
Reweighting-based calibrations may also reduce statistical fluctuations when
considering systematic uncertainties of the calibration, since events cannot
migrate across bin boundaries when applying weight variations.
It is worth noting that the application of weights to correct the distribution of
a particular observable will alter distributions over other observables through
correlations, while in the transportation set-up only distributions that are
explicitly corrected or have an explicit dependence on them are changed in the
process; whether this is desirable depends on the particular use-case.

This paper presents the continuous calibration of the ATLAS \dloner
flavour-tagging classification probabilities~\cite{FTAG-2019-07} (i.e.\ the
discriminant outputs) for \bjets.
A high-purity sample of \bjets is obtained by selecting events with an electron,
a muon, and exactly two jets, which yields a population of jets dominated by
\ttbar production and decay into \emuvvbb, following the optimized
requirements of previous flavour-tagging calibrations~\cite{FTAG-2018-01}.
Optimal calibration maps between the simulated and observed
distributions of \bjet classification probabilities are derived.

The paper is laid out as follows.
Section~\ref{sec:detector} introduces the ATLAS detector apparatus and
coordinate system, and the calibration methodology is described in
Section~\ref{sec:methods}.
Section~\ref{sec:samples} outlines the simulated and collected data samples
used in this study.
The data analysis and the estimation of uncertainties are described
in Sections~\ref{sec:analysis} and~\ref{sec:uncertainties}, respectively,
building largely on the work presented previously in Ref.~\cite{FTAG-2018-01}.
Section~\ref{sec:nonb_calibration} describes the calibration of the \dloner
distribution for background jets, and Section~\ref{sec:results} shows the
extracted calibration maps and data--simulation comparisons for \bjets.

% End of text imported from the .//intro.tex input file

% The next lines are included from the .//detector.tex input file
\section{ATLAS detector}
\label{sec:detector}

\newcommand{\AtlasCoordFootnote}{%
ATLAS uses a right-handed coordinate system with its origin at the nominal
interaction point (IP) in the centre of the detector and the \(z\)-axis along
the beam pipe.
The \(x\)-axis points from the IP to the centre of the LHC ring, and the
\(y\)-axis points upwards.
Polar coordinates \((r,\phi)\) are used in the transverse plane, \(\phi\) being
the azimuthal angle around the \(z\)-axis.
The pseudorapidity is defined in terms of the polar angle \(\theta\) as \(\eta =
-\ln \tan(\theta/2)\) and is equal to the rapidity $ y = (1/2) \ln[(E + p_z)/(E - p_z)] $ in the relativistic limit.
Angular distance is measured in units of \(\Delta R \equiv \sqrt{(\Delta y)^{2}
+ (\Delta\phi)^{2}}\).}

The ATLAS detector~\cite{PERF-2007-01} at the LHC covers nearly the entire solid angle around the collision point.\footnote{\AtlasCoordFootnote}
It consists of an inner tracking detector surrounded by a thin superconducting solenoid, electromagnetic and hadronic calorimeters,
and a muon spectrometer incorporating three large superconducting air-core toroidal magnets.

The inner-detector system (ID) is immersed in a \qty{2}{\tesla} axial magnetic field
and provides charged-particle tracking in the range \(|\eta| < 2.5\).
The high-granularity silicon pixel detector covers the vertex region and typically provides four measurements per track,
the first hit generally being in the insertable B-layer installed before Run~2~\cite{ATLAS-TDR-19,PIX-2018-001}.
It is followed by the silicon microstrip tracker, which usually provides eight measurements per track.
These silicon detectors are complemented by the transition radiation tracker (TRT),
which enables radially extended track reconstruction up to \(|\eta| = 2.0\).
The TRT also provides electron identification information
based on the fraction of hits (typically 30 in total) above a higher energy-deposit threshold corresponding to transition radiation.

The calorimeter system covers the pseudorapidity range \(|\eta| < 4.9\).
Within the region \(|\eta|< 3.2\), electromagnetic calorimetry is provided by barrel and
endcap high-granularity lead/liquid-argon (LAr) calorimeters,
with an additional thin LAr presampler covering \(|\eta| < 1.8\)
to correct for energy loss in material upstream of the calorimeters.
Hadronic calorimetry is provided by the steel/scintillator-tile calorimeter,
segmented into three barrel structures within \(|\eta| < 1.7\), and two copper/LAr hadronic endcap calorimeters.
The solid angle coverage is completed with forward copper/LAr and tungsten/LAr calorimeter modules
optimized for electromagnetic and hadronic energy measurements respectively.

The muon spectrometer comprises separate trigger and
high-precision tracking chambers measuring the deflection of muons in a magnetic field generated by the superconducting air-core toroidal magnets.
The field integral of the toroids ranges between \num{2.0} and \qty{6.0}{\tesla\metre}
across most of the spectrometer.
Three layers of precision chambers, each consisting of layers of monitored drift tubes, cover the region \(|\eta| < 2.7\),
complemented by cathode-strip chambers in the forward region, where the background is highest.
The muon trigger system covers the range \(|\eta| < 2.4\) with resistive-plate chambers in the barrel, and thin-gap chambers in the endcap regions.

The luminosity is measured mainly by the LUCID-2~\cite{LUCID2} detector, which records Cherenkov light produced in the quartz windows of photomultipliers located close to the beam pipe.

Events are selected by the first-level trigger system implemented in custom hardware,
followed by selections made by algorithms implemented in software in the high-level trigger~\cite{TRIG-2016-01}.
The first-level trigger accepts events from the \qty{40}{\MHz} bunch crossings at a rate below \qty{100}{\kHz},
which the high-level trigger reduces in order to record complete events to disk at about \qty{1}{\kHz}.

A software suite~\cite{SOFT-2022-02} is used in data simulation, in the reconstruction
and analysis of real and simulated data, in detector operations, and in the trigger and data acquisition
systems of the experiment.

% End of text imported from the .//detector.tex input file

% The next lines are included from the .//methods.tex input file
\section{Methodology} \label{sec:methods}

The aim of this study is to derive a \pt-dependent mapping \Tpt that moves the
flavour-tagging outputs of simulated jets such that, after this movement, the
probability density of the simulation matches that in data.
The amount that a jet's flavour-tagging score moves can be used to define a
distance metric between the original simulated distribution and the corrected
distribution.
In general, there are many mappings that will achieve closure between simulation
and data, but an optimal mapping is one that moves the simulation as little as
possible.
In the following section, these concepts are introduced more formally, and the
methods used to find the optimal solution are described.

\subsection*{Statement of the problem}

The goal here is to derive $\Tpt : \Qvec \to \Qvec$, where \Qvec is a space of
classifier outputs, \pu, \pc, and \pb, described in
Refs.~\cite{FTAG-2018-01, FTAG-2019-07, ATL-PHYS-PUB-2022-027}, such that after
the application of $T_{\pt}$ the probability density function (\enquote{density}) in
the corrected simulation is equivalent to that measured in data,
$\Tpthash \psim(\qvec | \pt) \approx \pdata(\qvec | \pt)$.
Throughout this paper, \qvec is an element of \Qvec, and the notation $M_\#
p(X)$ denotes the pushforward~\cite{villani2008optimal} of a distribution $p$
through a map $M$; i.e.\ $M$ acts on elements $x \in X$ while $M_\#$ operates
on probability distributions over $X$.
If samples are taken from some density $x \sim p(X)$, then samples taken from
the pushforward of $p$ through $M$ can be built by applying $M$ to each
sample, i.e.\ $M x \sim M_\# p(X)$.

In general, there are many mappings that achieve the desired closure, but an
\emph{optimal} transport map is one that, given a distance metric $c$ defined
over \Qvec, minimizes the total distance that the simulation must move,
\begin{equation*}
\Thatpt
= \arg \inf_{T_{\pt}}
\int_{\qvec \in \Qvec} c(\qvec , \Tpt~\qvec)
p_\mathrm{sim}(\qvec | \pt) d \qvec; \ \textrm{such that} \
\
(\Tpt)_\# \, \psim(\qvec | \pt) \approx \pdata(\qvec | \pt).
\end{equation*}
For Euclidean distance metrics of the form
$c^2(\vec x , \vec y) = (\vec x - \vec y)^2$, the optimal transport map is known to be
unique~\cite{villani2008optimal}.

\subsubsection*{Optimality property of maps}

The simulated samples used to analyse ATLAS data are based on a wealth of
knowledge of physical processes from a wide variety of sources accumulated over
many years.
Carefully tuned aspects of the simulation particularly relevant to jet
flavour tagging include high-fidelity approximations of soft and collinear
emissions within jets of partons as well as the subsequent hadronization
process~\cite{Sjostrand:2007gs, Bellm:2015jjp, Gleisberg:2008ta, Skands:2014pea,
ATL-PHYS-PUB-2014-021}, dedicated studies of the heavy-quark fragmentation
process~\cite{DELPHI:2011aa, Abbiendi:2002vt, Heister:2001jg, Abe:2002iq,
TOPQ-2017-19,STDM-2018-52} and heavy-hadron
decays~\cite{Lange:2001uf,ATL-PHYS-PUB-2022-035}, and a detailed understanding
of particle interactions with detector
components~\cite{Agostinelli:2002hh,SOFT-2010-01}.
The calibration procedure is an exercise in fine-tuning the simulation to
account for possible inconsistencies with observation at a detailed level;
deriving a transport map that minimally alters the output of the simulation
ensures that upstream predictions, which have been compared with (and tuned to
agree with) previous measurements, are corrected as little as possible.

\subsubsection*{Choice of calibration space}
This calibration is derived for the quantities $q_i = \logit~p_i$, defined
per component as
\begin{equation*}
\logit~p_i = \log~\frac{p_i}{1 - p_i + \epsilon},
\end{equation*}
where $\epsilon=10^{-10}$ is introduced to prevent divisions by zero, and the
index $i$ runs over the three flavours used in the ATLAS jet classifiers: $u$,
$c$, and $b$.
These \qvec are closely related to the likelihood ratios used to discriminate
between signal and background jets, as introduced in Eqs.~(\ref{eq:Db})
and~(\ref{eq:Dc}), and their components can take any real-numbered values.
Furthermore, up to a normalization constant the left inverse of $\logit~p_i$ is
the \sigmoid activation function\footnote{The \sigmoid activation function is
given by $p_i^\prime = 1/(1+\mathrm{e}^{-q_i})$.
To ensure that the probabilities sum to unity, the sigmoid transformation is
normalized using $p_i = p_i^\prime/\sum_i p_i^\prime$.}:
it is always possible to reconstruct the unique jet-flavour probabilities
$(p_b, p_c, p_u)$ from \qvec, meaning that a calibration of \qvec amounts to a
calibration of $(p_b, p_c, p_u)$.
Unlike the space of $(p_b, p_c, p_u)$, the \Qvec space covers $\mathbb{R}^3$,
and in this calibration, the \Qvec space is treated as Euclidean with the usual
quadratic distance between \qvec and $\vec q^\prime$,
$\sqrt{(\qvec - \vec q^\prime)^2}$.
Using this Euclidean distance means there is a unique optimal transport mapping.

For jet flavour tagging, ATLAS calibrations are parameterized by a jet's flavour
(i.e.\ calibrations are derived separately for \bjets, \cjets, and light-flavour
jets) and \pt ~\cite{FTAG-2018-01,FTAG-2019-02,FTAG-2020-08} in order to ensure
applicability across a wide range of analyses of LHC data.
The method introduced here follows that prescription and provides corrections
for \bjets as a function of \pt.
This procedure could easily be extended to calibrations parameterized by
additional jet properties that traditionally have been averaged over, which may
lead to better generalization properties in future work;
Ref.~\cite{Gambhir:2022dut} is recommended for a broader discussion of
calibration universality.

\subsubsection*{Deriving approximate calibration maps}

For metric spaces with a quadratic distance metric, the optimal transport map
between two continuous distributions is the gradient of a scalar convex
function~\cite{villani2008optimal}.
Partially input-convex neural networks (PICNNs) are flexible and generic
approximators of functions that are convex in a subset of their
arguments (i.e.\ \enquote{partially convex})~\cite{amos2017input}, and it is possible
to derive optimal transport maps for metric spaces that have a Euclidean transport
cost by using PICNNs~\cite{pmlr-v119-makkuva20a}.

Following this prescription, the optimal transport map between \pdata and
$\psim^\prime$ is derived via the loss function
\begin{equation*}
\mathcal L(f, g)
=
\frac{1}{N_\mathrm{data}} \sum_{\qvec, \pt \sim \pdata} f(\qvec , \pt)
+ \frac{1}{N_\mathrm{sim}} \sum_{\qvec, \pt \sim \psim^\prime}
\qvec \cdot \gradQ~g(\qvec , \pt) - f(\gradQ~g(\qvec , \pt), \pt),
\end{equation*}
where $f$ and $g$ are convex functions of \qvec but not necessarily of \pt,
and $N_\mathrm{data}$ and $N_\mathrm{sim}$ are respectively the number of
observed events and the number of simulated events, and \gradQ indicates the gradient with
respect to \qvec.
After minimizing $\mathcal L$ with respect to $f$ and maximizing it with respect to
$g$, the optimal choices $\hat f$ and $\hat g$ have the property that
\begin{equation*}
(\gradQ~\hat g)_\#\, \psim(\qvec | \pt) \approx \pdata(\qvec | \pt)
\end{equation*}
and
\begin{equation*}
(\gradQ~\hat f)_\#\, \pdata(\qvec | \pt) \approx \psim(\qvec | \pt),
\end{equation*}
i.e.\ the gradient of $\hat g$ pushes \psim forward to \pdata, and the gradient
of $\hat f$ pushes \pdata forward to \psim~\cite{Pollard:2021fqv}.
The desired calibration transport maps are then given by
$\Thatpt(\qvec) = \gradQ~\hat g(\qvec, \pt)$.

\subsubsection*{Background contamination and uncertainties}

While a purposely selected population of jets may have a high fraction of \bjets, there
is always some contamination from jets with other flavours.
In this study, the salient information needed to correctly account for this
background contamination is the fraction of \bjets at each point in $\Qvec
\times \pt$ space, $p(\mathrm{sig} | \qvec,~\pt)$.

A smooth signal jet fraction as a function of \pt, $\fsig = p(\mathrm{sig} |
\pt)$, is obtained via neural density ratio estimation
techniques~\cite{Cranmer:2015bka}:\ a classifier is trained to approximate the
signal-to-background ratio as a function of \pt,
$p(\mathrm{sig} | \pt) / p(\mathrm{bkg} | \pt)$, and then
$p(\mathrm{sig} | \pt)$ is solved for by setting
$p(\mathrm{bkg} | \pt) = 1 - p(\mathrm{sig} | \pt)$.

In order for the PICNN minimization procedure to result in the correct
mapping conditional on \pt, it is necessary for the \pt distributions in data
and simulation to be approximately equal.
To ensure that these are indeed identical, the conditional density
$\psim(\qvec | \pt)$ is approximated using conditional normalizing
flows~\cite{9089305}, which are neural solvers that can approximate continuous
multivariate distributions given samples from the target density.
Separate normalizing flows are used to approximate \psigpt and \pbkgpt.
Combined with \fsig, these yield the full prediction for
$p^\prime_\mathrm{sim}(\qvec,~\pt)$, which has the property
$p^\prime_\mathrm{sim}(\pt) = p_\mathrm{data}(\pt)$, defined as
\begin{equation}
p^\prime_\mathrm{sim}(\qvec,~\pt)
\equiv (\fsig\psigpt + (1 - \fsig)\pbkgpt)~\pdata(\pt).
\label{eq:pprime}
\end{equation}
Training batches for simulation are built by sampling from
$p^\prime_\mathrm{sim}$, which results in the desired, \pt-corrected
distribution~\cite{Algren:2023qnb}.
Given the data \pt distribution, $p^\prime_\mathrm{sim}$ can be separated into a
signal distribution $\psig^\prime$ and a background distribution $\pbkg^\prime$,
defined as:
\begin{align*}
\psig^\prime(\qvec,~\pt) \propto \psigpt~\fsig~\pdata(\pt) \\
\text{and~~~} \pbkg^\prime(\qvec,~\pt) \propto \pbkgpt~(1 - \fsig)~\pdata(\pt).
\label{eq:pprime_sig_bkg}
\end{align*}
Non-\bjet samples are effectively subtracted during the determination of
the calibration maps by leaving them in place for part of the training
procedure, following
\begin{equation*}
\Thatpt~\qvec = \begin{cases}
\gradQ~\hat g(\qvec) \quad &\text{for \bjets} \\
\qvec \quad &\text{for non-\bjets} \\
\end{cases}.
\end{equation*}
A more thorough treatment of this mechanism can be found in
Ref.~\cite{Pollard:2021fqv}.

Uncertainties in the predicted \bjet fractions, \fsig, and the
non-\bjet shape, \pbkgpt, are estimated by deriving a family of transport maps
for a range of variations of each of them.
For the purpose of illustrating the total uncertainty of the corrected
simulation, each variation is assumed to correspond to a $1\sigma$ point
of a normally distributed continuous parameter, and these parameters for
different variations are treated as fully uncorrelated.
Since the transport map for each variation is derived separately, a more careful
uncertainty treatment is possible in analyses that depend on this calibration by
using the full family of maps.
Systematic uncertainties are assessed by altering the simulated prediction in
various ways; these are described further in Section~\ref{sec:uncertainties}.
In this study, the flavour fraction \fsig is taken from simulation,
while in previous work these were constrained using background-enriched
sidebands.
The uncertainty in the shape of the non-\bjet background, \pbkgpt, is determined
via a \bjet-depleted control region and laid out in
Section~\ref{sec:nonb_calibration}.
Statistical uncertainties coming from the limited size of the observed data set
are estimated via bootstraps~\cite{efron1994introduction}.

\begin{figure}
\centering
\includegraphics[width=0.9\textwidth]{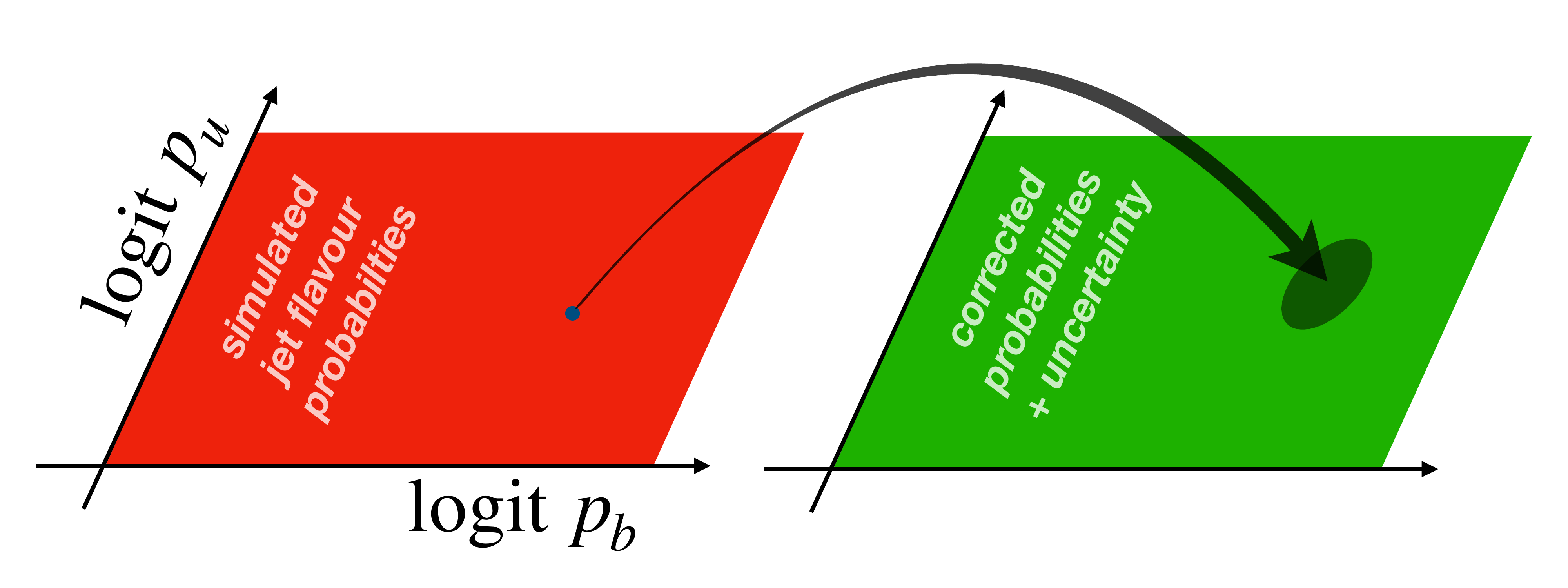}
\caption{
\label{fig:calibration-diagram}
Schematic of the optimal calibration maps derived in this study.
Only two of the three calibrated log-probabilities are shown.
The darkened region around the point of the arrow indicates that there is
some uncertainty in the transport mapping.
The transport maps are deterministic, so the uncertainty is calculated by
deriving transport maps between alternative simulations and the data.
}
\end{figure}

Figure~\ref{fig:calibration-diagram} shows a schematic of the calibration
process in the context of jet flavour tagging.

% End of text imported from the .//methods.tex input file

% The next lines are included from the .//samples.tex input file
\section{Samples of observed and simulated events}
\label{sec:samples}

This analysis uses the LHC $pp$ collision data recorded by the ATLAS detector
during the years 2015--2018, with a collision centre-of-mass energy of
$\sqrt{s} = 13$~TeV.
The total integrated luminosity is $140.1 \pm
1.2~\mathrm{fb}^{-1}$~\cite{DAPR-2021-01}.
Only collision events recorded during normal running of the entire ATLAS
detector are considered in this analysis~\cite{DAPR-2018-01}.
As detailed in Section~\ref{sec:analysis}, these data events were collected by
requiring at least one high-\pt electron or muon to be identified by the ATLAS
trigger system~\cite{TRIG-2016-01,TRIG-2018-05,TRIG-2018-01}.

A number of simulated event samples, corresponding to a Monte Carlo integration
of the relevant Standard Model calculations~\cite{Buckley:2011ms}, are used both
to predict the processes that populate the analysis selection region and to
assess uncertainties in the calibration of the simulation.
The production of \ttbar events was modelled using the
\POWHEGBOX[v2]~\cite{Frixione:2007nw,Nason:2004rx,Frixione:2007vw,Alioli:2010xd}
generator at NLO with the \NNPDF[3.0nlo]~\cite{Ball:2014uwa} set of parton distribution
functions~(PDFs) and the \hdamp parameter\footnote{The
\hdamp parameter is a resummation damping factor and one of the
parameters that controls the matching of \POWHEG matrix elements to
the parton shower and thus effectively regulates the
high-\pT radiation against which the \ttbar system recoils.} set
to 1.5\,\mtop~\cite{ATL-PHYS-PUB-2016-020}.
The events were interfaced to \PYTHIA[8.230]~\cite{Sjostrand:2014zea} to model
the parton shower, hadronization, and underlying event, with parameter values set
according to the A14 set of tuned parameters (tune)~\cite{ATL-PHYS-PUB-2014-021}
and using the \NNPDF[2.3lo] set of PDFs~\cite{Ball:2012cx}.
The decays of bottom and charm hadrons were performed by
\EVTGEN[1.6.0]~\cite{Lange:2001uf}.

Alternative samples were generated to assess the uncertainty in the \ttbar
prediction.
The uncertainty due to initial-state radiation (ISR) is estimated by
simultaneously varying the \hdamp parameter and the renormalization and
factorization scales (\muR and \muF), and choosing the Var3c up/down variants
of the A14 tune.
Final-state radiation (FSR) uncertainties are determined by scaling the
parton-shower time-like shower scale \muR in \PYTHIA up and down by a factor of
two.
The impact of using a different parton shower and hadronization model is
evaluated by comparing the nominal \ttbar sample with another event sample
also produced with the \POWHEGBOX[v2] generator using the \NNPDF[3.0nlo] PDF
set.
Events in the latter sample were interfaced with
\HERWIG[7.04]~\cite{Bahr:2008pv,Bellm:2015jjp}, which used the H7UE set of tuned
parameters~\cite{Bellm:2015jjp} and the \MMHT[lo] PDF
set~\cite{Harland-Lang:2014zoa}.
The decays of bottom and charm hadrons were again simulated using the
\EVTGEN[1.6.0] program.

The associated production of a top quark and \(W\) boson (\(tW\)) was
modelled by the \POWHEGBOX[v2] generator at NLO in QCD using the five-flavour
scheme and the \NNPDF[3.0nlo] set of PDFs.
The diagram removal scheme~\cite{Frixione:2008yi} was used to remove
interference and overlap with \ttbar production.
The events were interfaced to \PYTHIA[8.230], which used the A14
tune and the \NNPDF[2.3lo] PDFs~\cite{Ball:2012cx}.

The production of $W$+jets and \(Z+\)jets was simulated with the
\SHERPA[2.2.1]~\cite{Bothmann:2019yzt} generator using next-to-leading-order
(NLO) matrix elements (ME) for up to two partons, and leading-order (LO) matrix
elements for up to four partons, calculated with the
Comix~\cite{Gleisberg:2008fv} and
\OPENLOOPS~\cite{Buccioni:2019sur,Cascioli:2011va,Denner:2016kdg} libraries.
They were matched with the \SHERPA parton shower~\cite{Schumann:2007mg} using
the \MEPSatNLO
prescription~\cite{Hoeche:2011fd,Hoeche:2012yf,Catani:2001cc,Hoeche:2009rj}
and the set of tuned parameters developed by the \SHERPA authors.
The \NNPDF[3.0nnlo] set of PDFs was used and the samples were normalized to a
next-to-next-to-leading-order (NNLO) prediction~\cite{Anastasiou:2003ds}.

The \POWHEGBOX[v2] generator was used to simulate the \(WW\), \(WZ\) and
\(ZZ\)~\cite{Nason:2013ydw} production processes at NLO accuracy in \alphas.
The effect of singly resonant amplitudes and interference effects due to
\(Z/\gamma^*\) and same-flavour lepton combinations in the final state were
included, where appropriate.
Interference effects between \(WW\) and \(ZZ\) for same-flavour charged leptons
and neutrinos were ignored.
Events were interfaced to \PYTHIA[8.210]~\cite{Sjostrand:2014zea}
for the modelling of the parton shower, hadronization, and underlying
event, with parameters set according to the \AZNLO tune~\cite{STDM-2012-23}.
The \CT[10] PDF set~\cite{Lai:2010vv} was used for the hard-scattering
processes, whereas the \CTEQ[6L1] PDF set~\cite{Pumplin:2002vw} was used for the
parton shower.
\EVTGEN[1.2.0]~\cite{Lange:2001uf} was used to decay bottom and charm hadrons.

The effect of multiple interactions in the same and neighbouring bunch
crossings (\pileup) was modelled by overlaying the simulated hard-scattering
event with inelastic proton--proton (\(pp\)) events generated with
\PYTHIA[8.186]~\cite{Sjostrand:2007gs} using the \NNPDF[2.3lo] PDFs
and the A3 tune~\cite{ATL-PHYS-PUB-2016-017}.

% End of text imported from the .//samples.tex input file

% The next lines are included from the .//analysis.tex input file
\section{Analysis}
\label{sec:analysis}

A precise calibration of the \dloner flavour probabilities for \bjets requires an
event selection that yields a relatively pure \bjet sample and for which the
backgrounds are well understood.
In Section~\ref{sec:selection}, the object and event selection used to achieve
this are described; Section~\ref{sec:dataMC} shows the predicted event
kinematics compared with observed data after this selection.
Section~\ref{sec:fsig} discusses the neural estimates of \fsig and \psim built
from the MC predictions and the level of closure observed.
Finally, the PICNN architecture used to perform the calibration is
explained in Section~\ref{sec:PICNN}.

\subsection{Event selection}
\label{sec:selection}

A population of jets enriched in \bjets is obtained by applying a selection
targeting events from the dilepton \ttbar (\emuvvbb) process, which
provides a high-cross-section source of \bjets with relatively low contamination
from backgrounds.
The event selection follows earlier ATLAS calibrations of flavour tagging for
\bjets.
The selection is described briefly below, but a more detailed description may be
found in Ref.~\cite{FTAG-2018-01}.
The major change in the event selection is a requirement applied in the new
analysis on the invariant mass of lepton--jet pairs, $\mlj < 175~\GeV$;
previously the region $\mlj \ge 175~\GeV$ was used to define a
background-enriched control region to constrain jet flavour fractions \emph{in
situ}, but such events are vetoed in this analysis.

\subsubsection{Tracking and vertexing}

The ID is used to reconstruct tracks corresponding to charged
particles~\cite{PERF-2015-01,IDTR-2019-05}.
Primary vertices (PVs) are reconstructed by identifying points along the
LHC beamline consistent with a high density of charged particles, based on
reconstructed tracks~\cite{PERF-2015-08}.
At least one reconstructed PV is required in each event, and the PV with the
highest sum of the squared constituent track transverse momenta is considered
the \enquote{hard-scatter} vertex.

\subsubsection{Charged leptons}

Electrons and muons are used to preferentially select events with prompt
leptons.
Electron candidates are built from spatially compatible isolated electromagnetic
energy clusters and ID tracks~\cite{EGAM-2021-01}; they are required to have $\ET >
28$~GeV and $|\eta| < 2.47$ after their energies are corrected using
calorimeter shower shape information.
Candidates in the calorimeter barrel--endcap transition region
($1.37 < |\eta| < 1.52$) are removed to avoid large uncertainties in the
identification efficiency.
\Tight likelihood and isolation requirements are imposed to reduce the
contamination from non-prompt and fake electron candidates further.
Efficiency scale factors and energy calibration corrections derived in $Z\to ee$
events are applied to ensure the electron reconstruction and identification
performance in data is reflected in simulation.

Muon candidates are ID tracks geometrically matched to tracks in the muon
spectrometer~\cite{MUON-2018-03}. They must have $\pt > 28$~GeV and
$|\eta| < 2.5$, and a \Tight identification requirement is applied.
Furthermore, muons are required to be isolated from surrounding hadronic
activity in order to remove contamination from hadronic jets and heavy-flavour
decays, using the \Tight isolation criterion.
Muon identification efficiency and momentum response corrections are derived in
$Z \to \mu\mu$ events, and the simulation is adjusted to ensure the predicted
performance matches that observed in data~\cite{MUON-2022-01}.

\subsubsection{Hadronic jets}

Jets of hadronic particles are built from particle-flow candidates, which
use tracking and calorimetry information to ensure a good jet constituent
response for both the charged and neutral particles~\cite{PERF-2015-09}.
The \antikt combinatorial reconstruction algorithm is used with a radius
parameter $R = 0.4$~\cite{Cacciari:2008gp}, as implemented in
\Fastjet~\cite{Fastjet}.
The reconstructed jet transverse momenta are corrected to the particle-level jet
\pt using simulation~\cite{JETM-2018-05}.
The jet response is measured in multijet, $Z$+jets, and $\gamma$+jets data, and
the simulation is corrected to reflect the response in data events.
Jets are required to have reconstructed transverse momenta $\pt > 20$~GeV; they
must also fall within the fiducial volume of the inner tracker, $|\eta| < 2.5$.
A \enquote{jet vertex tagger} (JVT) is used to reduce the fraction of jets with large
energy fractions from interaction other than the hard-scatter collision
(\enquote{\pileup{}} interactions)~\cite{ATLAS-CONF-2014-018}.
The efficiency of the JVT selection in simulated jets from the hard-scatter
collision is 92\%, and for jets with $\pt > 60$~GeV the fraction of jets from
\pileup interactions is small enough that no JVT requirement is imposed.

\subsubsection{Jet flavour tagging}
For flavour tagging, reconstructed tracks with $\pt > 500$~MeV are geometrically
matched to hadronic jets using the \DeltaReta
measure.\footnote{$\DeltaReta \equiv \sqrt{(\Delta \eta)^2 + (\Delta \phi)^2}$}
The \DeltaReta matching requirement varies with jet \pt since the opening angle
of \bhadron decay products is wider for low-\pt \bhadrons than for those with
high \pt.
The matching criterion is $\DeltaReta(\mathrm{track,\ jet}) < 0.45$ at jet
$\pT = 20$~GeV, and decreases asymptotically
to $\DeltaReta < 0.25$ for very high \pt jets; tracks are assigned to
the closest jet in the event that more than one valid match is found.
A continuous calibration for the \dloner flavour-tagging
algorithm~\cite{FTAG-2019-07}, which uses a variety of low-level
inputs built from charged-particle tracks, is derived in this study.
The $b$- and \cjet identification discriminants, denoted by \dlonerb and \dlonerc,
are the $D_b$ and $D_c$ log-likelihood ratios respectively introduced in
Eqs.~(\ref{eq:Db}) and~(\ref{eq:Dc}) for the \dloner tagging algorithm; $f_c$ is set
to 0.018 for \dlonerb, while $f_b = 0.3$ for \dlonerc.

In simulation, jets are designated as $b$-jets, $c$-jets, $\tau$-jets, or light-flavour jets
depending on the nearby final-state heavy-flavour hadrons and \tauleptons.
Electroweakly decaying $b$- and $c$-hadrons and \tauleptons with $\pt > 5$~GeV
are matched to a reconstructed jet if they are within $\DeltaReta = 0.3$ of
the jet axis; priority is given to the nearest jet if more than
one match is found.
If a \bhadron is matched to a jet, the jet is labelled a \bjet; if no
\bhadron is matched, the jet is subsequently checked for charm hadrons and then
\tauleptons, and given the \cjet or $\tau$-jet designation.
If no heavy hadron or \taulepton is matched to a jet, it is considered a
light-flavour jet.
No special distinction is made for jets with more than one heavy hadron of the
same flavour, e.g.\ $bb$ or $cc$ jets; similarly, no attempt is made to separate
light-flavour jets based on their quark or gluon partonic origin.
It should also be noted that the fraction of $\tau$-jets after the following
event selection is much smaller than the uncertainty in the light-flavour jet fraction.
For the purposes of this calibration, $\tau$-jets, $c$-jets, and light-flavour jets are treated
coherently as a non-\bjet background population.

\subsubsection{Event requirements}

To select a sample enriched in $\ttbar \to \emuvvbb$ production, events are
required to contain exactly one selected electron candidate and exactly one
selected muon candidate.
In order to avoid double counting the same object across multiple signatures,
an object overlap-removal is applied among electrons, muons, and jets, following
the prescription in Section~6.1 of Ref.~\cite{FTAG-2018-01}.
After this, events with fewer than two prompt charged leptons make up a
negligible fraction of the total rate, and the majority contribution of
Drell--Yan $Z\to\ell\ell$ production is via the decays of two \tauleptons to one
electron and one muon.
At least one selected lepton candidate must match a corresponding object in the
ATLAS trigger system~\cite{TRIG-2018-05,TRIG-2018-01}, and efficiency scale
factors are applied to correct the trigger efficiency in simulation to
that in data.

In order to further improve the \bjet purity, additional selections are imposed
to reduce
the contribution from \ttbar{}+jets production, following the procedure in
Ref.~\cite{FTAG-2018-01}.
Events containing more than two selected jets are removed.
Leptons and jets are then paired by minimizing the sum of the squared
lepton--jet invariant masses over possible pairings:
$\min_{\alpha\beta}~(m_{j_1,\ell_\alpha}^2 +
m_{j_2,\ell_\beta}^2)$,
where $\alpha$ and $\beta$ index the two reconstructed leptons, and $j_1$ and
$j_2$ are the two reconstructed jets in the event.
Only events where both resulting lepton--jet pair invariant masses are less than
175~GeV, i.e.\ consistent with two top-quark decays, are considered.

\subsection{Comparisons of data and simulation}
\label{sec:dataMC}

\begin{figure}
\centering
\subfigure[]{\includegraphics[width=0.45\textwidth]{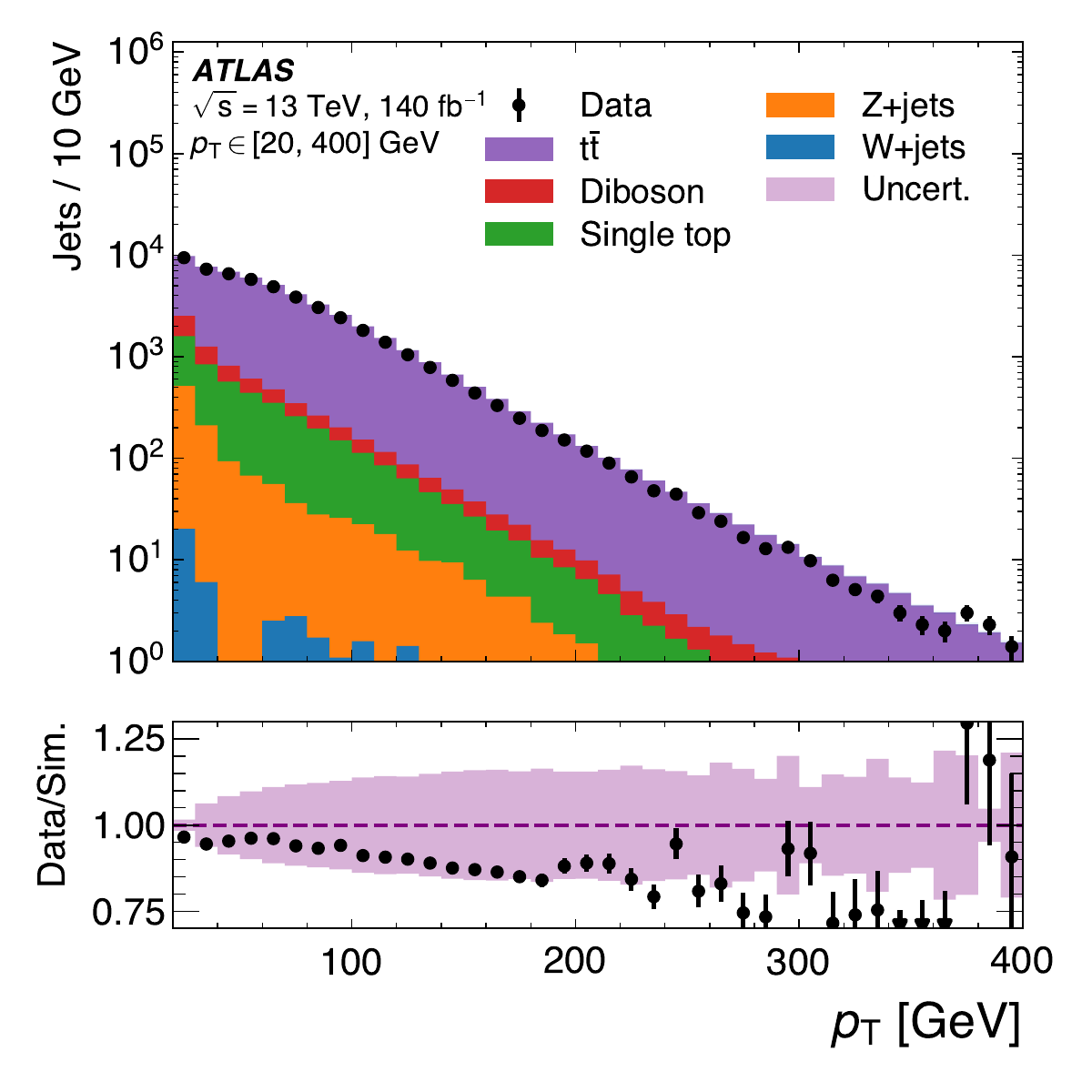}}
\subfigure[]{\includegraphics[width=0.45\textwidth]{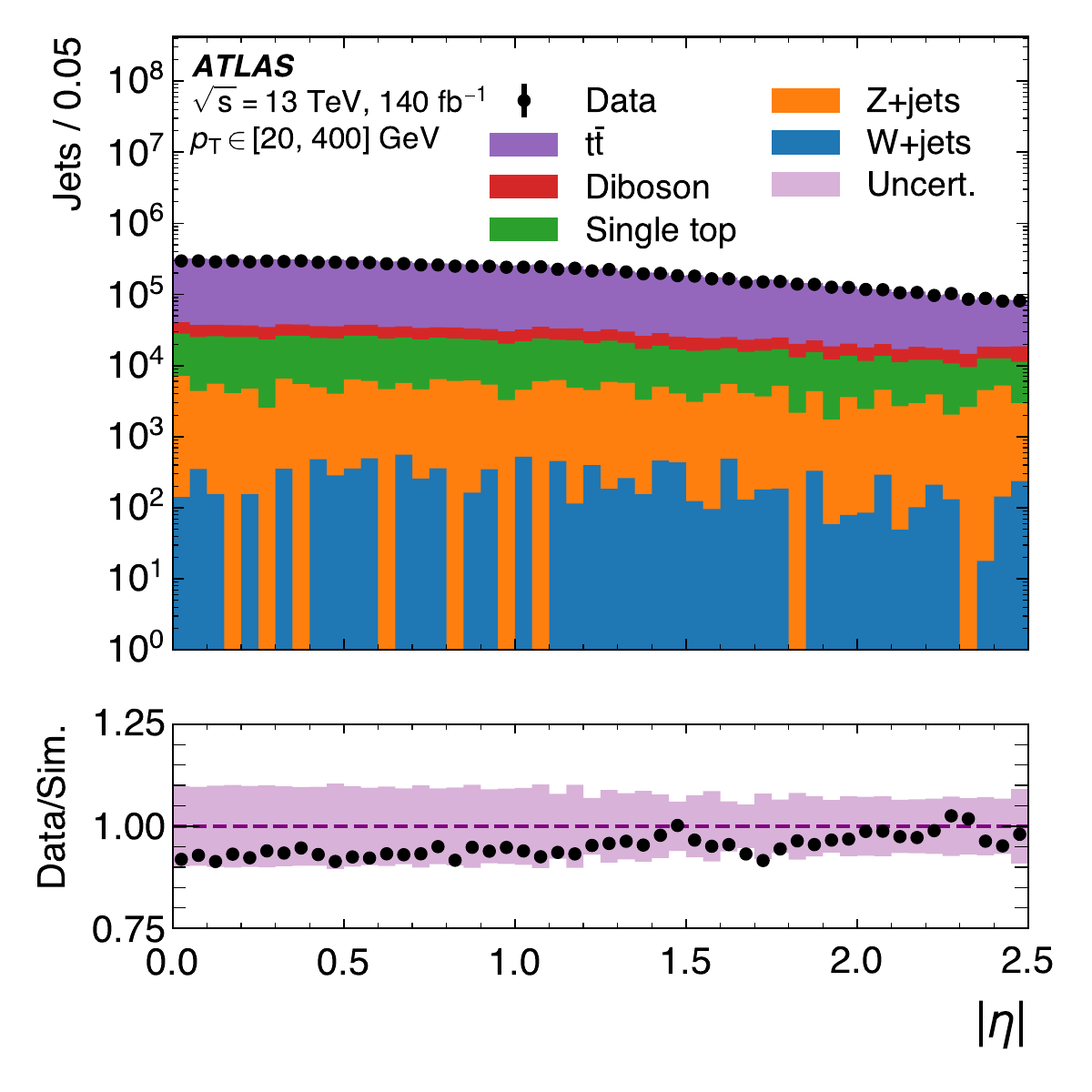}} \\
\subfigure[]{\includegraphics[width=0.45\textwidth]{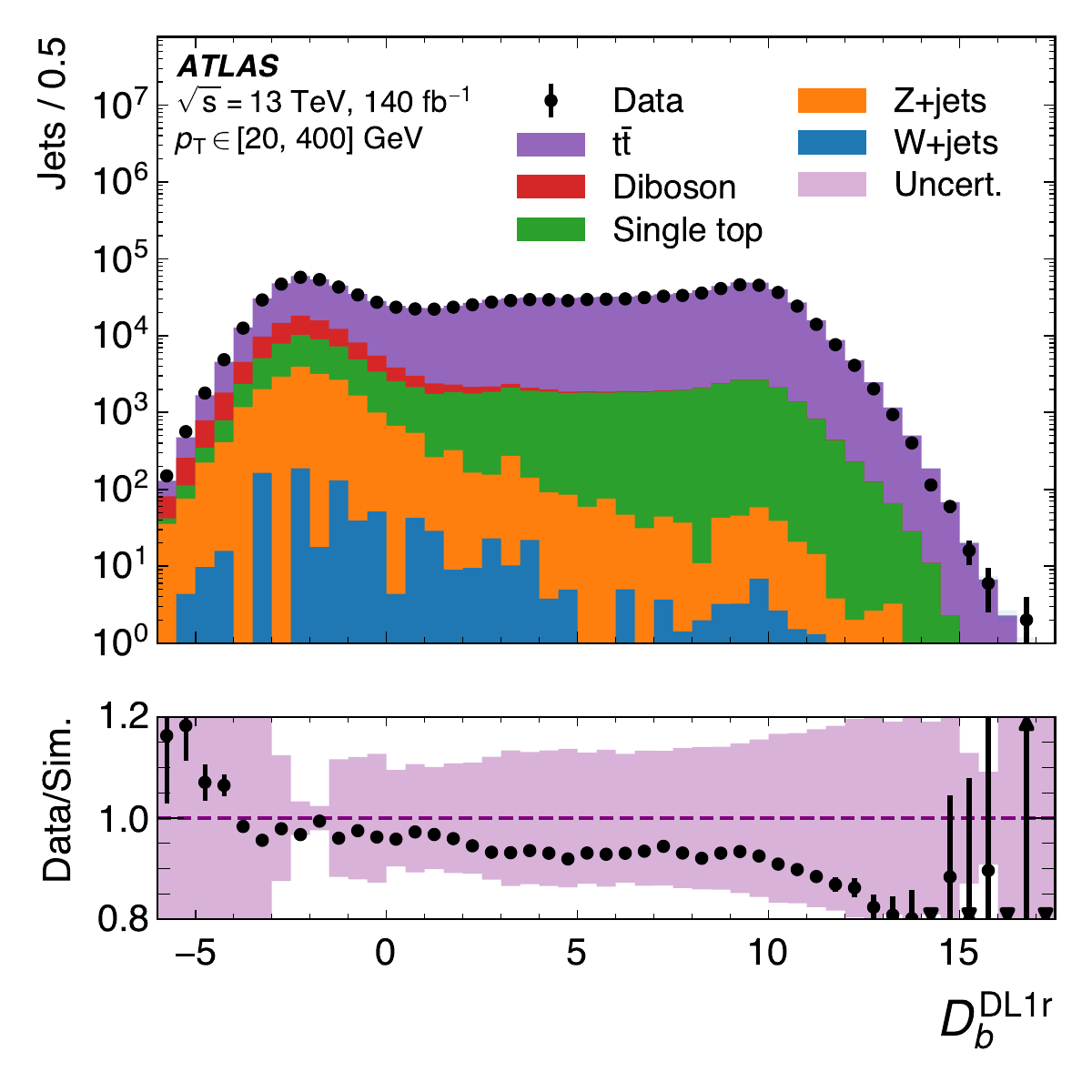}}
\subfigure[]{\includegraphics[width=0.45\textwidth]{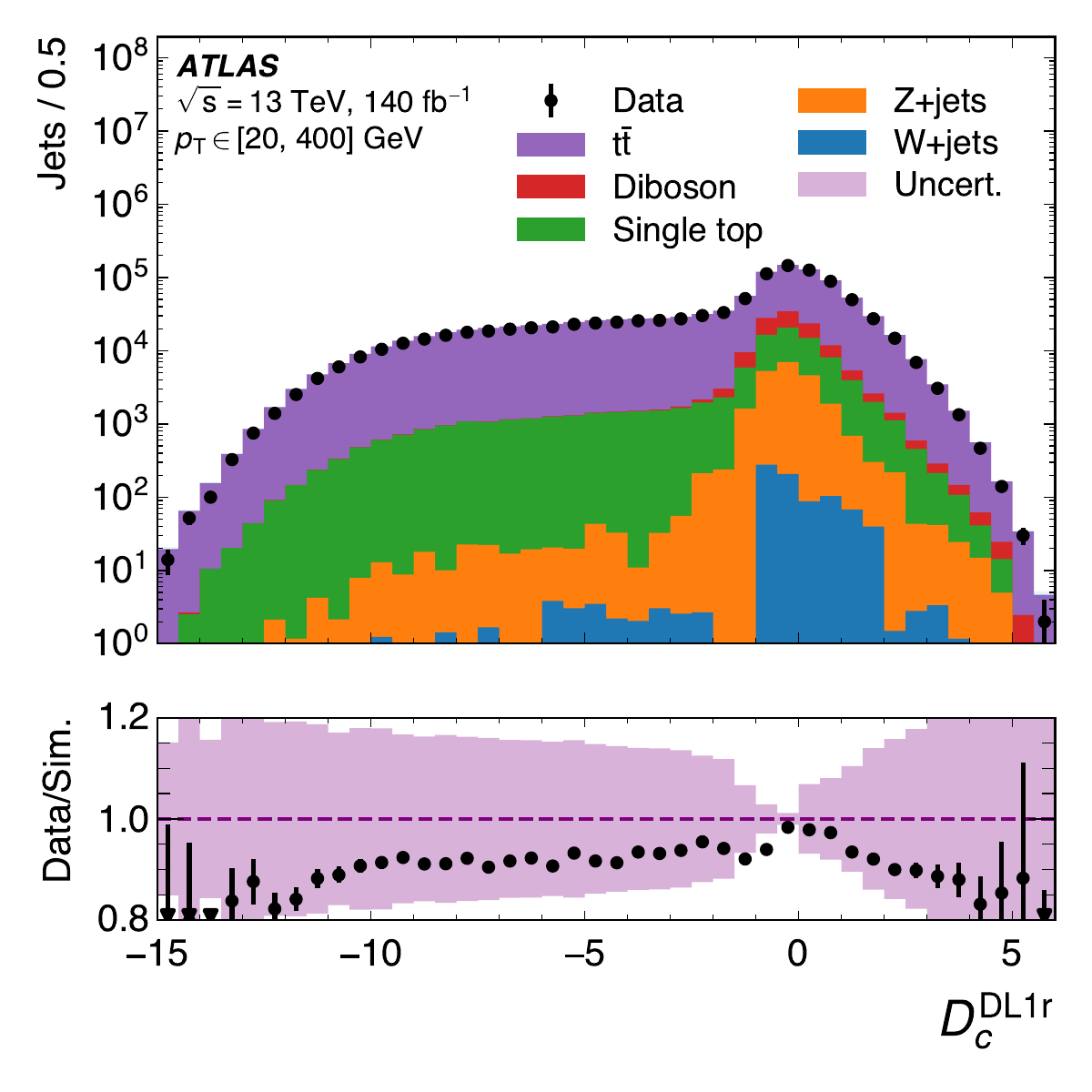}}
\caption{
Comparison of jet observables between the simulation's predictions and observed
data, with the predictions separated by process:\ the selected jets'
(a) \pt, (b) $|\eta|$, (c) \dlonerb and (d) \dlonerc.
Non-\bjet distributions were estimated following the procedure in
\cref{sec:nonb_calibration}; no correction is applied to the simulation for
\bjets.
The uncertainty bands denote the impact of all uncertainties described in
\cref{sec:uncertainties}.
}
\label{fig:jet-kinematics-process}
\end{figure}

\begin{figure}
\centering
\subfigure[]{\includegraphics[width=0.45\textwidth]{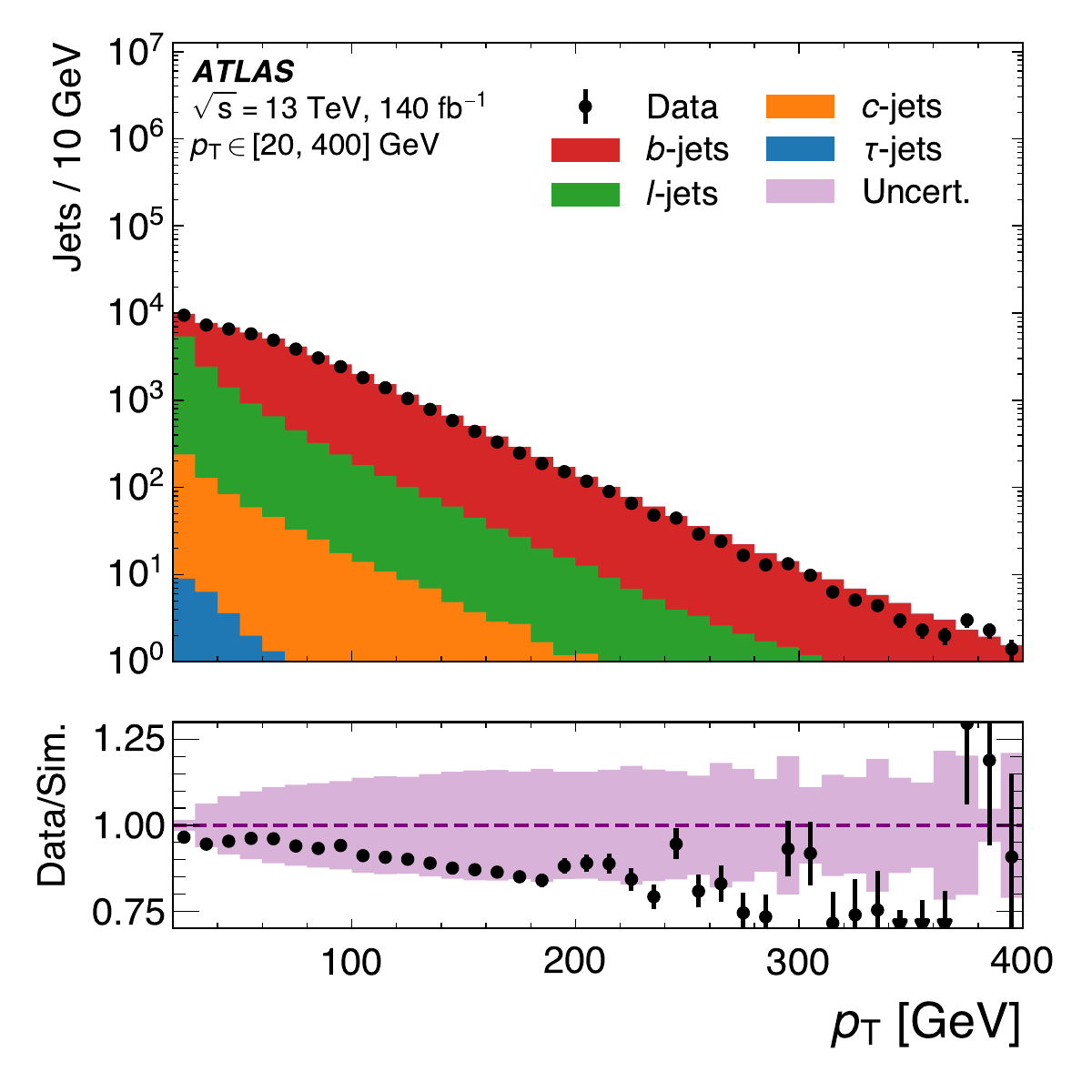}}
\subfigure[]{\includegraphics[width=0.45\textwidth]{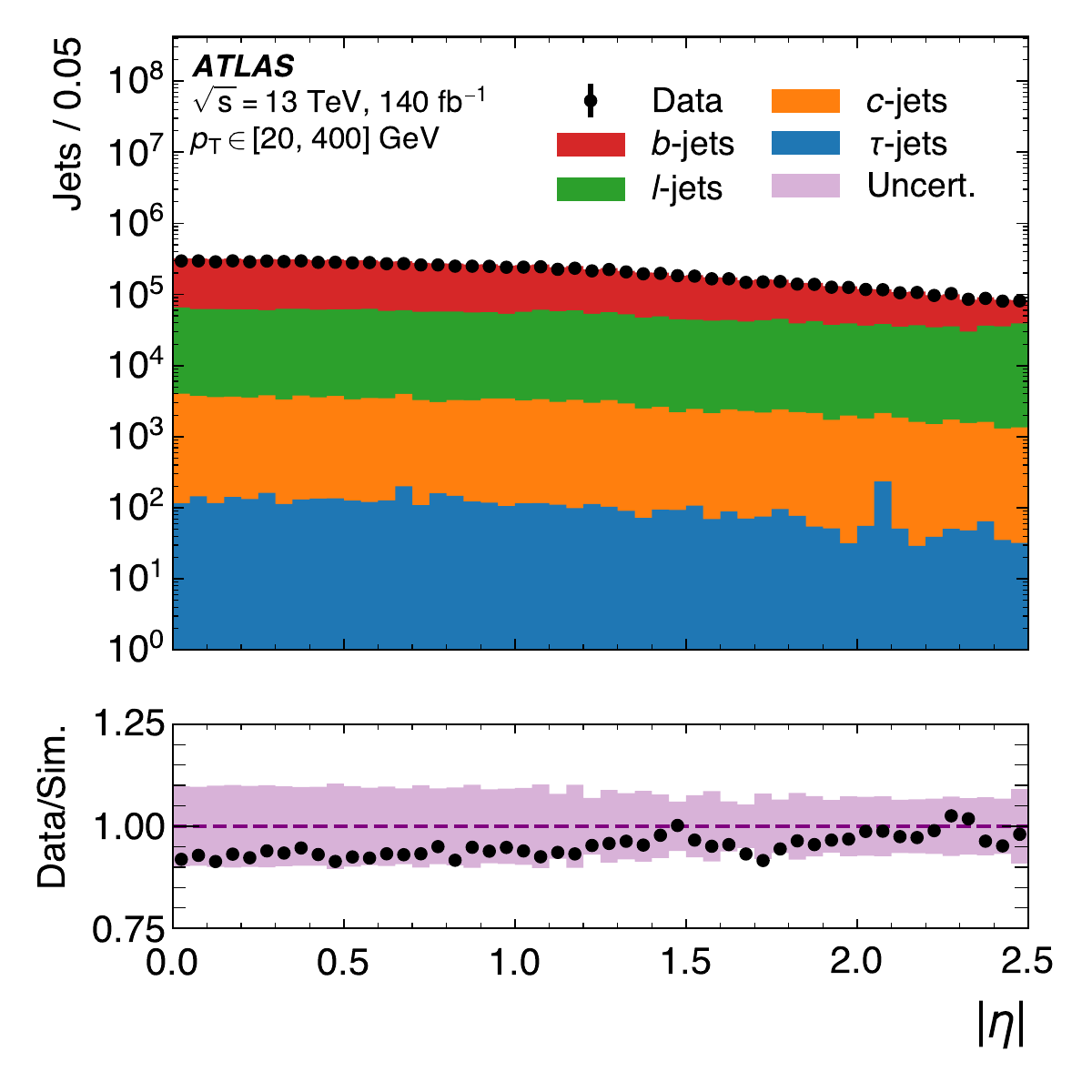}} \\
\subfigure[]{\includegraphics[width=0.45\textwidth]{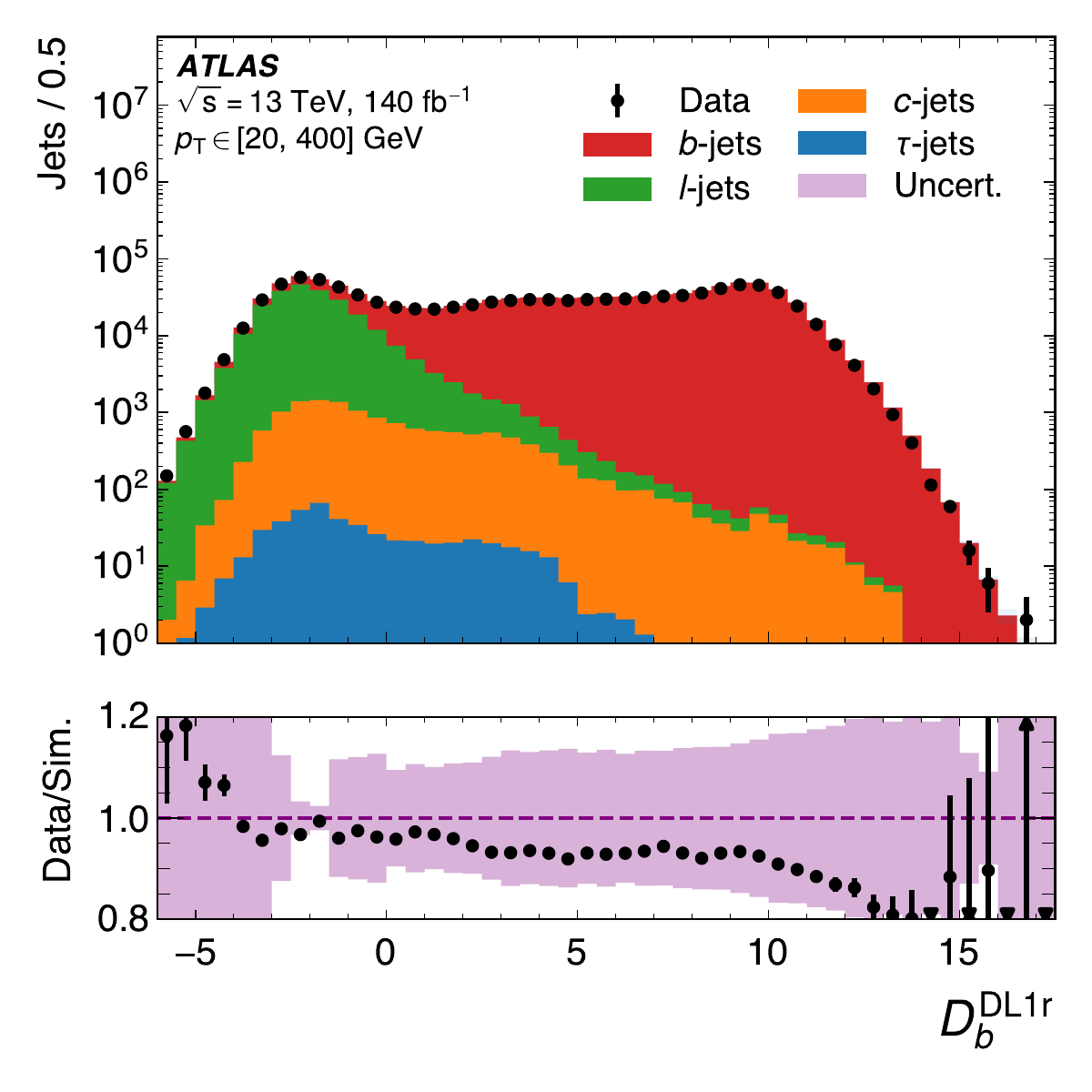}}
\subfigure[]{\includegraphics[width=0.45\textwidth]{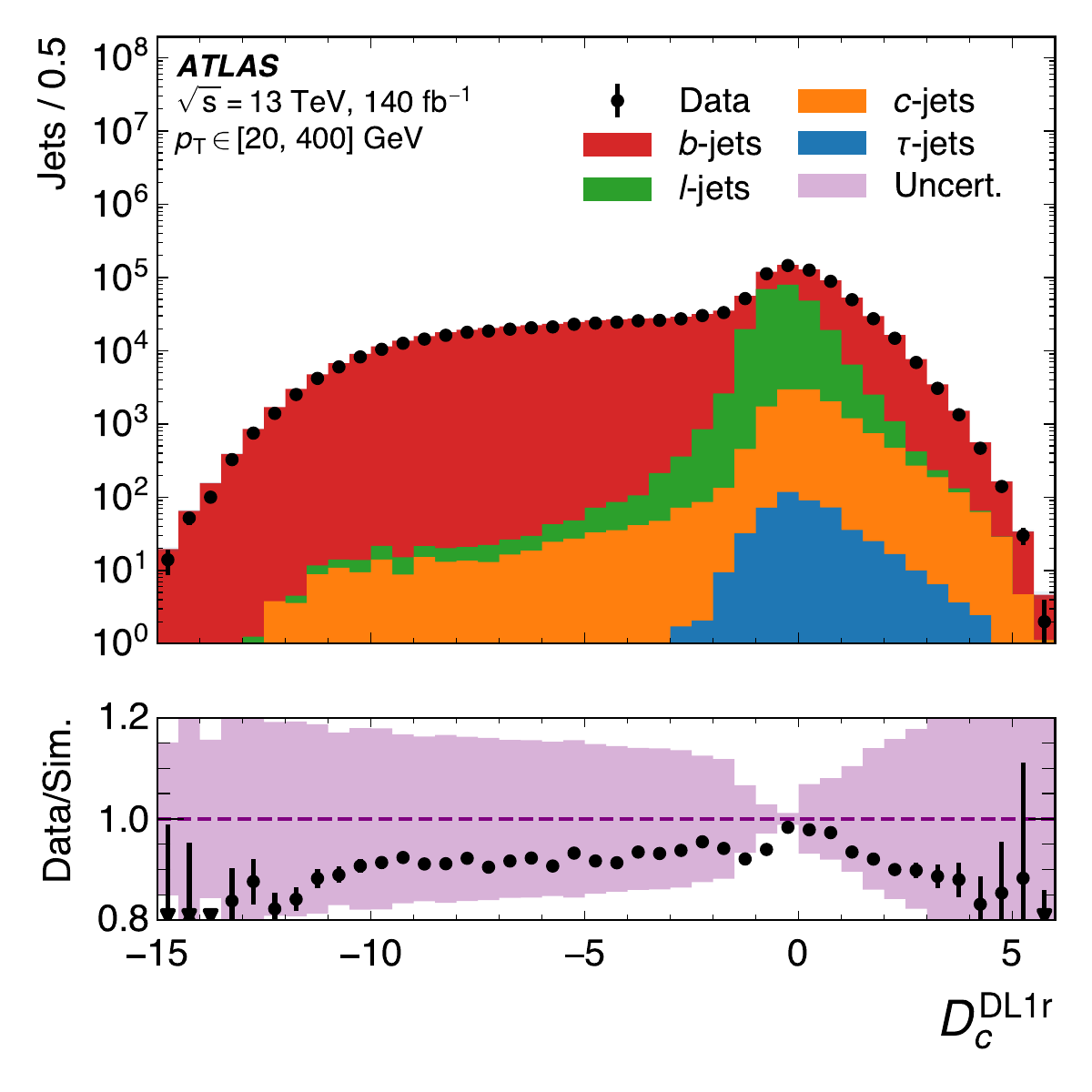}}
\caption{
\label{fig:jet-kinematics-flavour}
Comparison of jet observables between the simulation's predictions and observed
data, with the predictions separated by jet flavour:\ the selected
jets' (a) \pt, (b) $|\eta|$, (c) \dlonerb and (d) \dlonerc.
Non-\bjet distributions were estimated following the procedure in
\cref{sec:nonb_calibration}; no correction is applied to the simulation for
\bjets.
The uncertainty bands denote the impact of all uncertainties described in
\cref{sec:uncertainties}.
}
\end{figure}

For relevant jet kinematics, agreement between data and simulation is
observed within the assessed uncertainties, which are described in
Section~\ref{sec:uncertainties}.
\cref{fig:jet-kinematics-process} shows this agreement for the jet \pt, jet $\eta$,
\dlonerb, and \dlonerc after the full event selection described in
\cref{sec:selection}, with the prediction separated by production process.
\cref{fig:jet-kinematics-flavour} shows comparisons broken down by jet flavour.
It should be noted that the flavour-tagging observables are shown after correcting the
non-\bjet distributions, as described in Section~\ref{sec:nonb_calibration}, but
before applying any corrections to \bjets.
The mismodelling of the jet \pt spectrum in \ttbar events by these NLO
predictions was also observed previously by ATLAS~\cite{TOPQ-2018-26}; the
modelling uncertainties assessed in this study cover this discrepancy at the
$1\sigma$ level.

\subsection{Approximations of \fsig and $\psim(\qvec | \pt)$ }
\label{sec:fsig}

\begin{figure}
\centering
\subfigure[]{\includegraphics[width=0.75\textwidth]{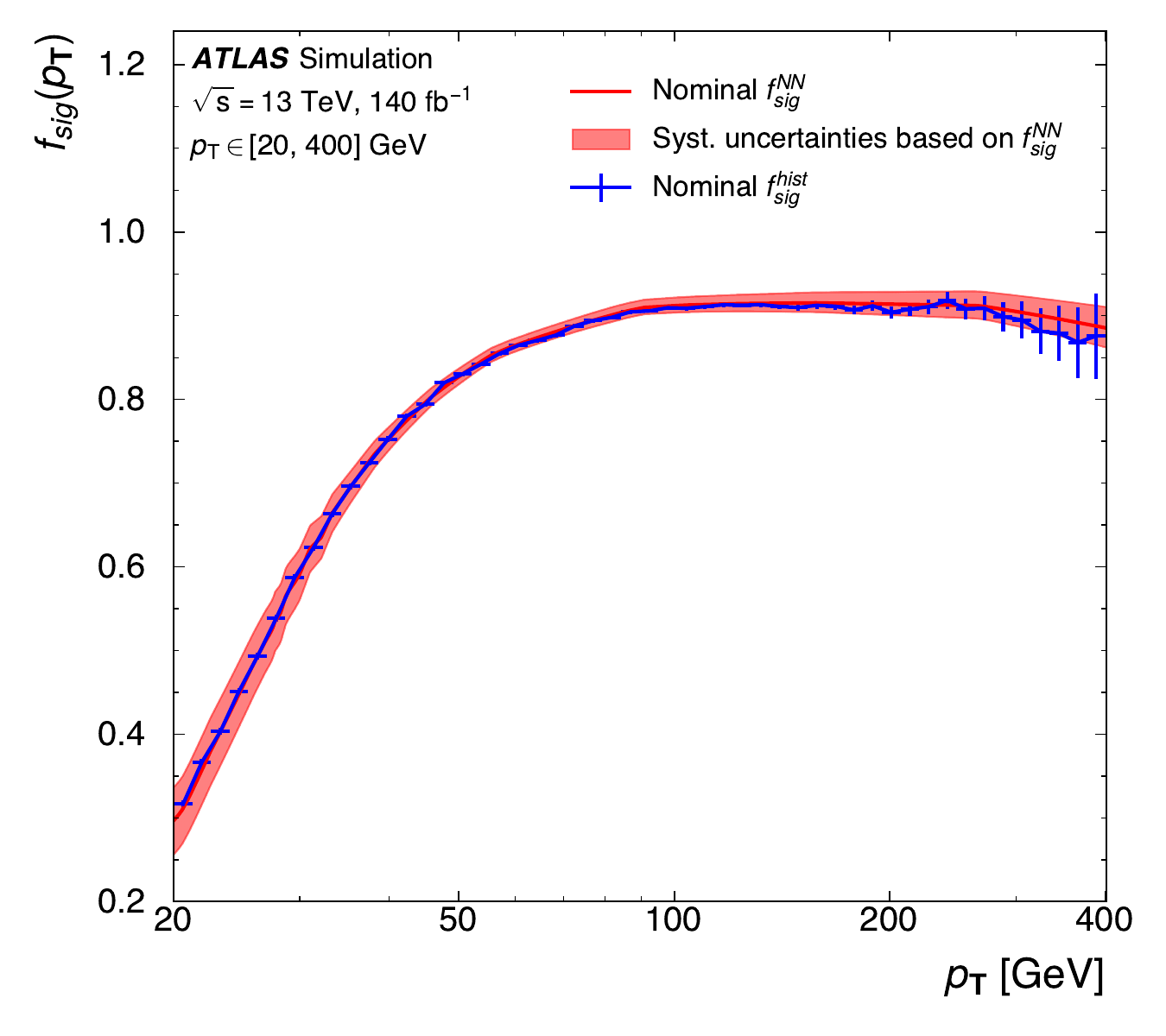}}
\caption{
The predicted \bjet purity, \fsig, after the event selection, is shown as a
function of jet \pt.
Both the binned MC ($f_{\text{sig}}^{\text{hist}}$) and neural purity ($f_{\text{sig}}^{\text{NN}}$) estimates are shown, with good
agreement observed between the two.
The uncertainty bands denote the impact of all uncertainties described in
\cref{sec:uncertainties}.
}
\label{fig:bjet-purity}
\end{figure}

As described in \cref{sec:methods}, key ingredients to the derivation of
the \qvec optimal calibration maps are continuous approximations of
the \bjet fraction $f_\mathrm{sig}(\pt)$ and the jet-flavour probability
distributions $\psim(\qvec | \pt)$ as a function of the jet \pt.
The former is derived using a neural density ratio estimate trained on jets from
the simulation, and the latter is estimated using conditional
normalizing flows.
\cref{fig:bjet-purity} shows the predicted \bjet purity for the neural density ratio estimate,
compared for validation purposes with a binned estimate of the same ratio obtained directly from MC events.

\begin{figure}
\centering
\subfigure[]{\includegraphics[width=0.3\textwidth]{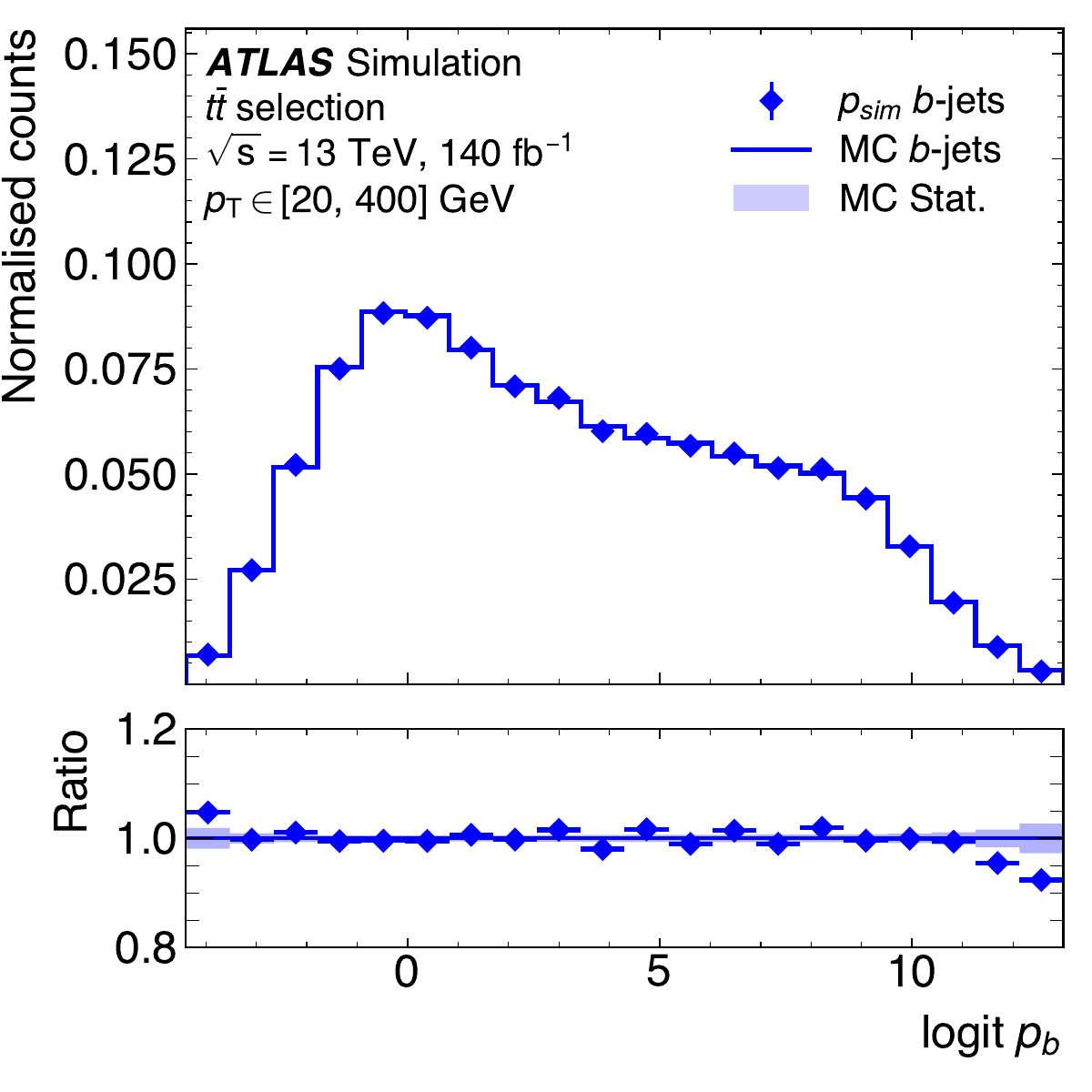}}
\subfigure[]{\includegraphics[width=0.3\textwidth]{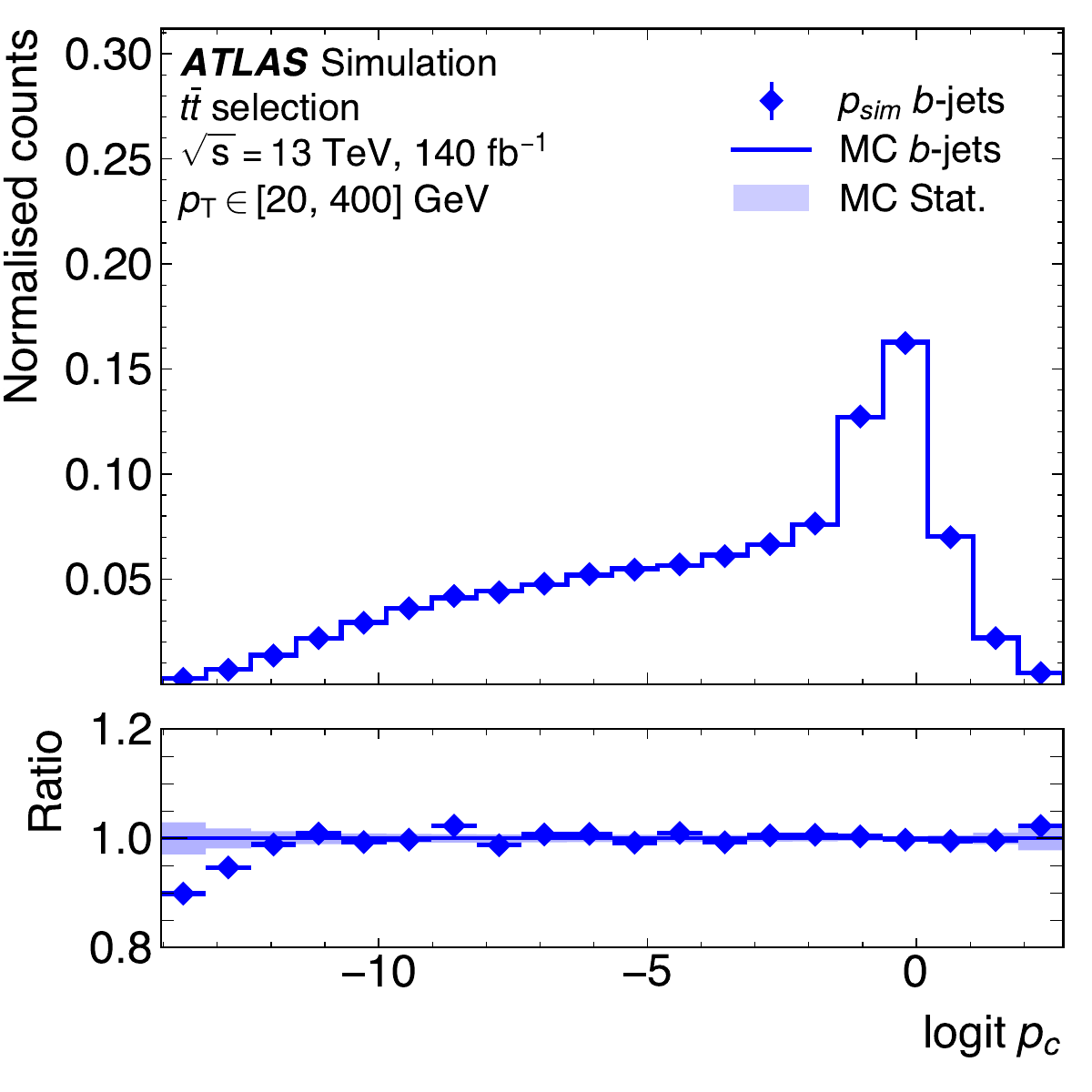}}
\subfigure[]{\includegraphics[width=0.3\textwidth]{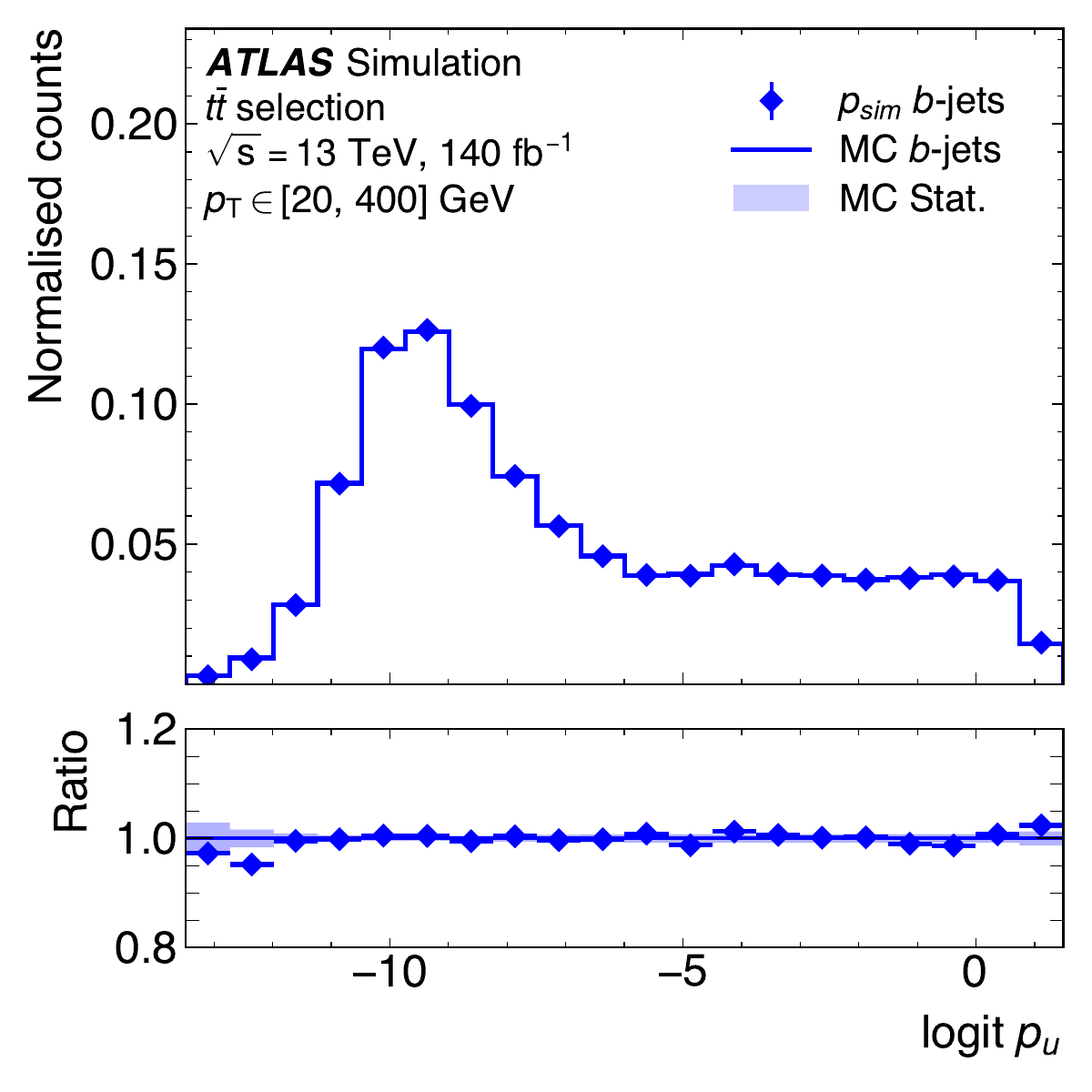}}
\caption{
\label{fig:flow_closure_bjets_logit}
Comparison of binned 1D marginals of \psig between the simulation's
predictions (MC) and the neural estimates ($p_{sim}$) integrated over jet
\pt in the \ttbar region:\ (a) \logitpb, (b) \logitpc, and (c) \logitpu.
Only statistical uncertainties from the limited size of the sample used to
fill the histograms are shown in the error bars.
}
\end{figure}

\begin{figure}
\centering
\subfigure[]{\includegraphics[width=0.3\textwidth]{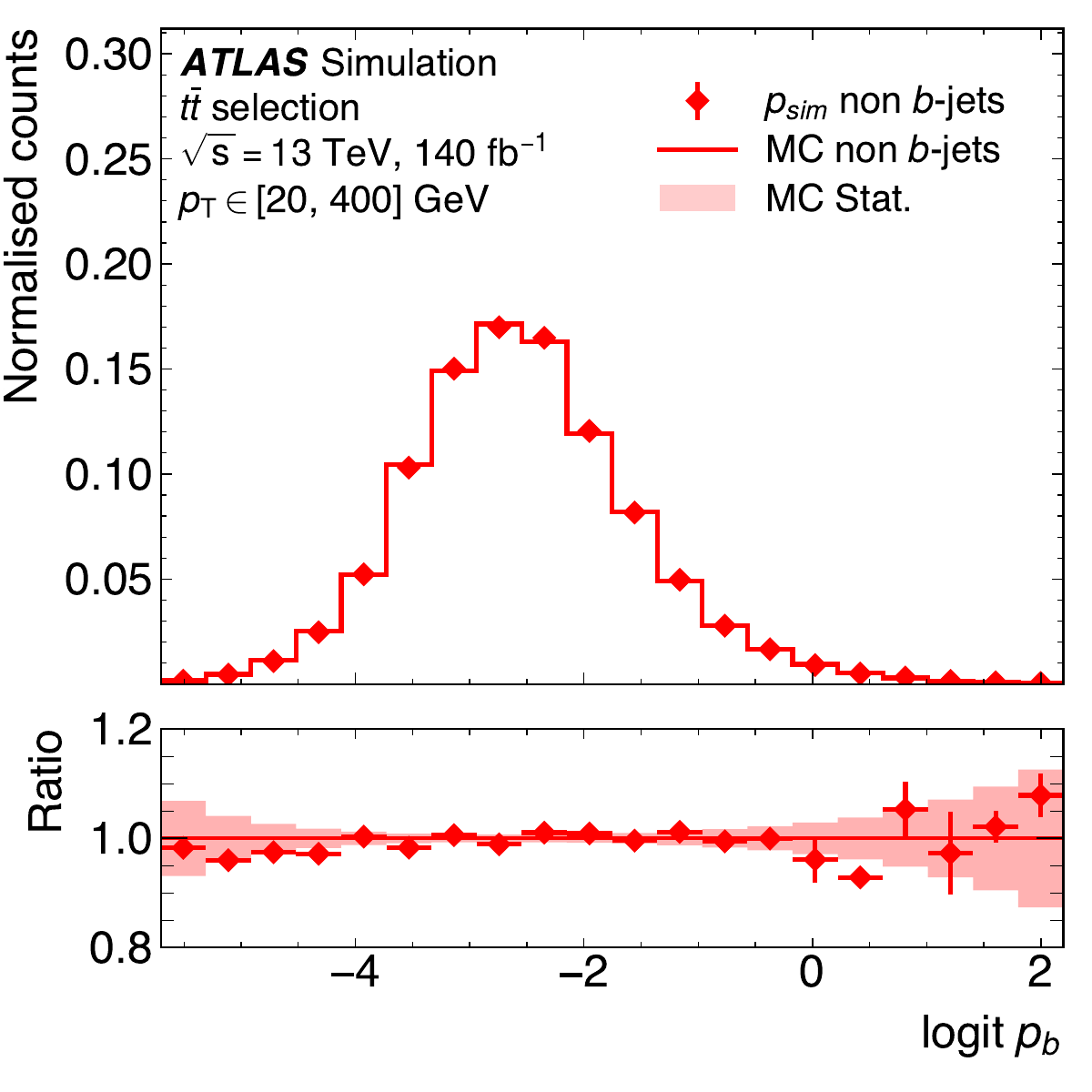}}
\subfigure[]{\includegraphics[width=0.3\textwidth]{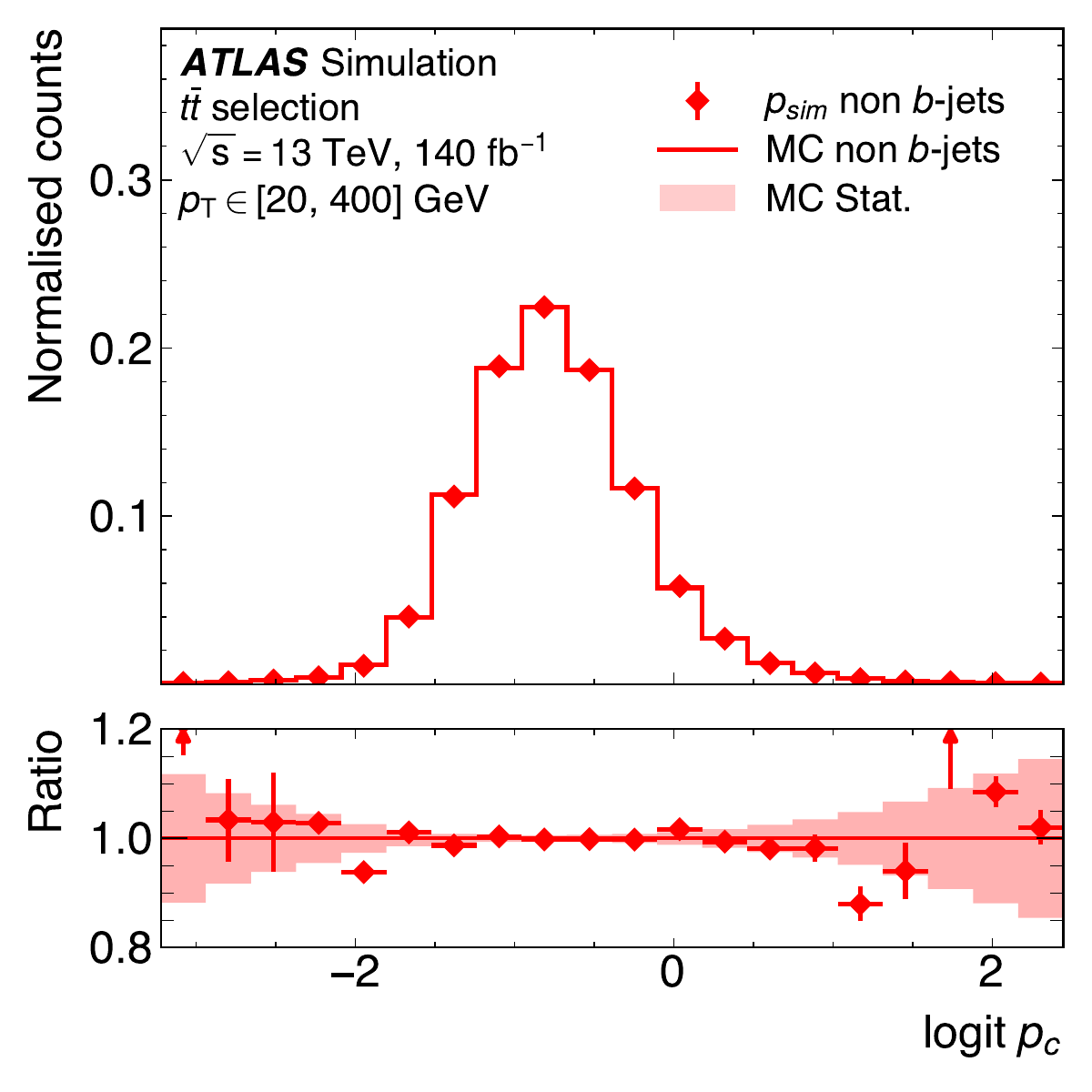}}
\subfigure[]{\includegraphics[width=0.3\textwidth]{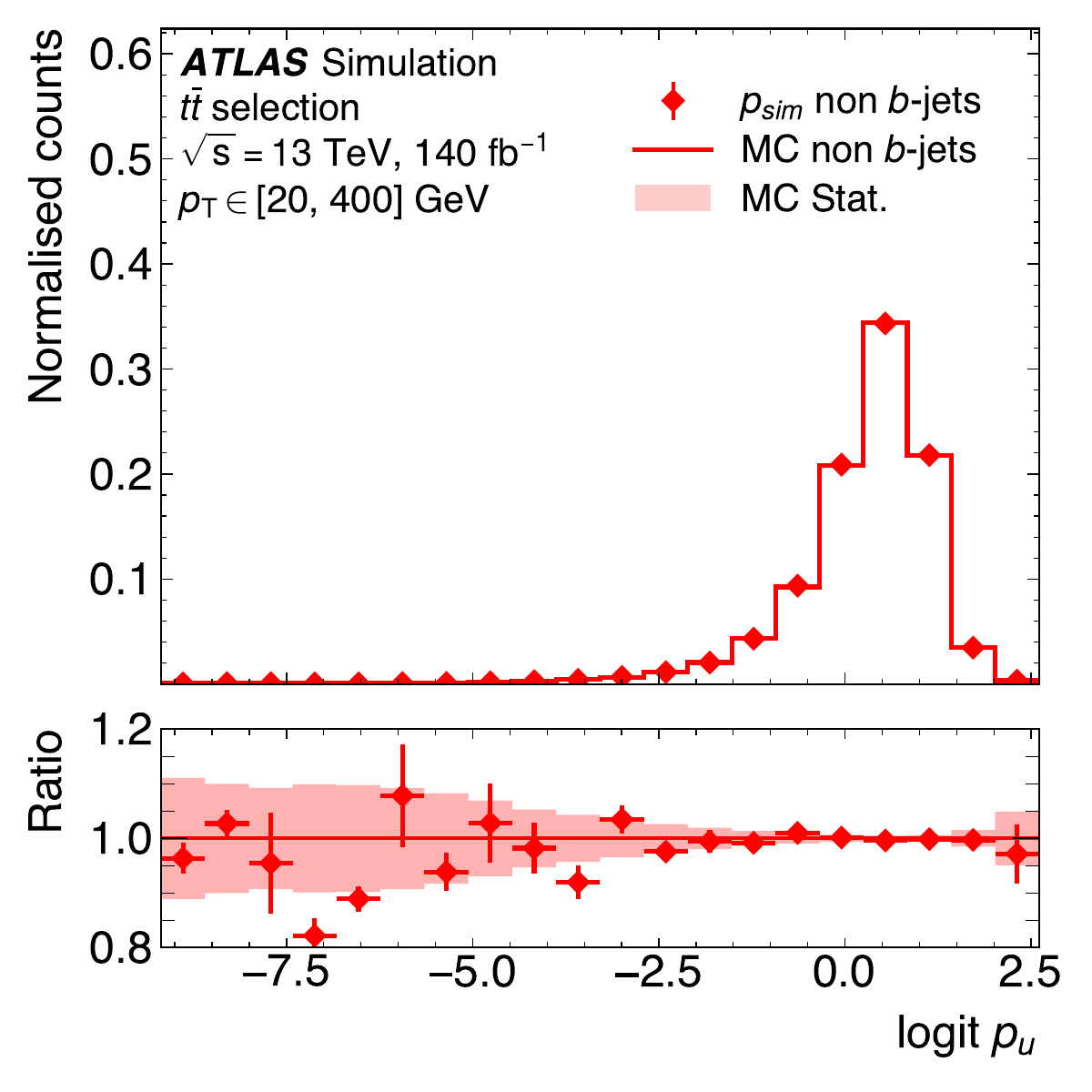}}
\caption{
\label{fig:flow_closure_nonbjets_logit}
Comparison of binned 1D marginals of \pbkg between the simulation's
predictions (MC) and the neural estimates ($p_{sim}$) integrated over jet
\pt in the \ttbar region:\ (a) \logitpb, (b) \logitpc, and (c) \logitpu.
Only statistical uncertainties from the limited size of the sample used to
fill the histograms are shown in the error bars.
}
\end{figure}

\begin{figure}
\centering
\subfigure[]{\includegraphics[width=0.3\textwidth]{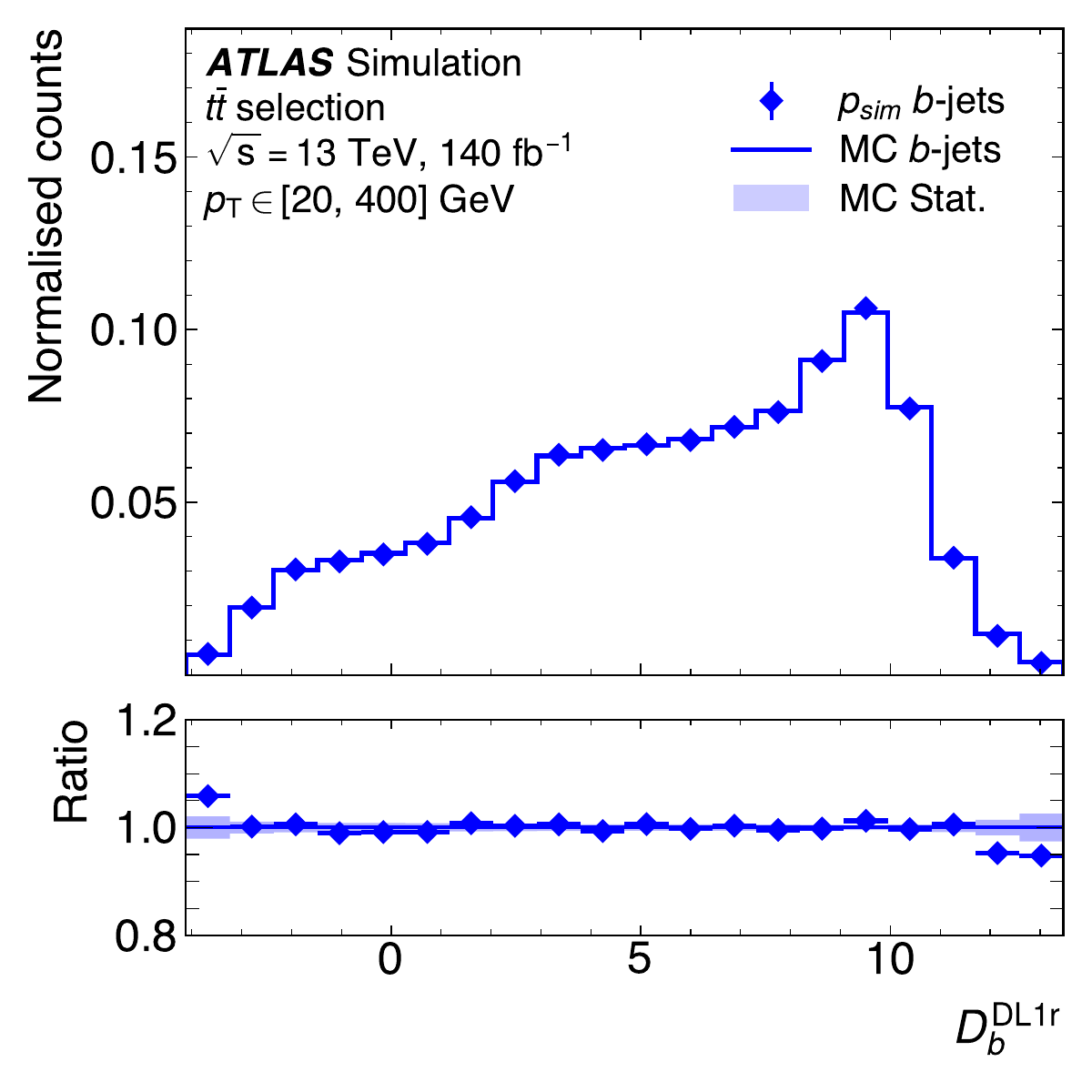}}
\subfigure[]{\includegraphics[width=0.3\textwidth]{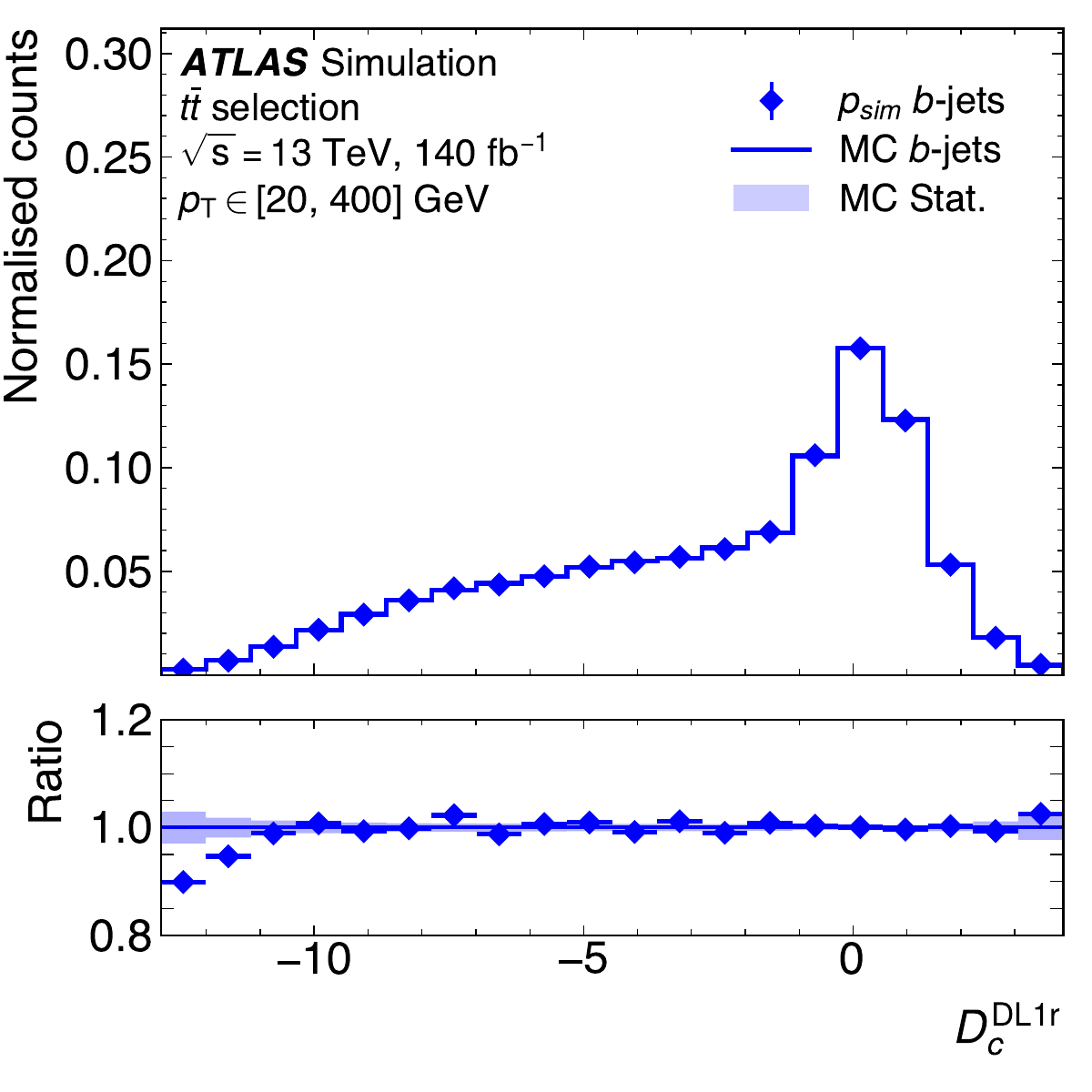}} \\
\subfigure[]{\includegraphics[width=0.3\textwidth]{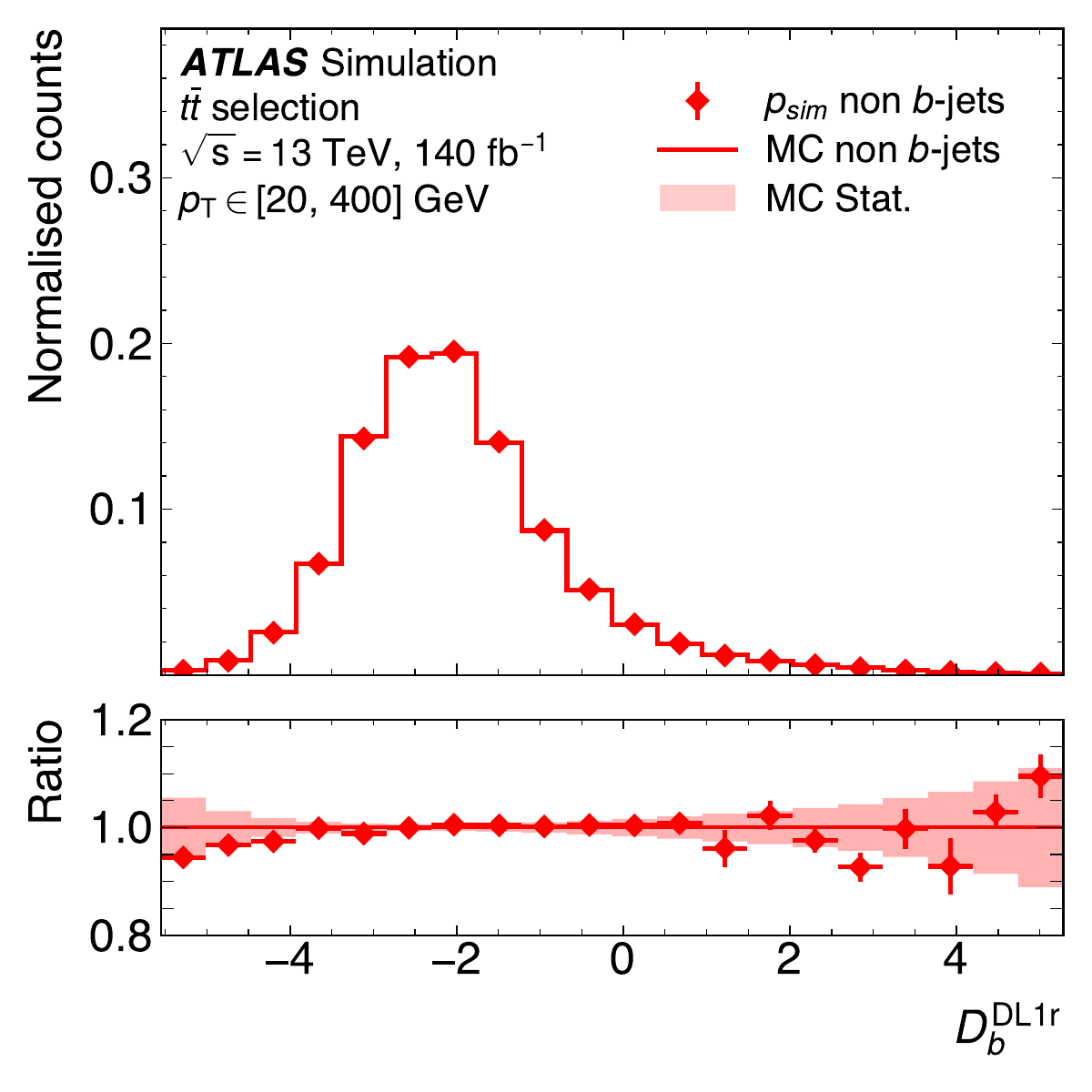}}
\subfigure[]{\includegraphics[width=0.3\textwidth]{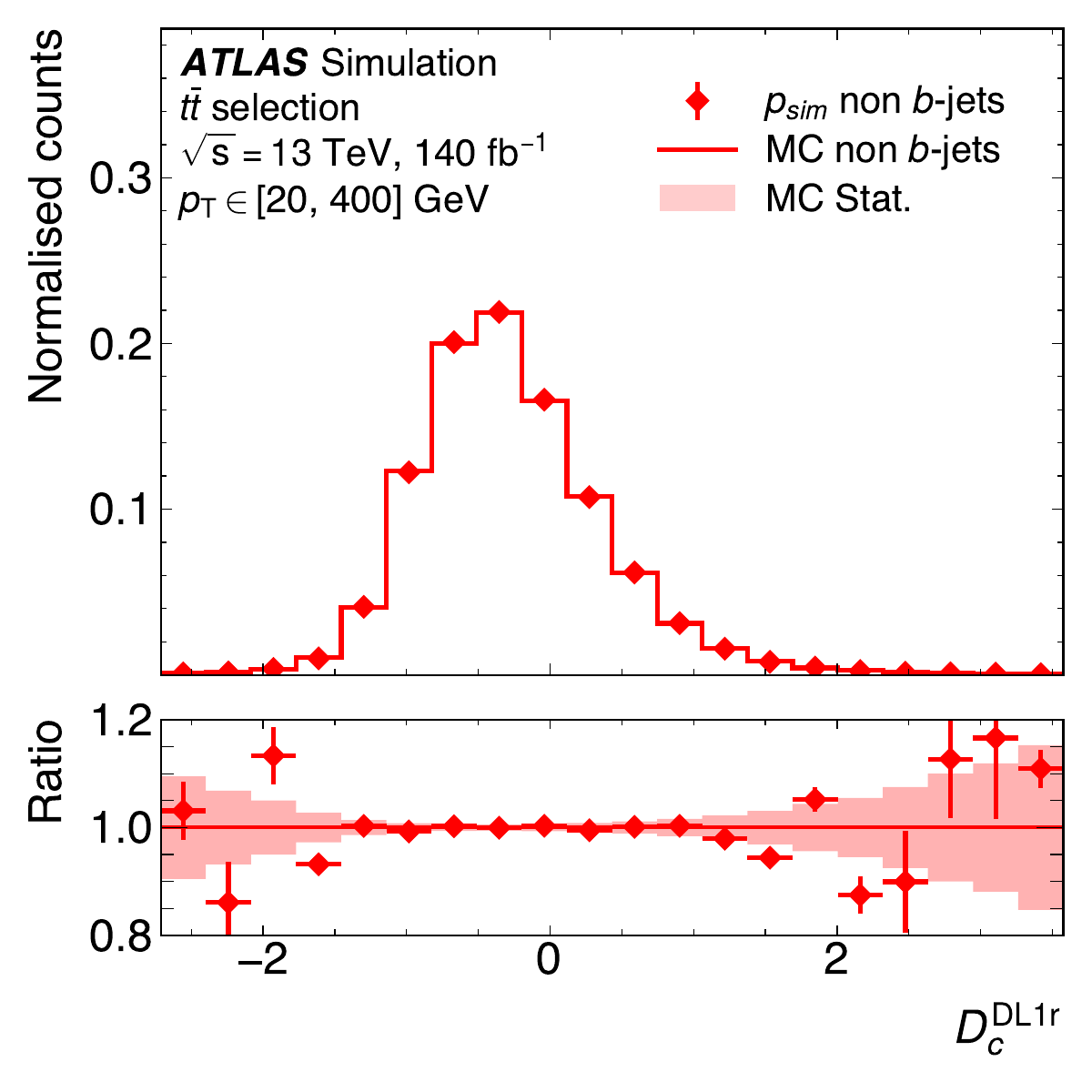}} \\
\caption{
\label{fig:flow_closure_dl1r}
Comparison of binned 1D marginals between the simulation's predictions (MC)
and the neural estimates ($p_{sim}$) integrated over jet \pt for (a)
\dlonerb for \psig, (b) \dlonerc for \psig, (c) \dlonerb for \pbkg, and (d)
\dlonerc for \pbkg in the \ttbar region.
Only statistical uncertainties from the limited size of the sample used to
fill the histograms are shown in the error bars.
}
\end{figure}

Two conditional normalizing flows are used to parameterize $\psim(\qvec | \pt)$
for \bjets and non-\bjets, denoted by \psigpt and \pbkgpt respectively.
The \cjet and $\tau$-jet fractions are predicted to be very small (1.9\% and
0.2\%, respectively), so they are treated together with the light-flavour jets in
\pbkg.
Comparisons between MC and neural estimates in the space of 1D marginal distributions
of $\psim(\qvec, \pt)$ are shown in
Figures~\ref{fig:flow_closure_bjets_logit},~\ref{fig:flow_closure_nonbjets_logit},
and~\ref{fig:flow_closure_dl1r}.
In all cases, good closure is observed.


\subsection{NN training set-ups}

The \texttt{PyTorch} library~\cite{torch} is used to train the various neural networks (NN) in this analysis.
Sample preparations, manipulations, and visualizations are performed
with \texttt{NumPy}~\cite{numpy}, \texttt{SciPy}~\cite{2020SciPy-NMeth},
\texttt{Matplotlib}~\cite{matplotlib}, \texttt{Pandas}~\cite{reback2020pandas},
\texttt{Seaborn}~\cite{seaborn}, and \texttt{SciKit-HEP}~\cite{scikit-hep}.

\subsubsection*{Multi-layer perceptron for \fsig}

To approximate $\fsig$, a multi-layer perceptron composed of 6 hidden
layers and 512 nodes per layer is used with the \texttt{LeakyReLU} activation
function.
Minimum-finding is performed with the \texttt{ADAMW} optimizer~\cite{ADAMW} with
$\beta_0 = 0$, $\beta_1 = 0.9$ and a \texttt{Cosine
Annealing}~\cite{CosineAnnealingLR} learning rate schedule with an initial
learning rate of $5 \times 10^{-5}$ which decreases to $10^{-7}$.
The batch size is 1024.

\subsubsection*{Normalizing flows for \psigpt and \pbkgpt}

Two conditional normalizing flows with identical hyperparameters are employed
to approximate \psigpt and \pbkgpt.
The base distribution is a uniform density over the unit cube.
Neural spline flows~\cite{durkan2019neuralsplineflows} with 8 stacks of rational
quadratic transforms, each with 12 bins, were found to provide good closure with
the simulation.
The training is performed over batches of 1024 samples with the \texttt{ADAMW}
optimizer with the same settings as in the training of $\fsig$.
The learning rate schedule used is the \texttt{Cosine
Annealing}~\cite{CosineAnnealingLR} scheduler with an initial learning rate of $10^{-4}$ that decreases to $10^{-7}$ over a total of 500 epochs.

\subsubsection*{PICNNs for $f$ and $g$}
\label{sec:PICNN}

The network architecture used for the calibration consists of two equally sized
PICNNs~\cite{amos2017input,cpflows,Algren_2024}, denoted by $f$ and $g$, which are
each formed of parallel convex and non-convex pathways~\cite{Pollard:2021fqv};
after training, $\gradQ~g$ yields the optimal transport map
$\hat{T}_{\pt}$.
Both networks have a depth of 4 layers with 128 (8) neurons per layer in the
convex (non-convex) pathway.
After each convex layer except for the final layer, a zeroed \softplus function\footnote{Softplus is defined as
$\softplus(x)=\log(1+\mathrm{e}^x)$.
For stability reasons, the zeroed softplus function, $\softplus^0 (x) =
\softplus(x) - \softplus(0)$, is used.} is used as the activation function, and for the non-convex layers, the
\texttt{CELU}~\cite{celu} activation function is used.
The \softplus function is also applied to enforce positive weights in the convex
pathway.
The \texttt{ADAMW} optimizer with $\beta_0 = 0$ and
$\beta_1 = 0.9$ is used for minimum-finding, and a \texttt{Cosine
Annealing}~\cite{CosineAnnealingLR} learning rate schedule is
chosen with an initial learning rate of $10^{-4}$ which decreases to $10^{-7}$.
The networks are trained for 500 epochs, each consisting of 512 training
steps comprising 10 training batches for $g$ and 4 iterations for $f$; training
batches are built from 1024 jets.

% End of text imported from the .//analysis.tex input file

% The next lines are included from the .//analysis_nonb.tex input file
\section{Background shape estimate}
\label{sec:nonb_calibration}

The shape of the \qvec distribution in background jets is dependent on
a correct description of several aspects of the simulation that are difficult to
model, such as the rates of light-hadron decays-in-flight and photon
conversions; the detector resolution and efficiency of charged-particle
tracking, which are also impacted by radiation damage over time; and the
modelling of interactions between jet constituents and the detector
material~\cite{ATLAS-CONF-2018-006}.
Indeed, in regions dominated by non-\bjets, a large discrepancy between data and
simulation is observed, and this mismodelling of the non-\bjets would bias the
\bjet calibration.
To minimize this bias, a correction of the \qvec shape versus \pt for non-\bjets
is derived in such a region enriched in light-flavour jets and subsequently
applied to non-\bjets in the \ttbar phase space where the main \bjet calibration
is performed.

A region primarily composed of \zjets events is selected to yield a
background-rich population of jets.
The event selection largely follows that laid out in Ref.~\cite{FTAG-2019-02}
but is summarized briefly here.
Two same-flavour charged light leptons ($ee$ or $\mu\mu$ pairs) are required,
and the invariant mass $m_{\ell\ell}$ of this pair of leptons is required to lie
near the $Z$-boson mass peak, $81 < m_{\ell\ell} / \GeV < 101$.
The leading lepton must have $\pt > 28$~GeV, and the single-lepton triggers
must fire, as already described in Section~\ref{sec:selection}.
Due to poor modelling of events with low $Z$-boson transverse momentum, the \pt
of the dilepton system must be greater than 50~\GeV.
At least one jet must be reconstructed with $\pt > 20~\GeV$ and $|\eta| <
2.5$.

\begin{figure}
\centering
\subfigure[]{\includegraphics[width=0.75\textwidth]{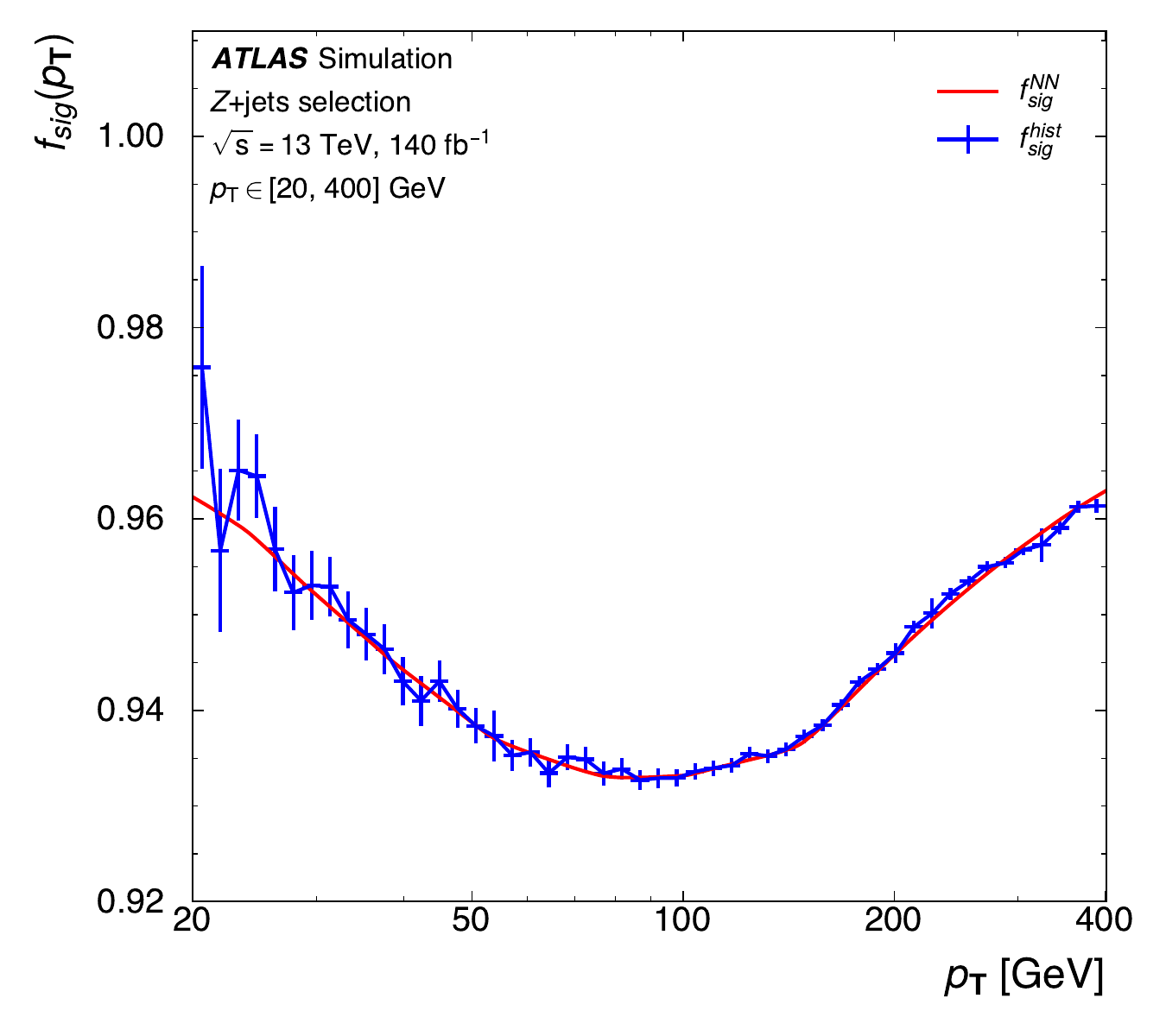}}
\caption{
The predicted non-\bjet purity, \fsig, after the \zjets event selection, as
a function of jet \pt.
The predicted purity is shown from the binned
($f_{\text{sig}}^{\text{hist}}$) and neural ($f_{\text{sig}}^{\text{NN}}$)
estimates.
The uncertainty bars denote the statistical uncertainty on the binned
estimate.
}
\label{fig:non-bjet-purity}
\end{figure}

For this non-\bjet calibration, \bjets are treated as a background and are
effectively subtracted to extract the transport maps for non-\bjets, following
the description in Section~\ref{sec:methods}.
MC simulation predicts that the fraction of non-\bjets is above 93\% across the
full \pt range probed in the \zjets phase space.
The NN-based estimate used to describe the non-\bjet fraction vs jet \pt is
shown in Figure~\ref{fig:non-bjet-purity} and compared for validation purposes
to its binned estimate.
The ratio of charm jets to light-flavour jets in the \zjets phase space is
similar to that in the \ttbar phase space, so these components are treated
together in the background shape estimate:\ in \zjets events approximately 8\%
of non-\bjets are charm jets, while this fraction is $\approx 6\%$ in \ttbar
events.
This difference in charm-jet fraction has a very small impact on the non-\bjet
shape compared to the assessed uncertainties.
The non-\bjet corrections are derived as optimal transportation maps between
simulation and data for the leading jet (ordered by \pt) in each event.

In some regions, particularly at low jet \pt and high $\logitpb$, the predicted
\bjet yield exceeds the data yield; this can occur if, for example, the $Z+b$
production cross-section is too large in simulation.
Under these conditions, there is no transport map for non-\bjets that results
in closure between corrected simulation and data.
Since this portion of the \Qvec space is ultimately dominated by uncertainties
other than that on the shape of the non-\bjet background, a simple approach is
used to enforce the existence of a solution:\ it is assumed that all jets in
such regions are \bjets.
A NN discriminator is trained to output a density ratio
estimate of $r_b = f_b(\pt) \psim^b(\qvec | \pt) / \pdata(\qvec | \pt)$,
where $f_b(\pt)$ is the predicted \bjet fraction and $\psim^b(\qvec | \pt)$ is
the predicted \bjet distribution in the \zjets phase space.
If $r_b$ is less than unity, jets in data are probabilistically
down-sampled by $1 - r_b$ to subtract the $b$-jet background.
Wherever this ratio is greater than unity (i.e.~the \bjet yield is predicted
to be above the data yield), all data jets are removed.
This amounts to enforcing $f_b(\pt) \psim^b(\qvec | \pt) = \pdata(\qvec | \pt)$,
i.e.~assuming all jets in data in such cases are \bjets.

\begin{figure}
\centering
\subfigure[]{\includegraphics[width=0.32\textwidth]{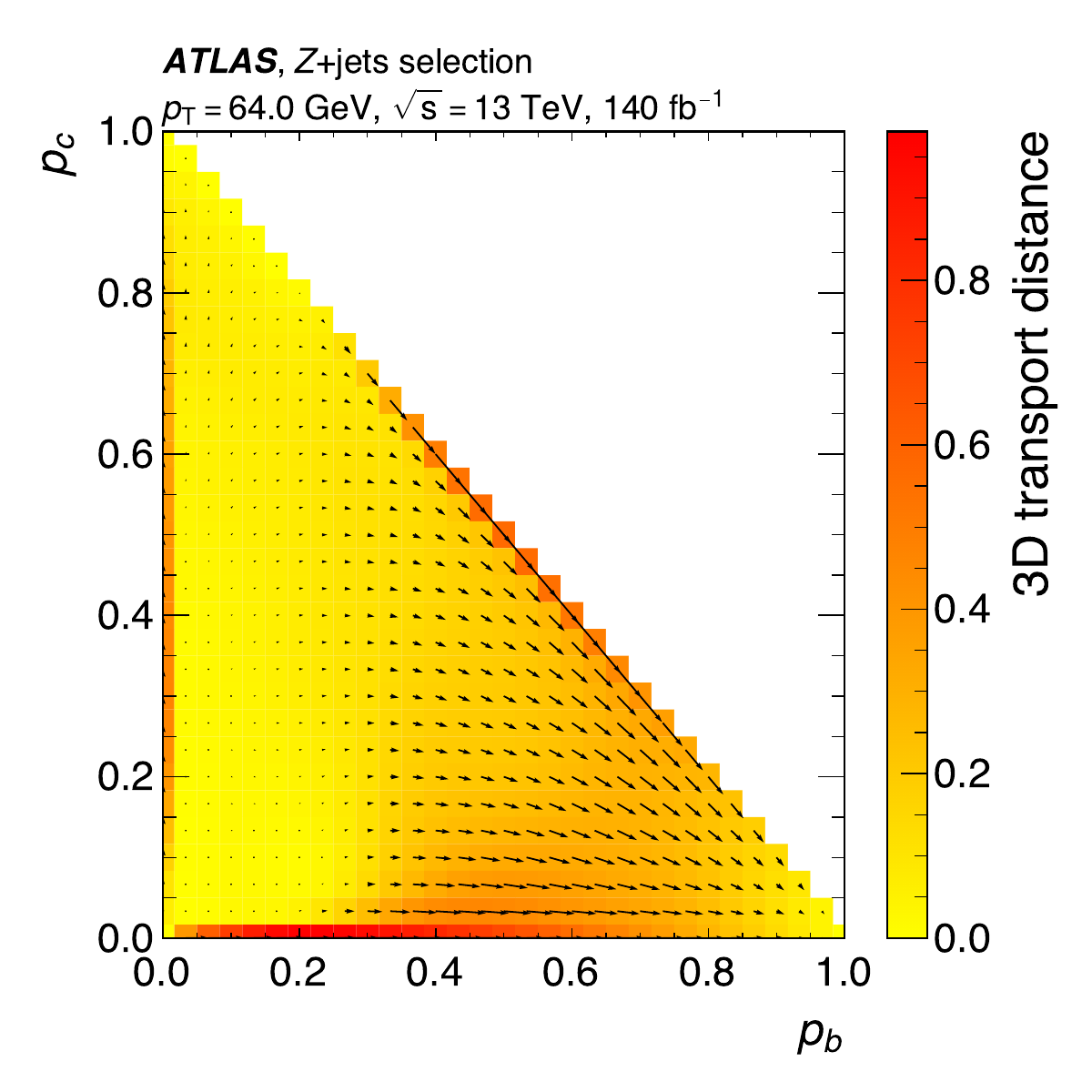}}
\subfigure[]{\includegraphics[width=0.32\textwidth]{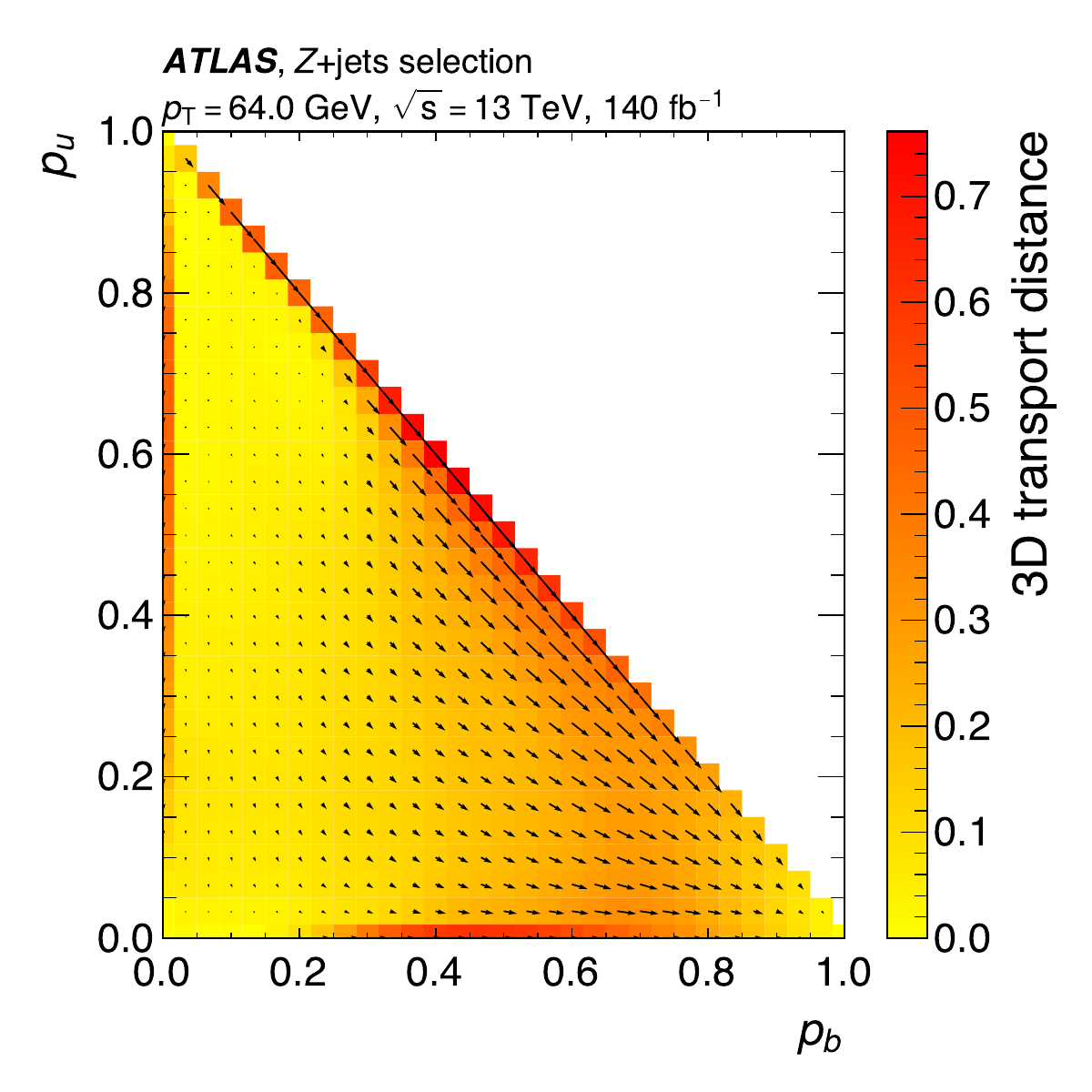}}
\subfigure[]{\includegraphics[width=0.32\textwidth]{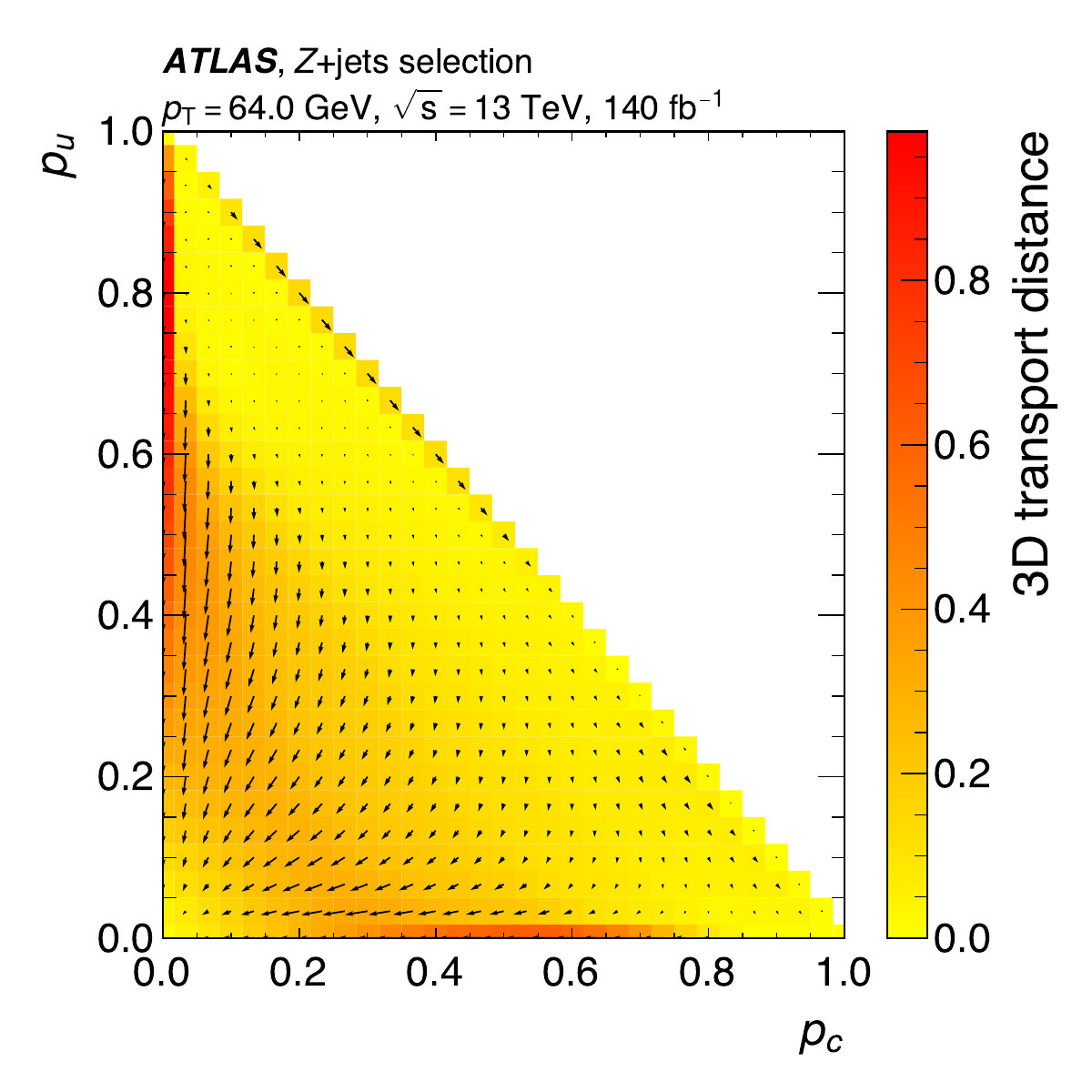}}
\caption{
\label{fig:transpmaps_zjets}
2D slices of the calibration transport maps for non-\bjets with $\pt =
64~\GeV$:\ (a) $p_b$ vs $p_c$, (b) $p_b$ vs $p_u$, and (c) $p_c$ vs $p_u$.
Arrow directions and lengths correspond to the displacement of a non-\bjet
at the arrow's origin, $\hat T$, needed to obtain closure between the data
and simulation such that $\hat T_\#~\psim(\qvec | \pt) \approx
\pdata(\qvec | \pt)$.
The length of the arrows was normalized to avoid overlap and is
indicated by the colour of the background.
For each 2D map, the missing dimension is determined by requiring
$\sum_i p_i = 1$.
}
\end{figure}

\begin{figure}
\centering
\subfigure[]{
\includegraphics[width=0.3\textwidth]{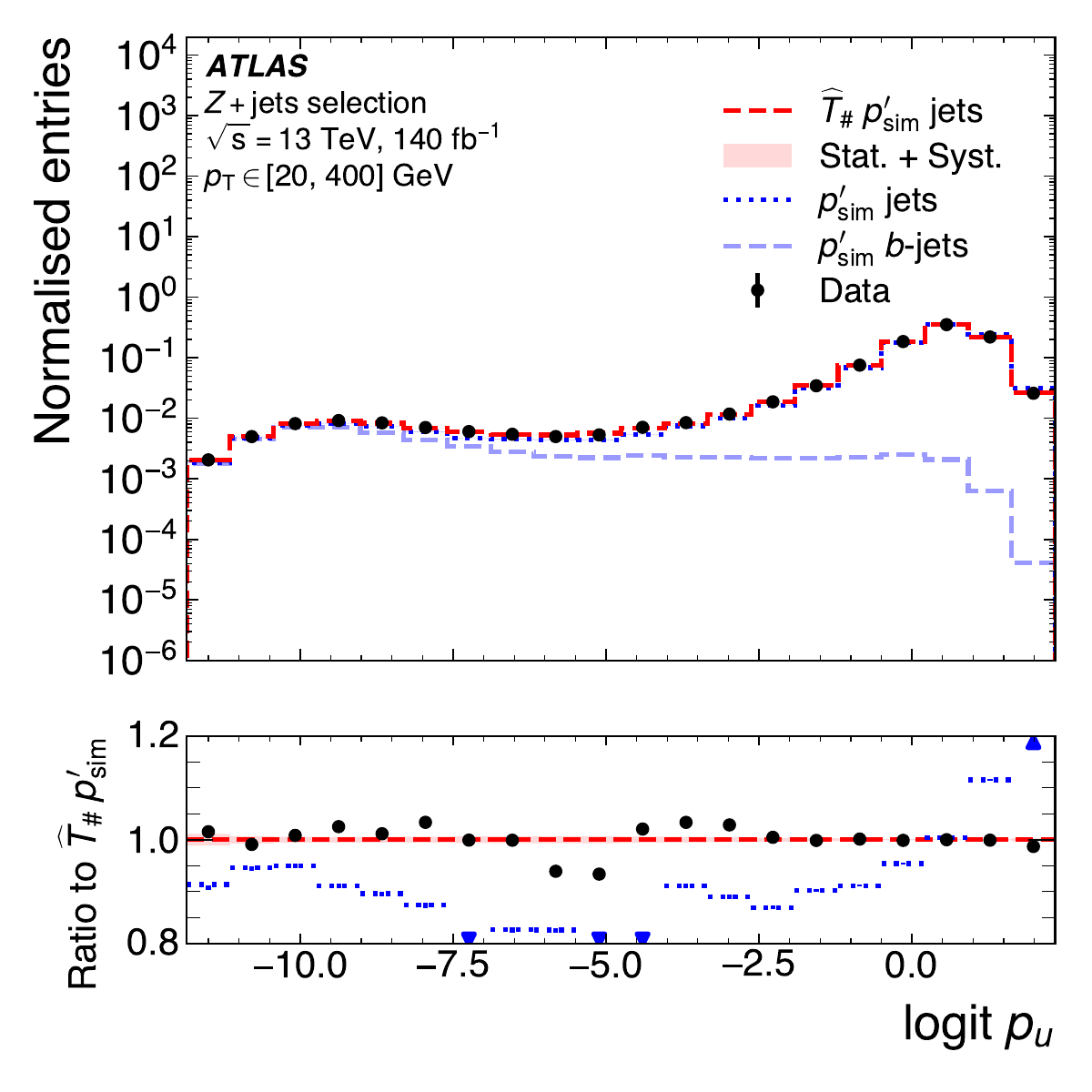}
}
\subfigure[]{
\includegraphics[width=0.3\textwidth]{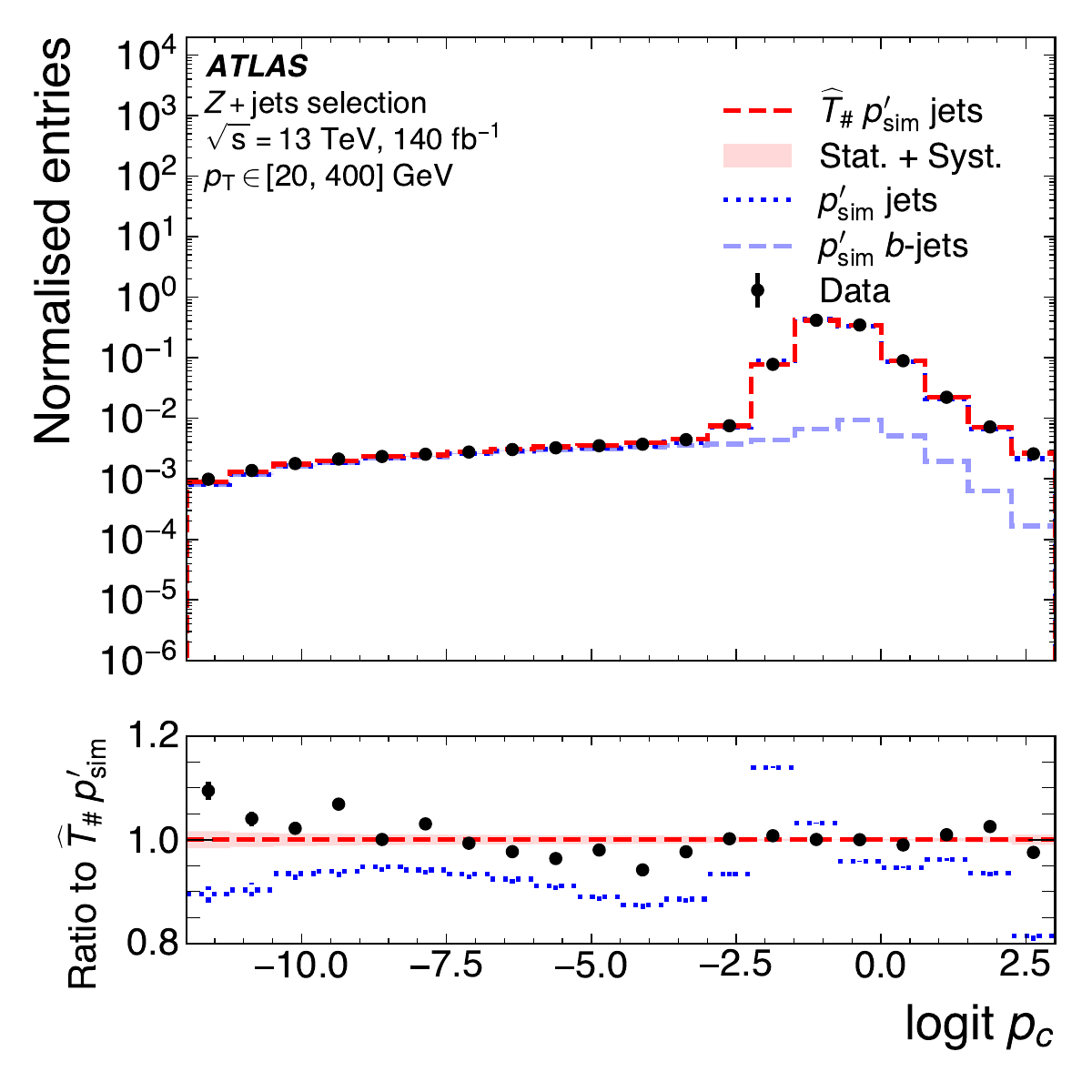}
}
\subfigure[]{
\includegraphics[width=0.3\textwidth]{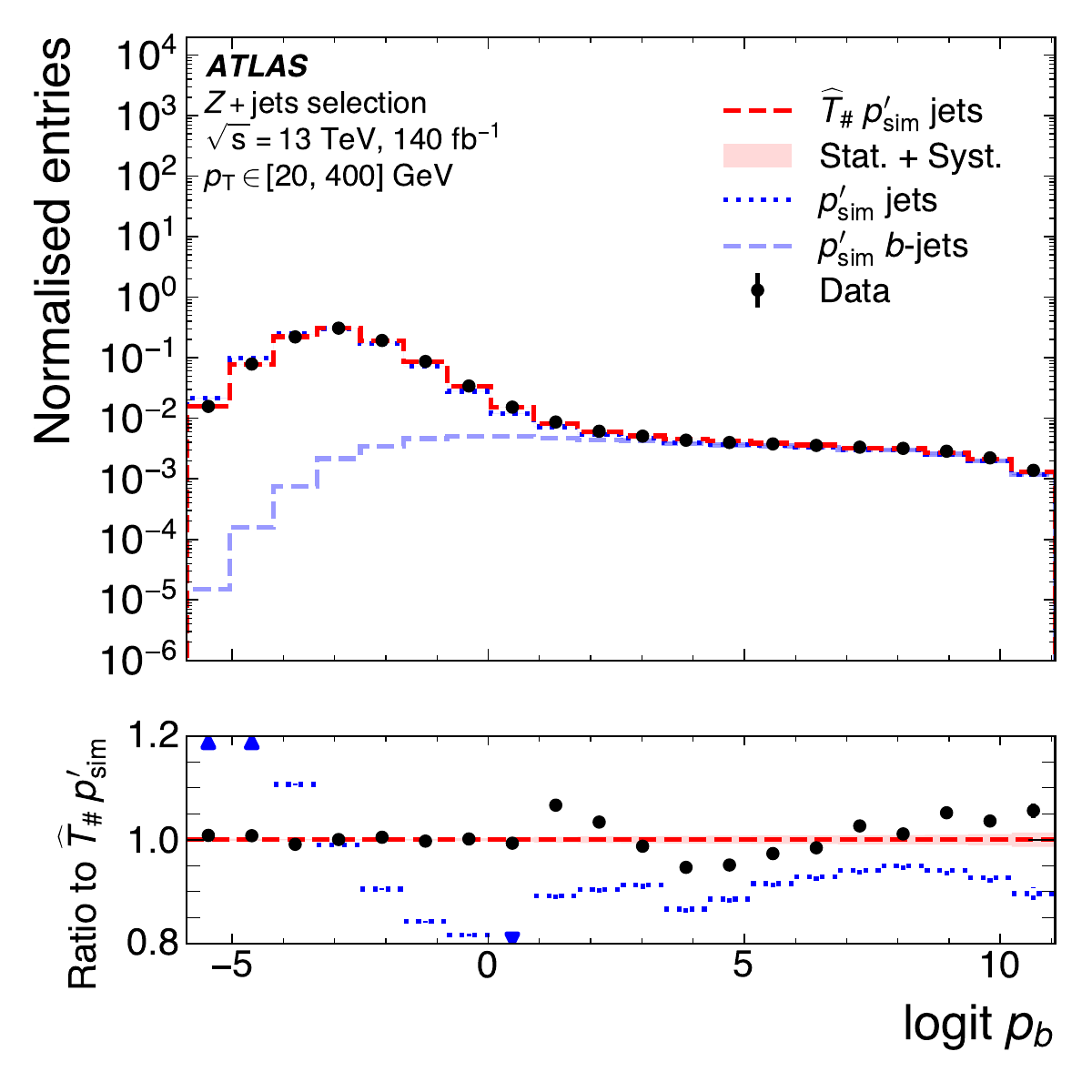}
}
\caption{
Comparisons between the observed data, uncalibrated simulation,
$\psim^\prime$, and corrected simulation, $\hat T_\# \psim^\prime$, of the
\zjets selection in
(a) \logitpu, (b) \logitpc, and (c) \logitpb.
Both the non-\bjet and \bjet components of the simulation are included in
the predicted yields.
Only statistical uncertainties from the limited sample size of the data and
simulation used to fill the histograms are shown in the error bars.
In the \bjet calibration, the full difference between $\psim^\prime$ and the
corrected non-\bjets shape is taken as a systematic uncertainty.
Blue triangles in the ratio panel indicate that the ratio is beyond the
limits of the y-axis range.
}
\label{fig:logit_calib_zjets_logy}
\end{figure}

\begin{figure}
\centering
\subfigure[]{
\includegraphics[width=0.4\textwidth]{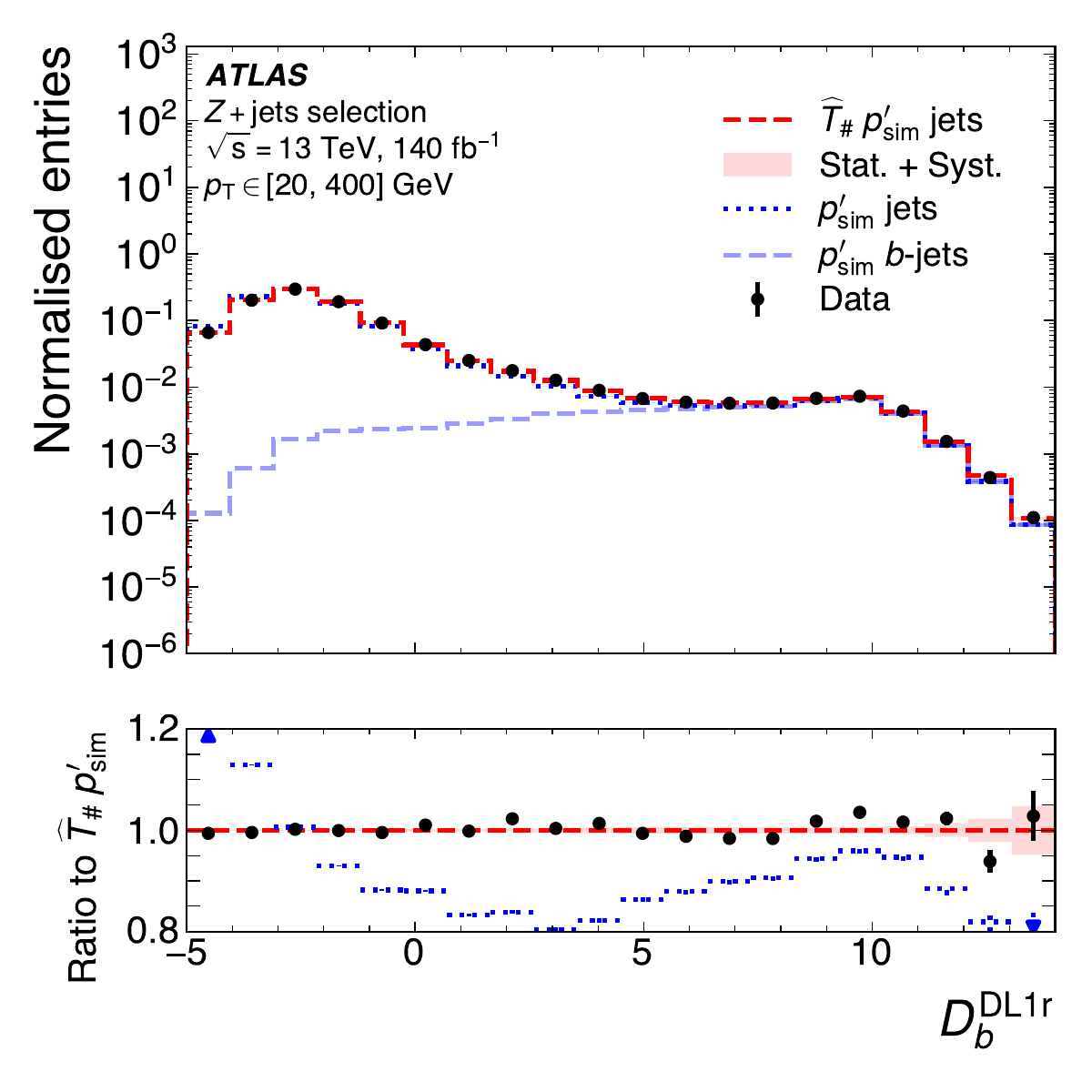}
}
\subfigure[]{
\includegraphics[width=0.4\textwidth]{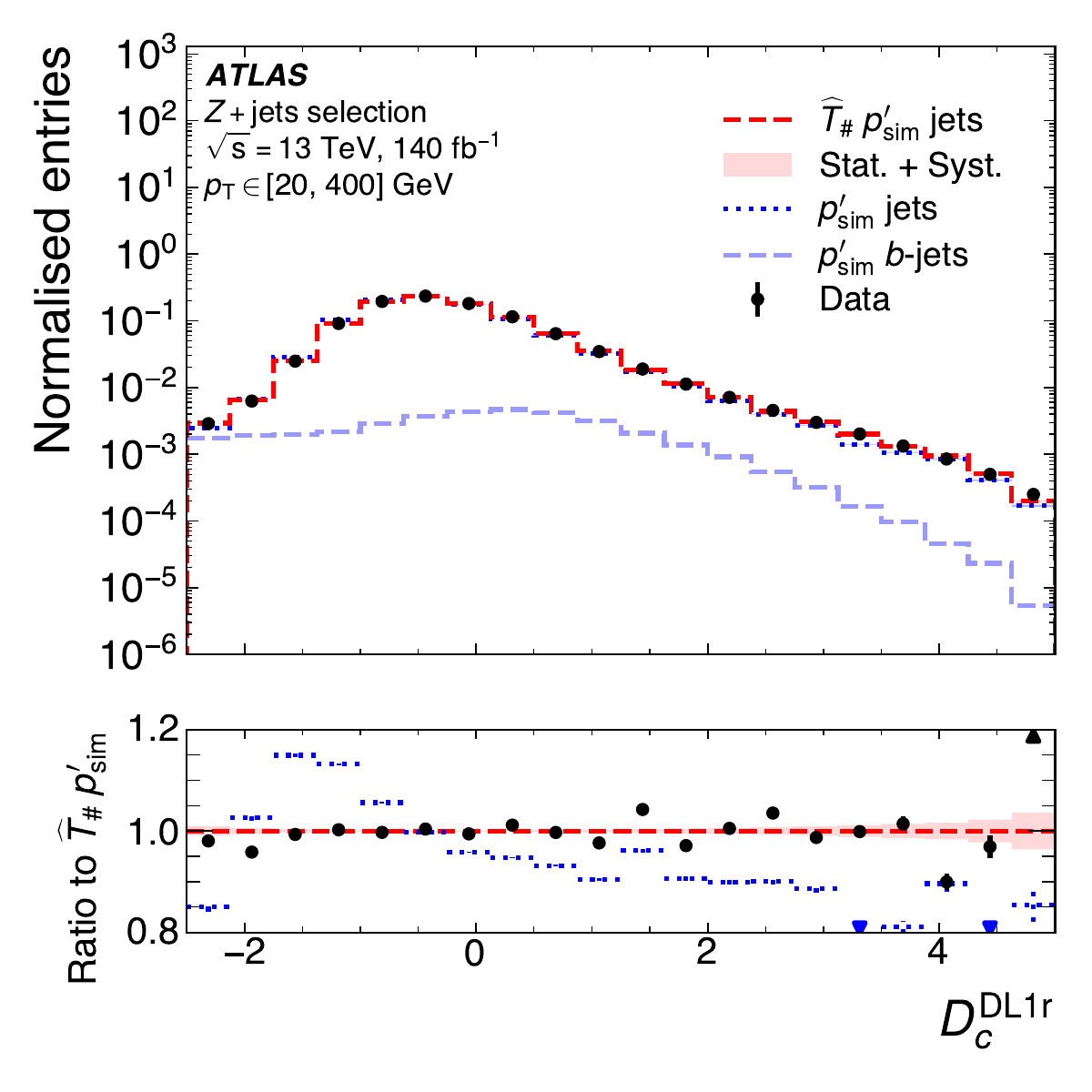}
}
\caption{
Comparisons between the observed data, uncalibrated simulation,
$\psim^\prime$, and corrected simulation, $\hat T_\# \psim^\prime$, of the
\zjets selection in
(a) \dlonerb and (b) \dlonerc.
Both the non-\bjet and \bjet components of the simulation are included in
the predicted yields.
Only statistical uncertainties from the limited sample size of the data and
simulation used to fill the histograms are shown in the error bars.
In the \bjet calibration, the full difference between $\psim^\prime$ and the
corrected non-\bjets shape is taken as a systematic uncertainty.
Blue triangles in the ratio panel indicate that the ratio is beyond the
limits of the y-axis range.
}
\label{fig:dl1rb_calib_zjets_logy}
\end{figure}

\cref{fig:transpmaps_zjets} shows the 2D transport maps in $\vec p$ space for
non-\bjets, evaluated at the mean observed jet \pt in the \ttbar phase space,
$\langle p_{\mathrm{T, data}}\rangle_{\ttbar}=64$~GeV; to obtain these maps, the
$p_i$ are converted into $q_i$, the calibration is applied to obtain an updated
set $q^\prime_i$, and then these are translated back into altered $p^\prime_i$
to determine the displacement with respect to the original $p_i$.
Figures~\ref{fig:logit_calib_zjets_logy} and~\ref{fig:dl1rb_calib_zjets_logy}
show comparisons between simulated and observed flavour observables in the \zjets
region after the non-\bjet corrections are applied.
Good agreement is observed between the data and corrected simulation in the
regions for which the non-\bjet background is important in the \ttbar phase
space, as can be seen in Figure~\ref{fig:jet-kinematics-flavour}.
As discussed in Section~\ref{sec:uncertainties}, the full difference between the
distributions of \pbkgpt for uncalibrated and calibrated non-\bjets in the
\ttbar phase space is taken as a systematic uncertainty of the \bjet calibration
maps.


% End of text imported from the .//analysis_nonb.tex input file

% The next lines are included from the .//uncertainties.tex input file
\section{Uncertainties} \label{sec:uncertainties}
Several uncertainties in the optimal calibration maps are estimated by deriving
alternative maps and comparing them with the nominal calibration.
Statistical uncertainties are approximated by sampling the observed data with
replacement (\enquote{bootstrapping}) 33 times and rederiving the calibration maps --
one for each bootstrap.
Analyses utilizing this calibration can estimate the impact of its statistical
uncertainty on their results by applying each alternative calibration,
comparing these outcomes with the nominal, and considering the combined impact of
these uncorrelated variations divided by $\sqrt{N_\mathrm{bootstraps}}$.
The statistical uncertainties, as presented in Section~\ref{sec:results}, in
most cases are small compared to the total uncertainty.

Systematic uncertainties are assessed by altering the simulation in a number of
ways that impact \fsig and $\pbkg(\qvec | \pt)$ in Eq.~(\ref{eq:pprime}).
Each alternative simulation hypothesis results in a new calibration map, and, as
with the statistical uncertainties, each set of corrections can be applied
to estimate the corresponding uncertainty in the parameters of interest in a
particular analysis.
The systematic uncertainties assessed for \fsig are those used in previous
flavour-tagging calibration measurements using similar event selection
criteria~\cite{FTAG-2018-01}.
Notably, $\psig(\qvec | \pt)$ is \emph{not} varied:\ a single
normalizing-flow approximation of this density is used for all systematic
variations.

Rather than training a new NN approximation of \fsig for each variation, only
a correction $\Delta$ to the nominal $\logit \fsig$ is approximated with
a secondary NN:
\begin{equation*}
\logit f_\mathrm{sig}^\mathrm{variation}(\pt)
= \logit f_\mathrm{sig}^\mathrm{nominal}(\pt) + \Delta^\mathrm{variation}(\pt)
\end{equation*}
This was found to converge more reliably than retraining an entirely new
approximation of \fsig for each variation.

The uncertainties with the most significant impact on \fsig are evaluated in
this measurement.
Of the assessed uncertainties, those that had more than a 2\% impact on
the total \fsig systematic uncertainty in any jet \pt region were retained.
The retained systematic uncertainties are the ISR, FSR, and alternative
parton-shower model uncertainties for \ttbar production, as described in
Section~\ref{sec:dataMC}; the jet energy scale (JES) uncertainties from the
$\eta$-intercalibration procedure, the jet flavour response for gluon- and
\bjets, the jet flavour composition, and the jet-\pileup subtraction
procedure; and uncertainties in the jet energy resolution (JER)
eigenvector~\cite{JETM-2018-05} for the leading jet.
Uncertainties that were determined to have a negligible impact
include all other JES and JER uncertainties; uncertainties due to parton
distribution functions; uncertainties from charged-lepton triggering,
reconstruction, identification, and isolation efficiencies; and uncertainties
related to charged-lepton momentum resolutions.
The uncertainty due to limited-size simulated samples was also determined to be
negligible.
It was found that the neglected uncertainties did not meaningfully change the
\fsig uncertainty covariance when it was calculated for ten \pt bins.

In most cases, systematic uncertainties comprise two variations, \enquote{up} and
\enquote{down}, which are symmetrized around the nominal prediction:\ the magnitude of
the uncertainty $v$ is set to $|v| = |v_\mathrm{up} - v_\mathrm{down}| / 2$.
This symmetrized variation is generally assumed to correspond to the standard
deviation of a nuisance parameter prior or constraint.

An additional uncertainty in the estimate of the non-\bjet background is
derived by correcting $\pbkg(\qvec | \pt)$ in the $Z$+jets phase space,
as described in Section~\ref{sec:nonb_calibration}.
The difference between the $b$-jet calibration maps before and after applying
these corrections to non-\bjets in the \emuvvbb measurement
region is taken as a systematic uncertainty.
For each other systematic variation, the expected impact on the non-\bjet
background shape was found to be negligible compared to this difference.

In order to illustrate the total uncertainty of the calibrated simulation in
this study, each uncertainty's impact on the prediction is assessed, and
the total uncertainty is calculated in the Gaussian approximation as the
quadrature sum of all uncertainties.

% End of text imported from the .//uncertainties.tex input file

% The next lines are included from the .//results.tex input file
\section{Calibration results for \bjets}
\label{sec:results}

With continuous estimates of the \bjet purity as a function of jet \pt, \fsig,
and with the \bjet and background jet \qvec densities conditional on \pt,
\psigpt and \pbkgpt, the optimal calibration maps of the jet flavour
probabilities are derived from the simulated and observed jets.
In the following figures, $\psim^\prime(\qvec, \pt)$ is obtained from a
simulated sample 30 times larger than the data sample.

\subsection{Extracted optimal calibration maps and comparisons with data}

\begin{figure}
\centering
\subfigure[]{\includegraphics[width=0.32\textwidth]{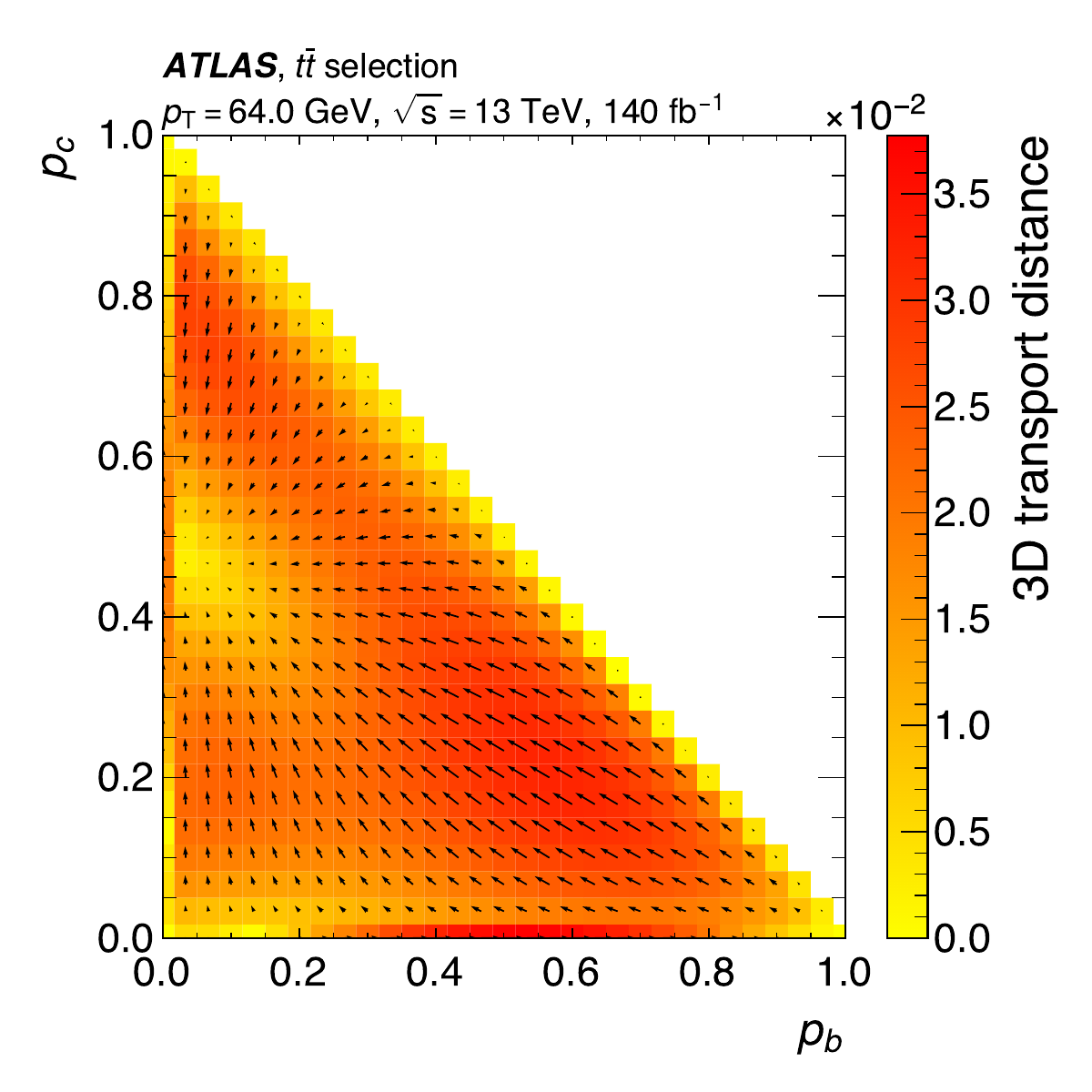}}
\subfigure[]{\includegraphics[width=0.32\textwidth]{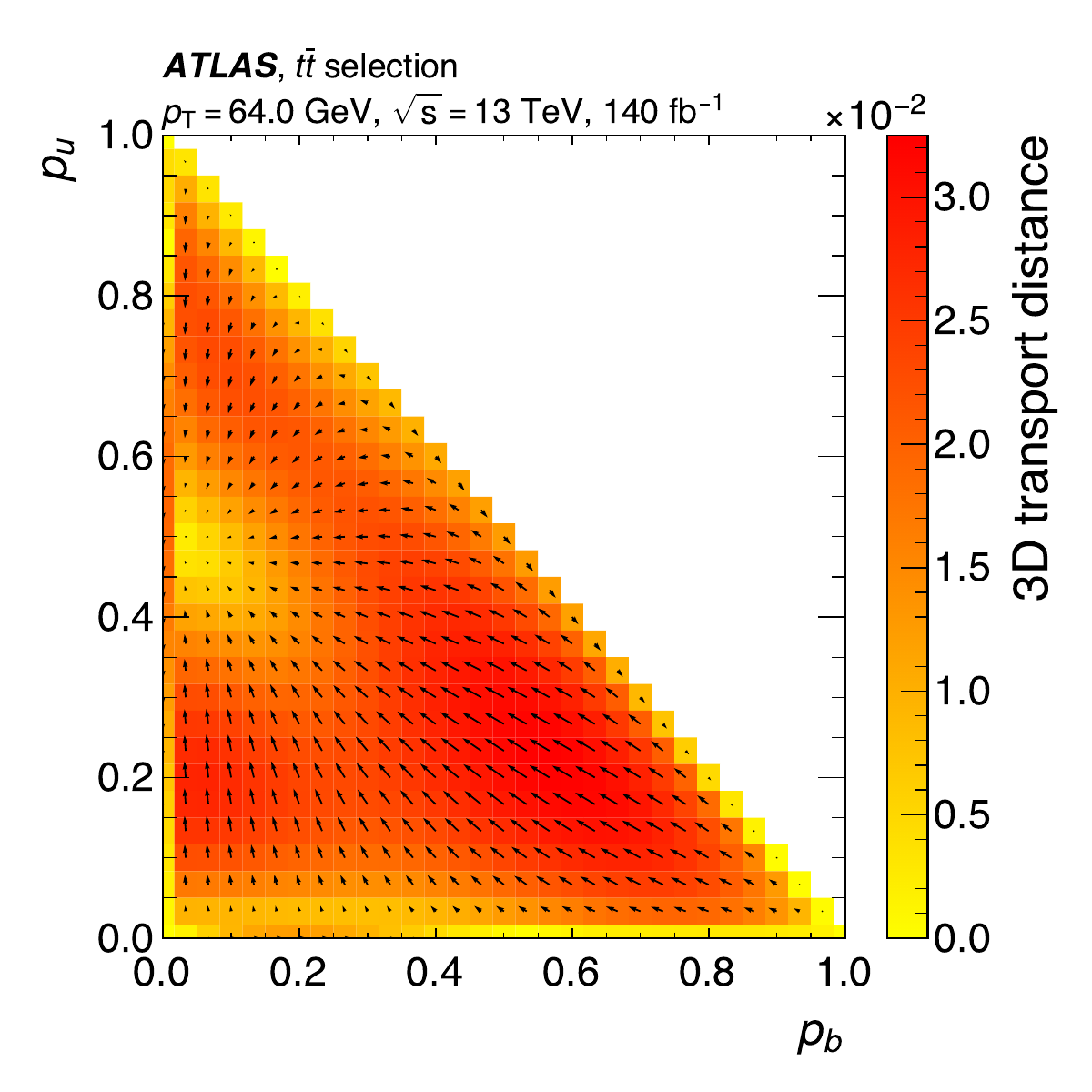}}
\subfigure[]{\includegraphics[width=0.32\textwidth]{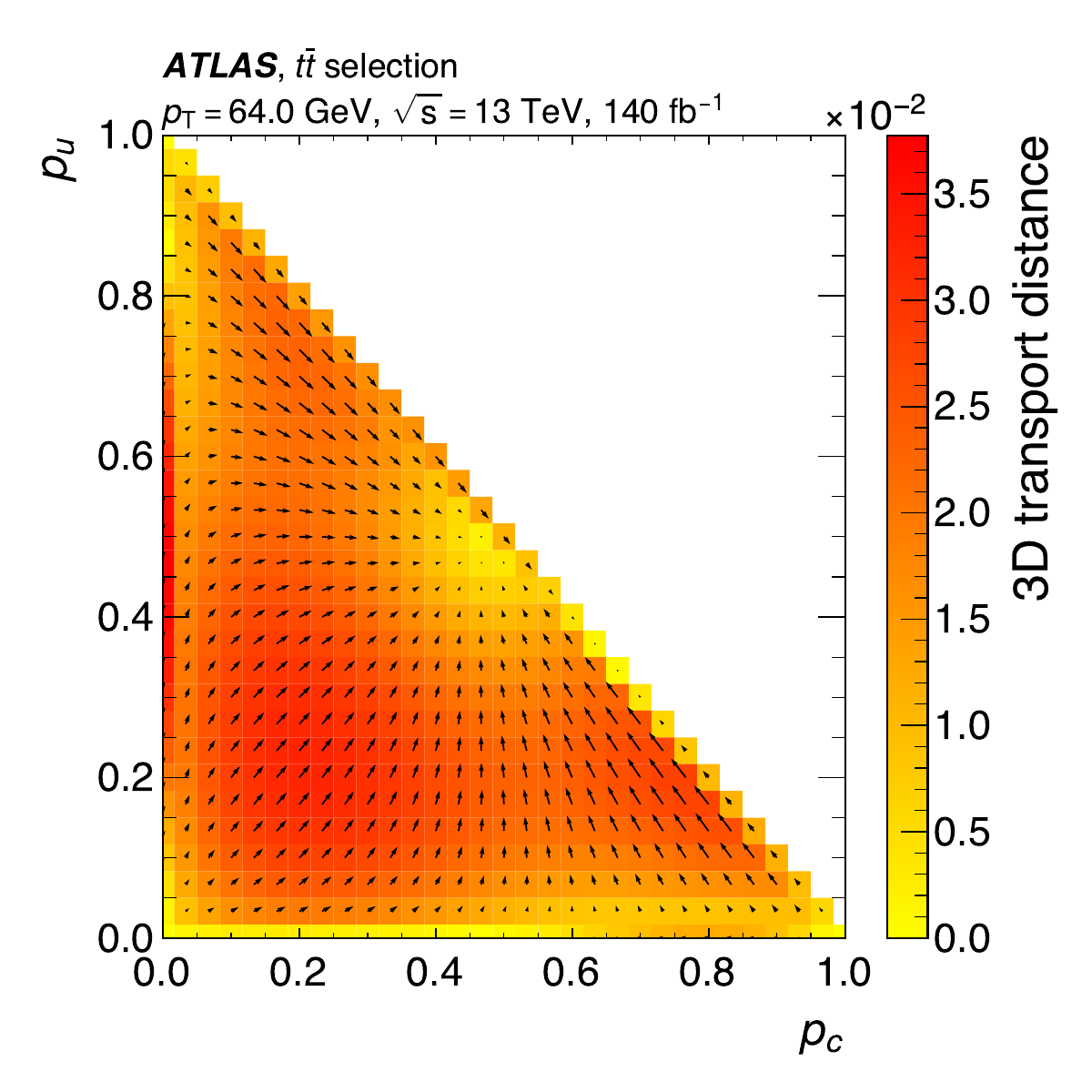}}
\caption{
\label{fig:transpmaps}
2D slices of the calibration transport maps for \bjets
with the mean jet \pt in the \ttbar phase space,
$\langle p_{\mathrm{T, data}}\rangle_{\ttbar}=64$~GeV:\ (a) \pb vs
\pc, (b) \pb vs \pu, and (c) \pc vs \pu.
Arrow directions and lengths correspond to the displacement of a \bjet at
the arrow's origin, $\hat T$, needed to obtain closure between the data and
simulation such that $\hat T_\#~\psim(\qvec | \pt) \approx \pdata(\qvec |
\pt)$.
The length of the arrows was normalized to avoid overlap and is
indicated by the colour of the background.
For each 2D map, the missing dimension is determined by requiring
$\sum_i p_i = 1$.
}
\end{figure}

\cref{fig:transpmaps} shows the 2D transport maps for \bjets evaluated at the
mean \pt in data in the \ttbar region, $\langle p_{\mathrm{T,
data}}\rangle_{\ttbar}=64$~GeV.
As in Section~\ref{sec:nonb_calibration}, the $p_i$ are converted into $q_i$,
the calibration applied, and the resulting $q^\prime_i$ translated back into
$p^\prime_i$ to calculate the calibration displacement with respect to the
original $p_i$.

\begin{figure}
\centering
\includegraphics[width=0.50\textwidth]{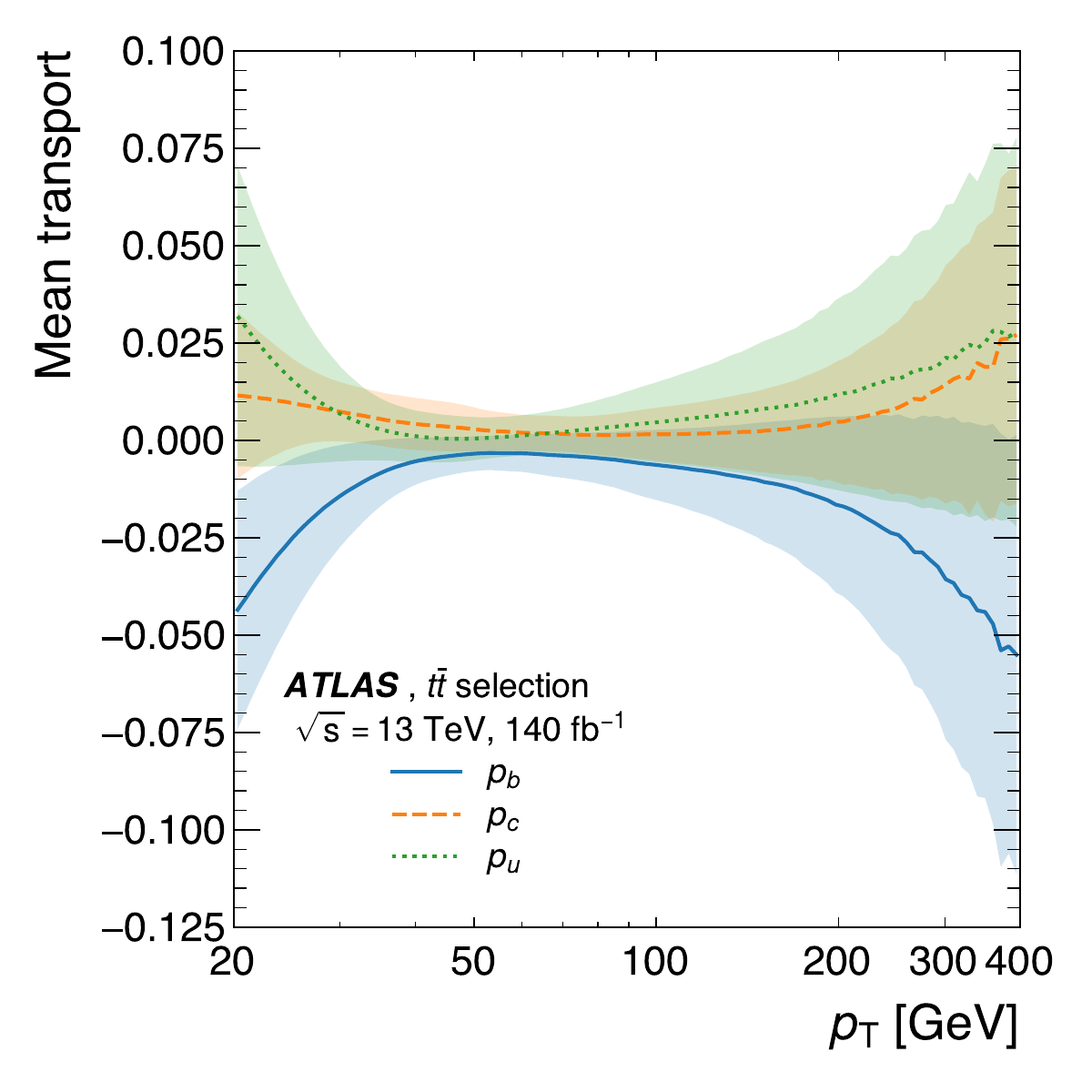}
\caption{
The mean and standard deviation of the 1D transport displacements $\delta
q_i$ as a function of \pt for \bjets in the \ttbar phase space.
The solid lines and shaded areas respectively indicate the mean and standard
deviation of the 1D displacements as a function of \pt.
Only the nominal transport map is shown, and dimensions in \qvec space that
are not shown are marginalized over $\psim^\prime$.
}
\label{fig:transport_displacement}
\end{figure}

Similarly, the mean 1D displacements for the nominal calibration,
$\delta \qvec_i \equiv (\Thatpt\, \qvec - \qvec)_i, i \in (b, c, u)$,
for each of the three components is shown in
Figure~\ref{fig:transport_displacement} as a function of \pt.
In the range $30 < \pt / \GeV < 100$ the average correction is modest
($|\delta q_i| \lesssim 0.1$) but grows to be as large as
$|\delta q_i| \sim 0.5{-}0.75$ at low and high \pt.
In particular, the mean \logitpb correction is negative for all \pt in the
calibration range, while for \logitpu and \logitpc the mean correction is always
positive: the tagging algorithm slightly overstates its separating power between
$b$- and non-\bjets in data, especially at low- and high-\pt.

\begin{figure}
\centering
\subfigure[]{\includegraphics[width=0.32\textwidth]{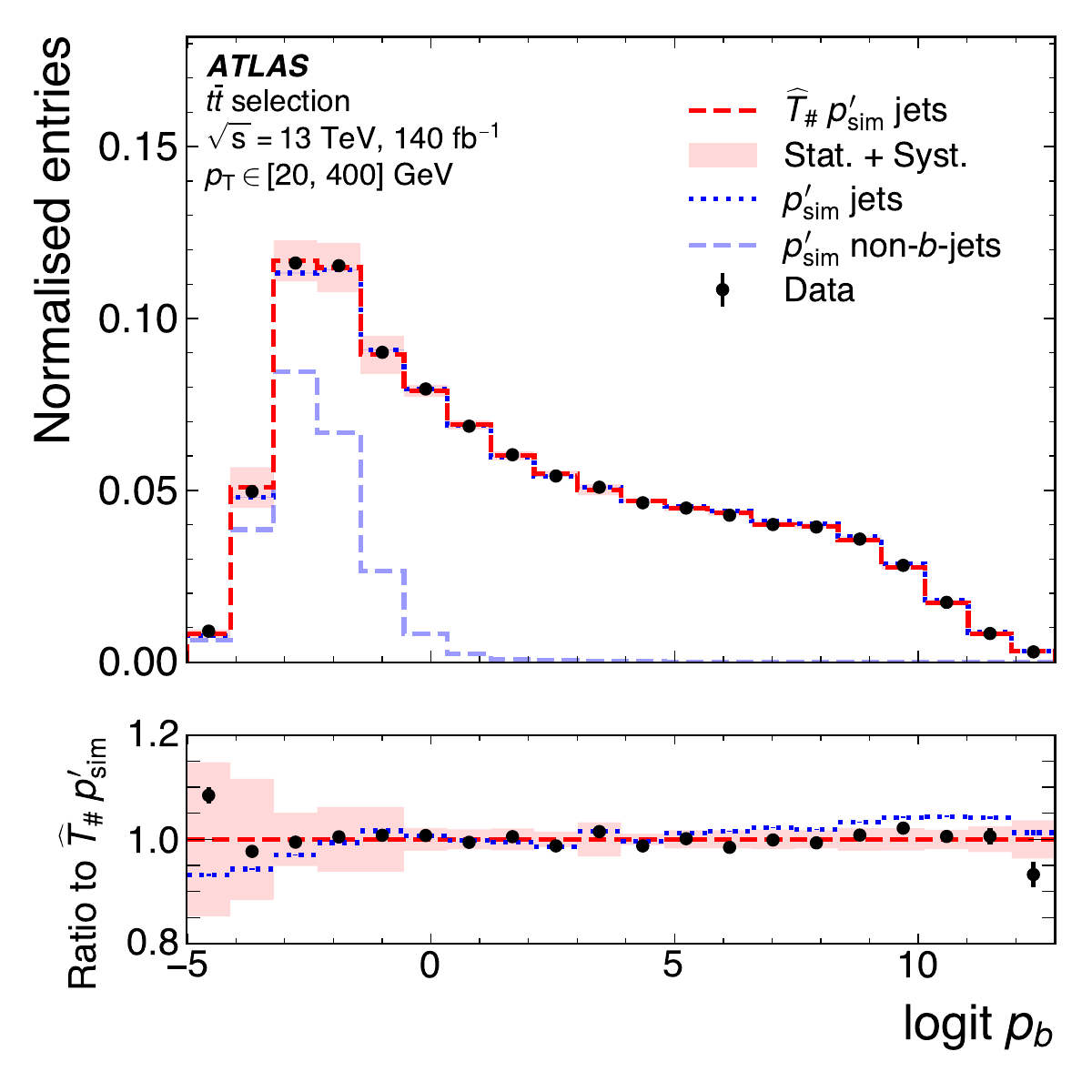}}
\subfigure[]{\includegraphics[width=0.32\textwidth]{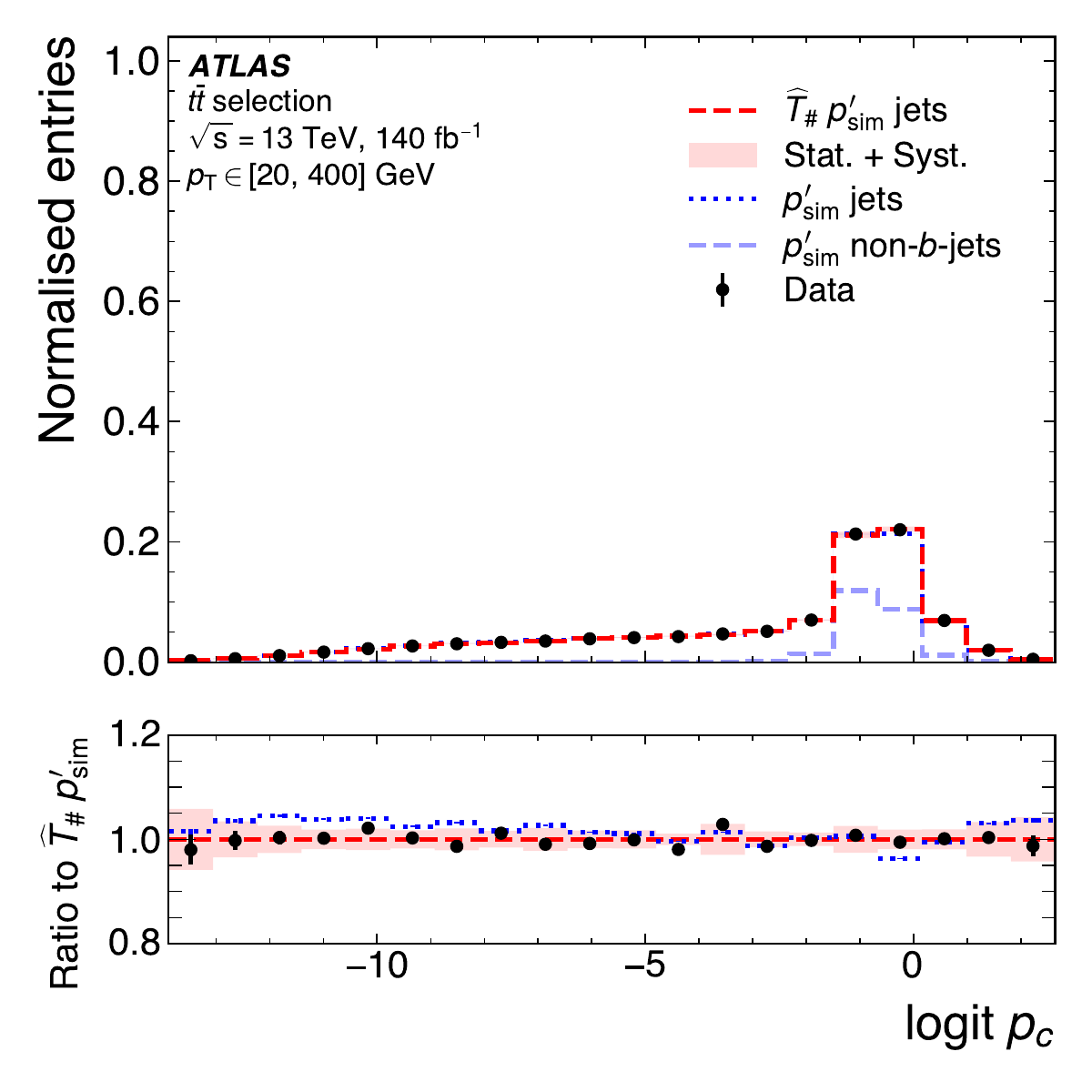}}
\subfigure[]{\includegraphics[width=0.32\textwidth]{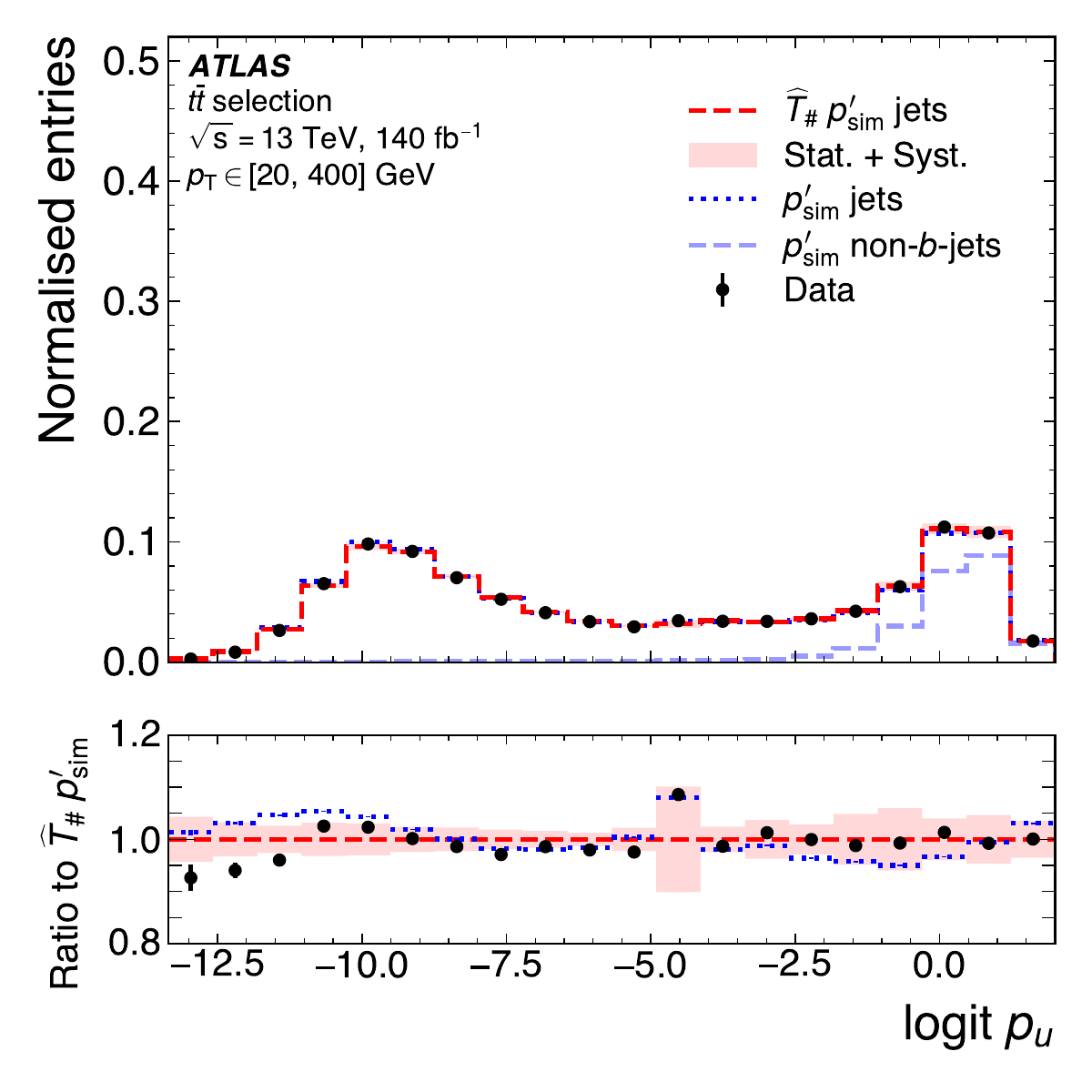}}
\caption{
\label{fig:closure_logits}
Comparisons between the observed data, uncalibrated
simulation, $\psim^\prime$, and corrected simulation, $\hat T_\#
\psim^\prime$, for (a) \logitpb, (b) \logitpc, and (c) \logitpu.
Both the non-\bjet and total (non-\bjet + \bjet) normalized yields are shown
for simulation.
The uncertainty band of the corrected simulation denotes the impact of
all uncertainties described in \cref{sec:uncertainties} on the calibration maps.
All marginals are shown for the \pt-corrected density
$\psim^\prime(\qvec,\pt) = \psim(\qvec | \pt) \pdata(\pt)$.
Non-\bjets in simulation were corrected using transport maps derived in
\cref{sec:nonb_calibration}.
}
\end{figure}

\begin{figure}
\centering
\subfigure[]{\includegraphics[width=0.32\textwidth]{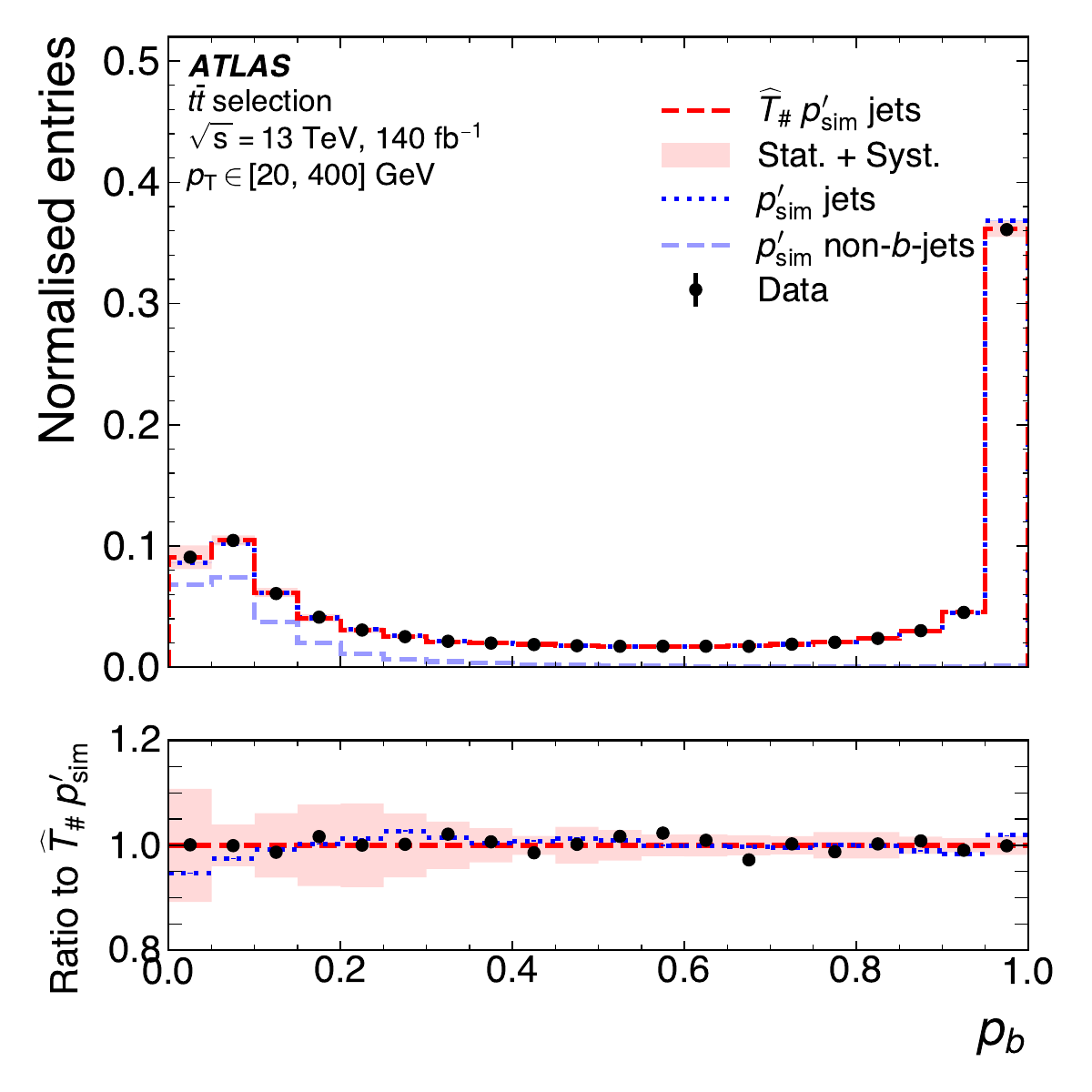}}
\subfigure[]{\includegraphics[width=0.32\textwidth]{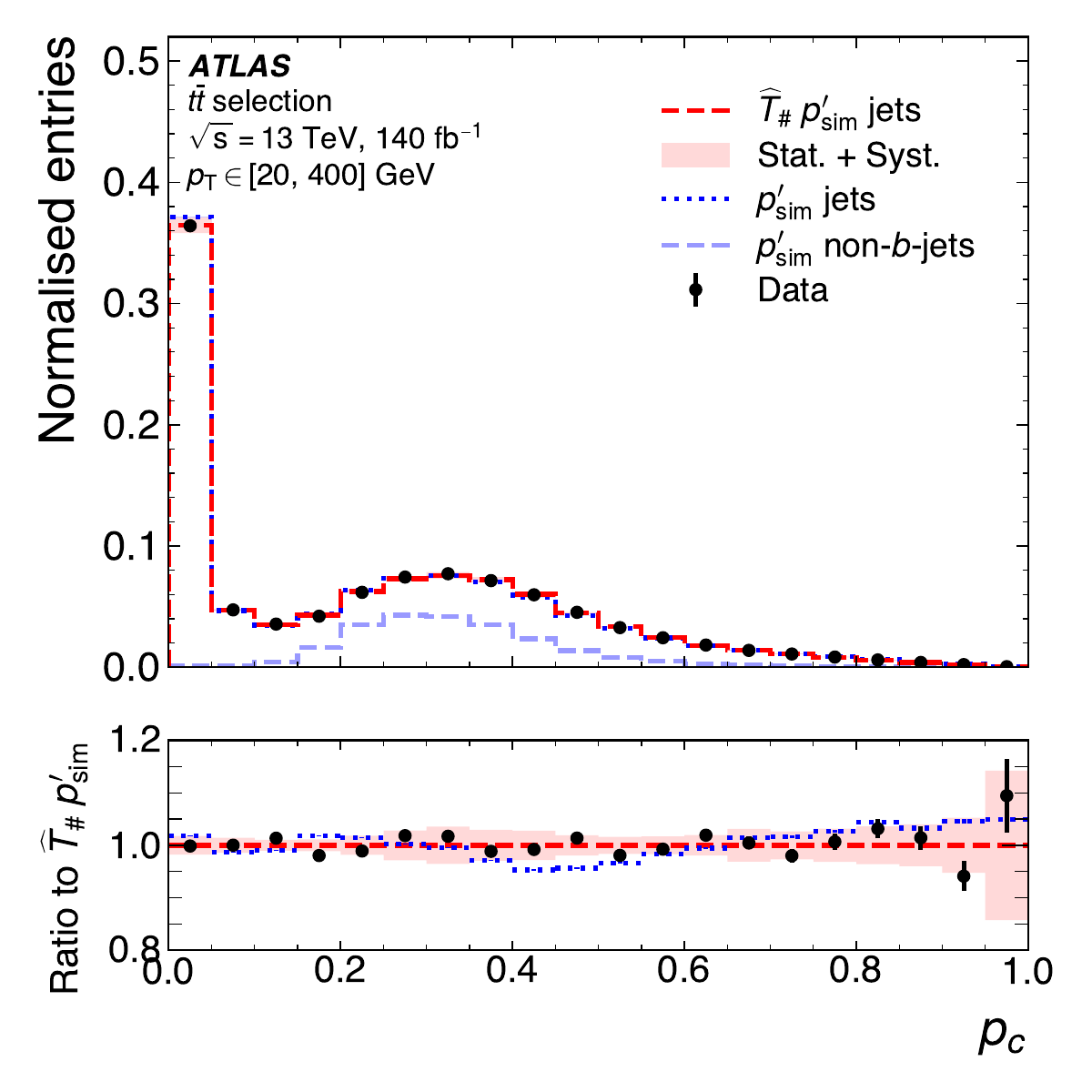}}
\subfigure[]{\includegraphics[width=0.32\textwidth]{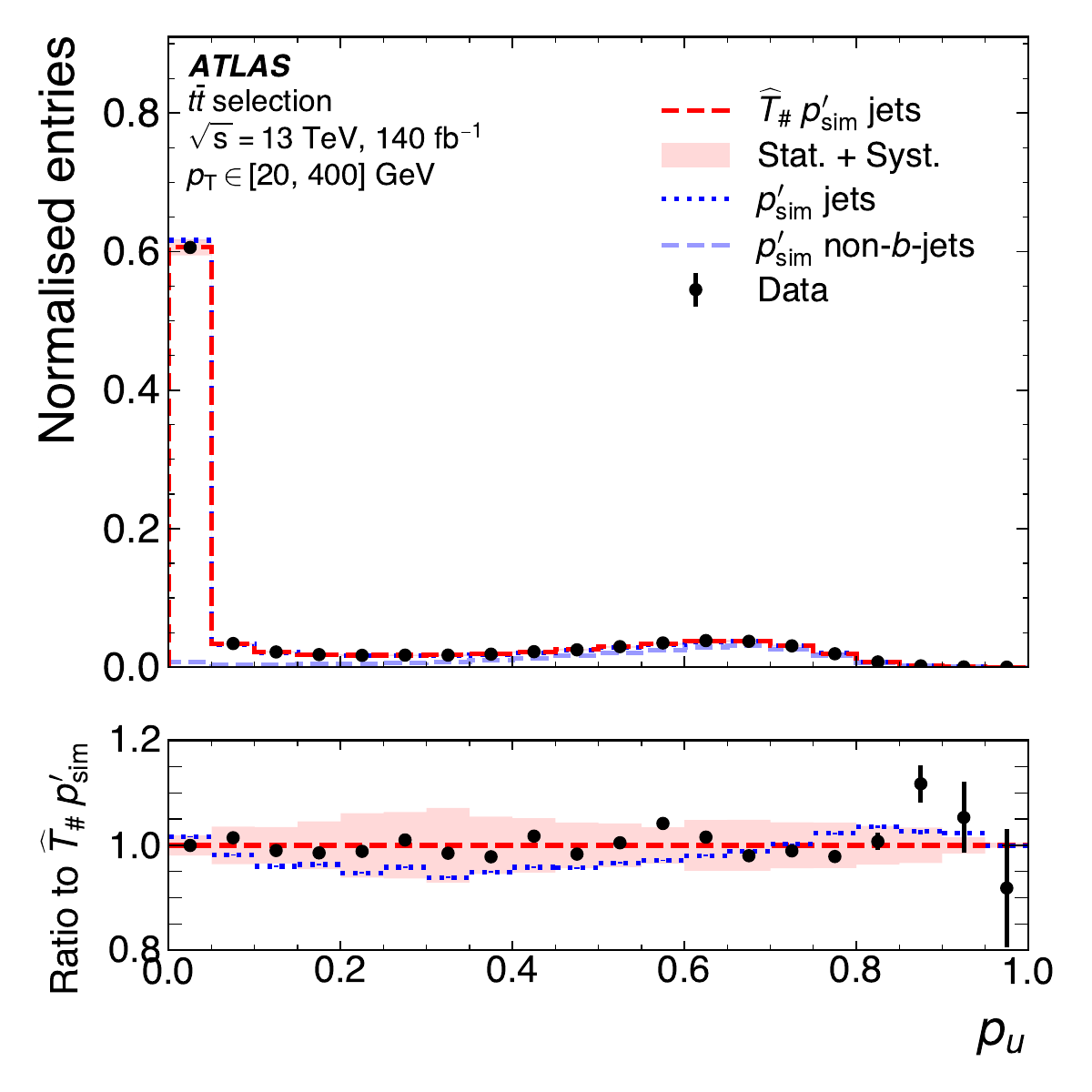}}
\caption{
Comparisons between the observed data, uncalibrated
simulation, $\psim^\prime$, and corrected simulation, $\hat T_\#
\psim^\prime$, for (a) \pb, (b) \pc, and (c) \pu.
Both the non-\bjet and total (non-\bjet + \bjet) normalized yields are shown
for simulation.
The uncertainty band of the corrected simulation denotes the impact of
all uncertainties described in \cref{sec:uncertainties} on the calibration maps.
All marginals are shown for the \pt-corrected density
$\psim^\prime(\qvec,\pt) = \psim(\qvec | \pt) \pdata(\pt)$.
Non-\bjets in simulation were corrected using transport maps derived in
\cref{sec:nonb_calibration}.
}
\label{fig:closure_probs}
\end{figure}

\begin{figure}
\centering
\subfigure[]{
\includegraphics[width=0.4\textwidth]{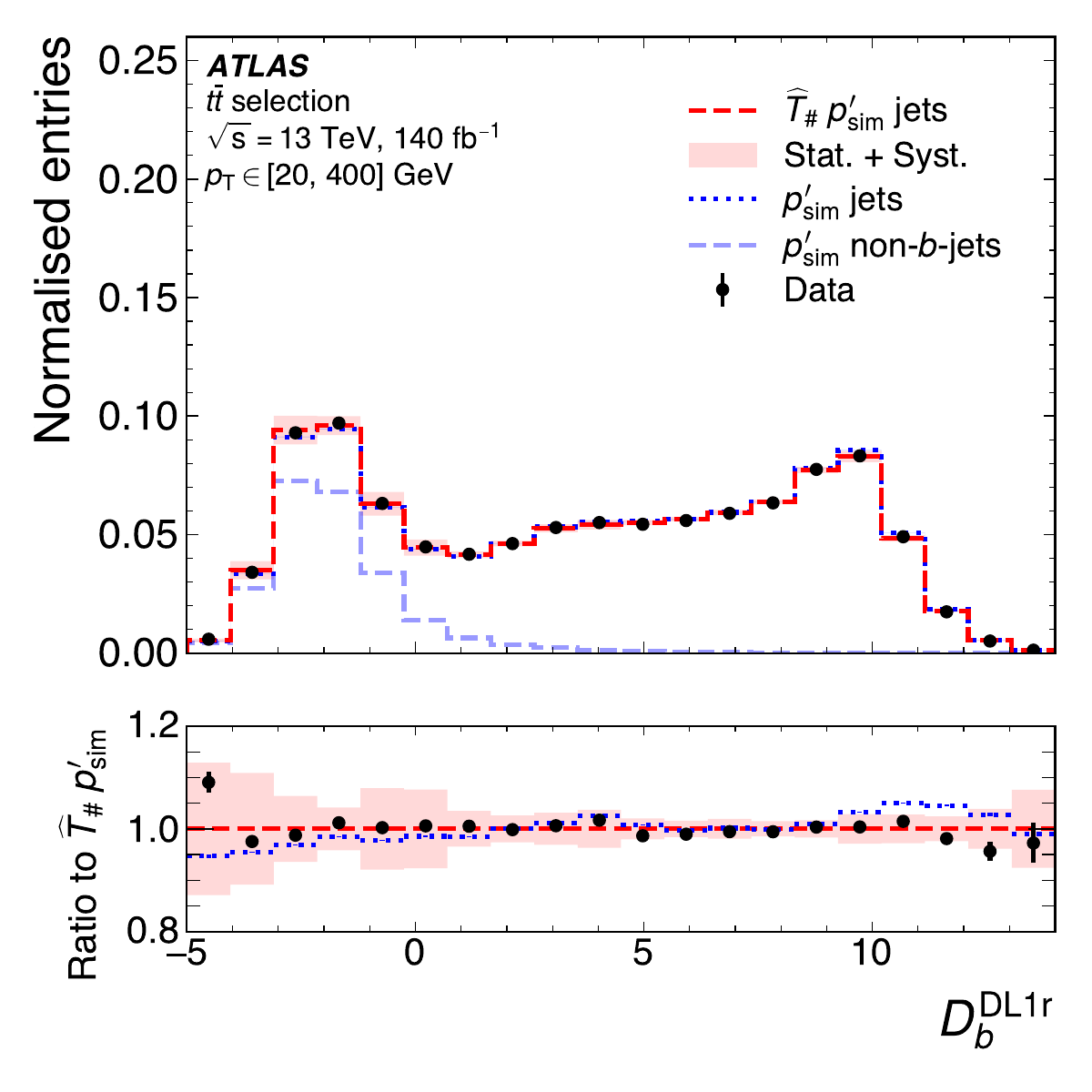}
}
\subfigure[]{
\includegraphics[width=0.4\textwidth]{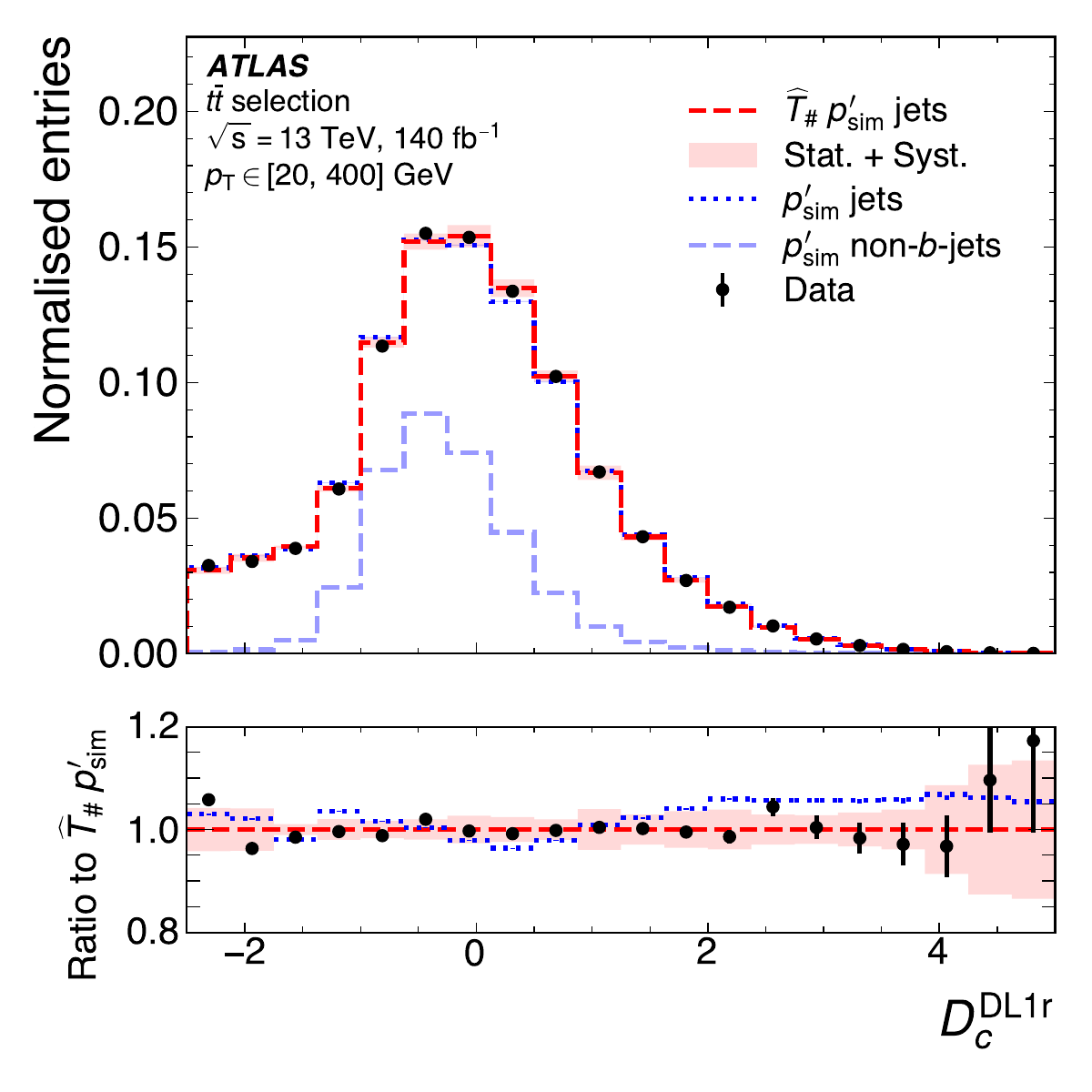}
}
\caption{
Comparisons between the observed data, uncalibrated
simulation, $\psim^\prime$, and corrected simulation, $\hat T_\#
\psim^\prime$, for (a) \dlonerb and (b) \dlonerc.
Both the non-\bjet and total (non-\bjet + \bjet) normalized yields are shown
for simulation.
The uncertainty band of the corrected simulation denotes the impact of
all uncertainties described in \cref{sec:uncertainties} on the calibration
maps.
All marginals are shown for the \pt-corrected density
$\psim^\prime(\qvec,\pt) = \psim(\qvec | \pt) \pdata(\pt)$.
Non-\bjets in simulation were corrected using transport maps derived in
\cref{sec:nonb_calibration}.
}
\label{fig:closure_dloner_main}
\end{figure}

Figures~\ref{fig:closure_logits},~\ref{fig:closure_probs},
and~\ref{fig:closure_dloner_main} compare distributions from jets in data and
simulation, before and after calibrations are applied, for \qvec, the
flavour-tagging probabilities, and \dloner $b$- and $c$-tagging variants,
respectively.
In all cases, the calibrated distributions are in agreement with the data.


\subsection{Closure tests for common \dlonerb operating points}
\label{subsec:dlonerb-closure}

\begin{figure}
\centering
\subfigure[]{
\includegraphics[width=0.4\textwidth]{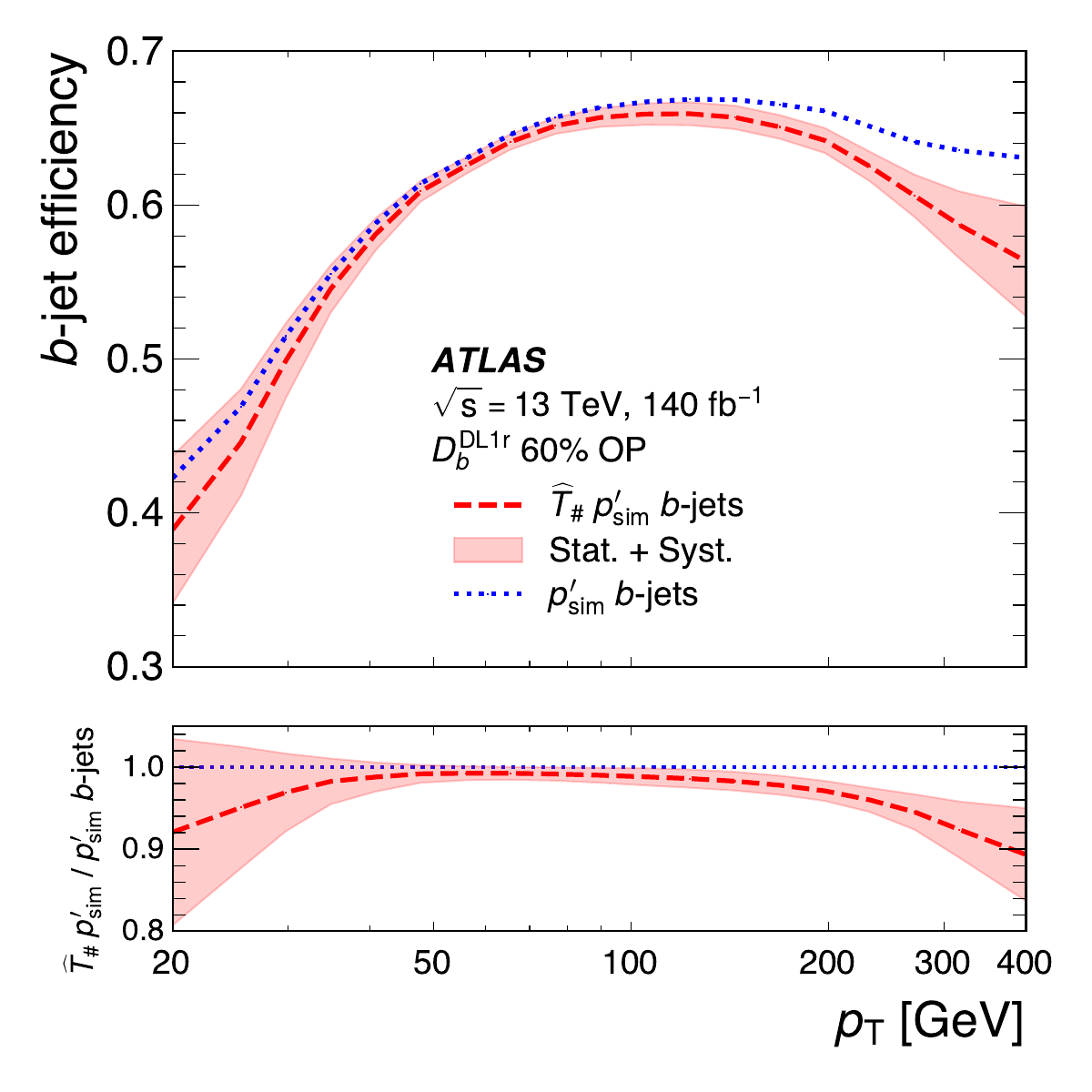}
}
\subfigure[]{
\includegraphics[width=0.4\textwidth]{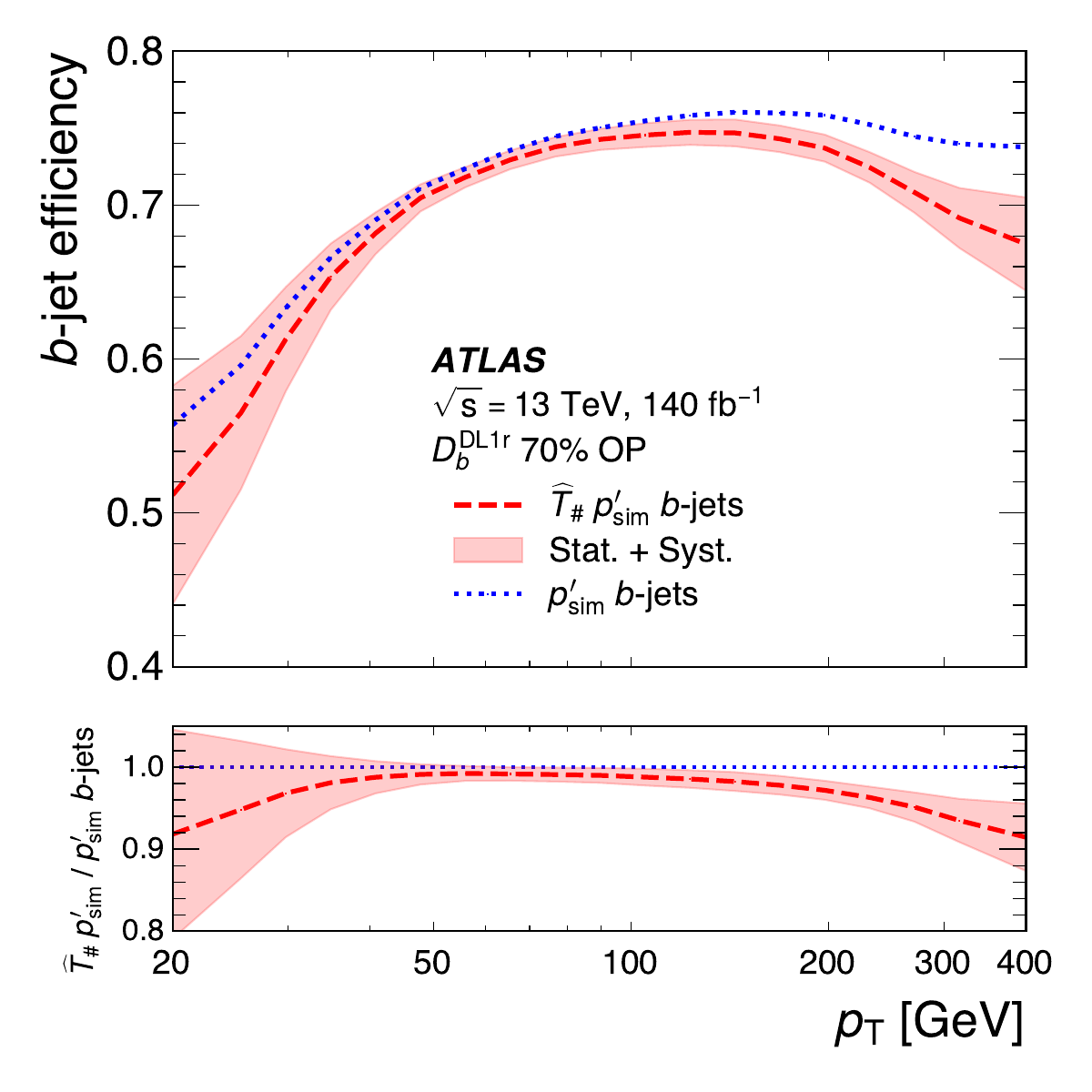}
} \\
\subfigure[]{
\includegraphics[width=0.4\textwidth]{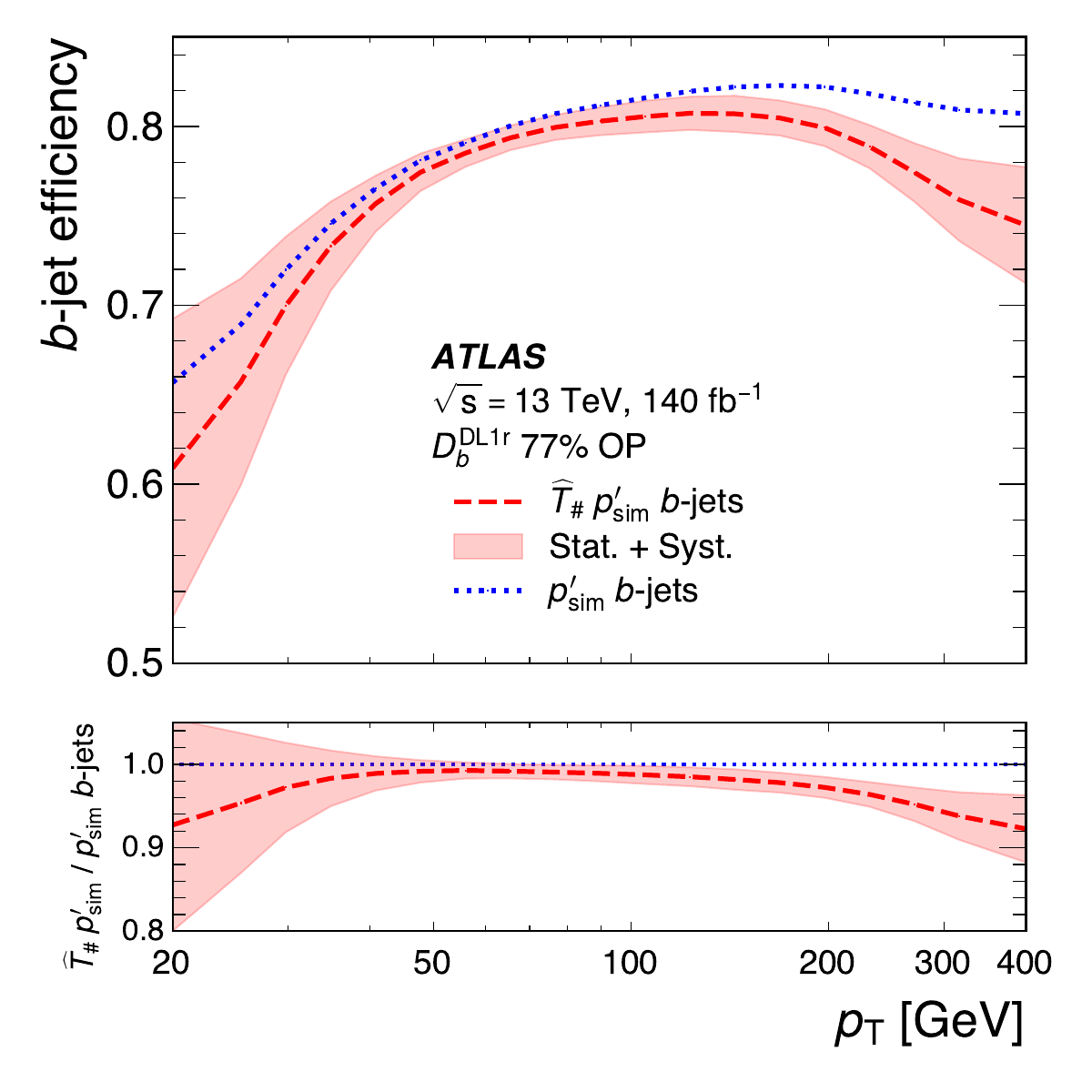}
}
\subfigure[]{
\includegraphics[width=0.4\textwidth]{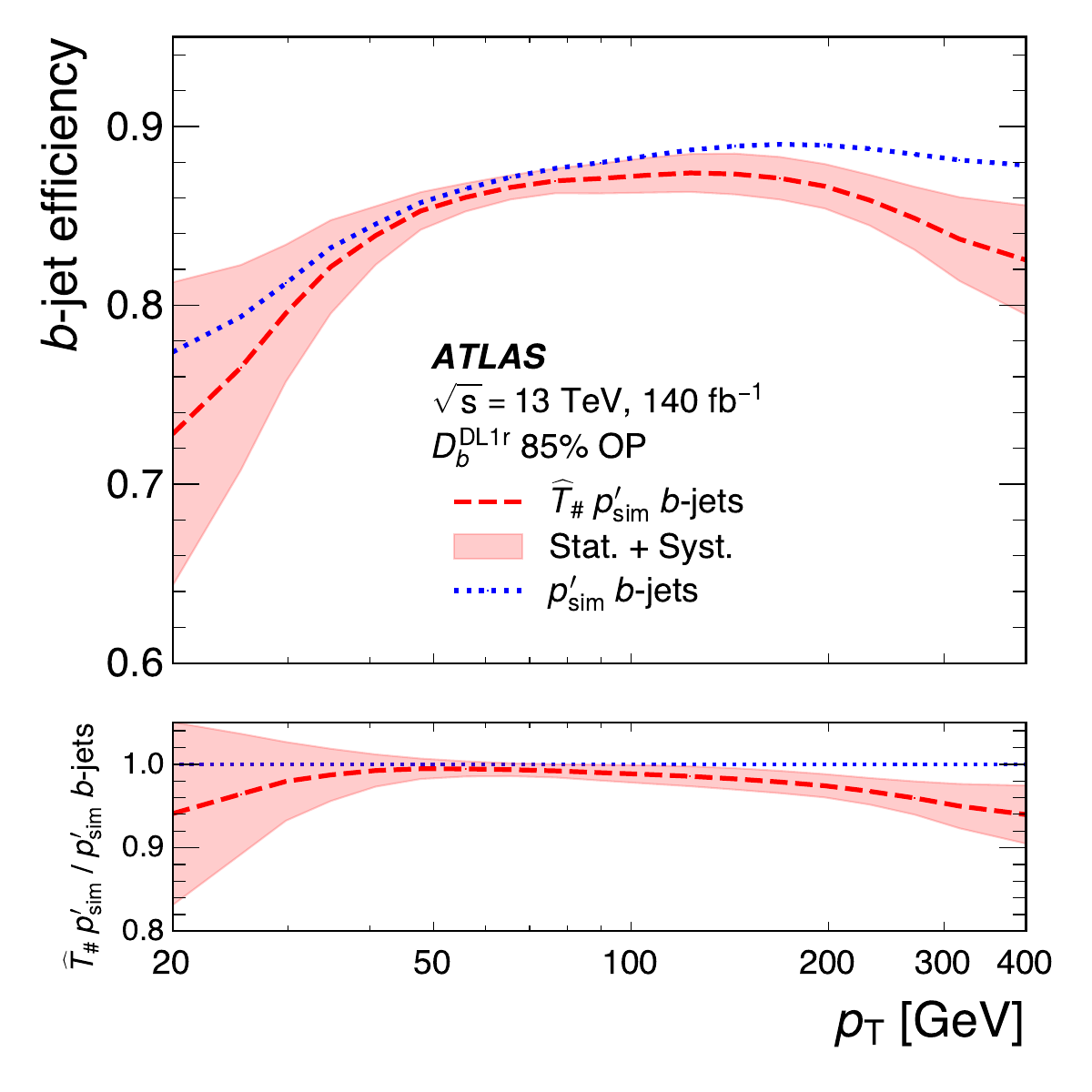}
}
\caption{
\label{fig:Floweffs_dlonerb}
Comparison between \dlonerb \bjet efficiencies predicted by the nominal
simulation, $\psim^\prime$, and the calibrated simulation, $\hat T_\#
\psim^\prime$.
The \dlonerb (a) 60\%, (b) 70\%, (c) 77\%, and (d) 85\% efficiency operating
points are shown.
The uncertainty band of the calibrated simulation denotes the impact of
all uncertainties described in \cref{sec:uncertainties} on the calibration
maps.
}
\end{figure}

\begin{figure}
\centering
\subfigure[]{
\includegraphics[width=0.4\textwidth]{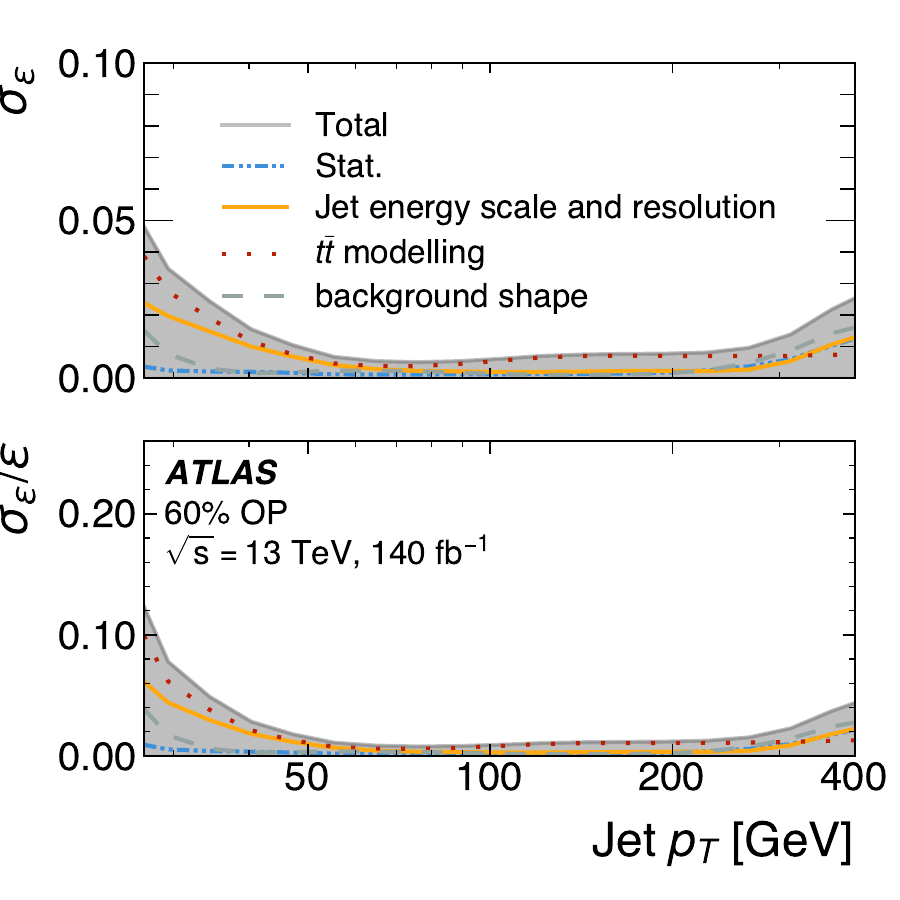}
}
\subfigure[]{
\includegraphics[width=0.4\textwidth]{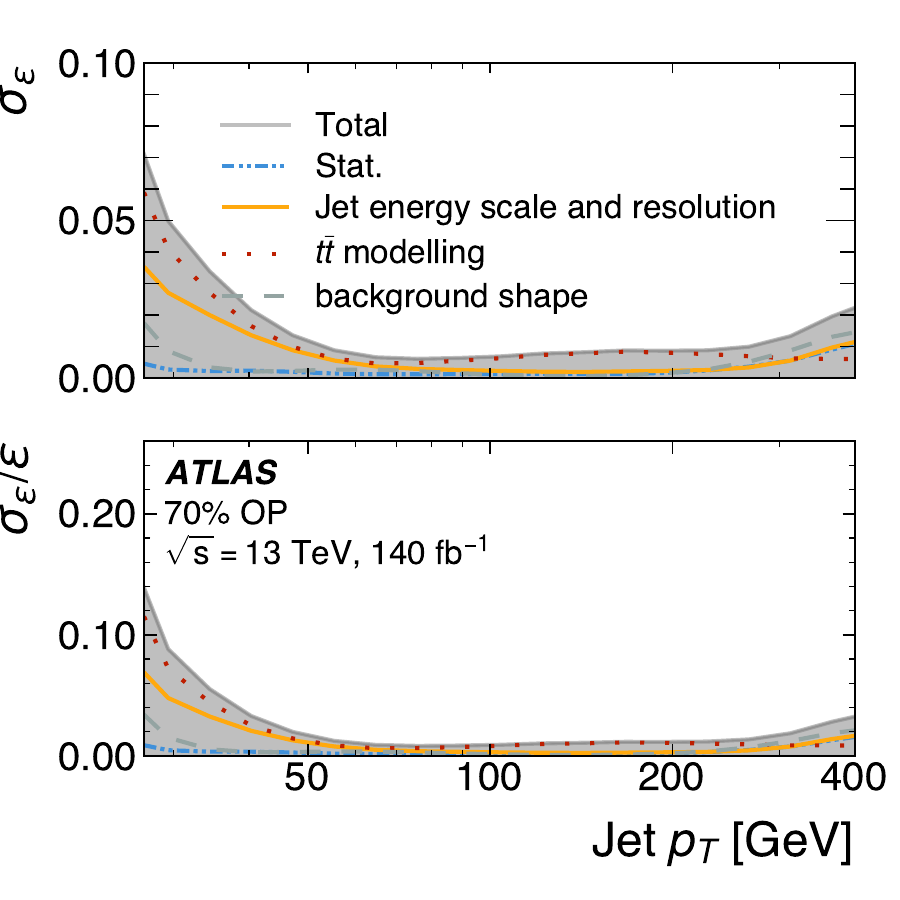}
} \\
\subfigure[]{
\includegraphics[width=0.4\textwidth]{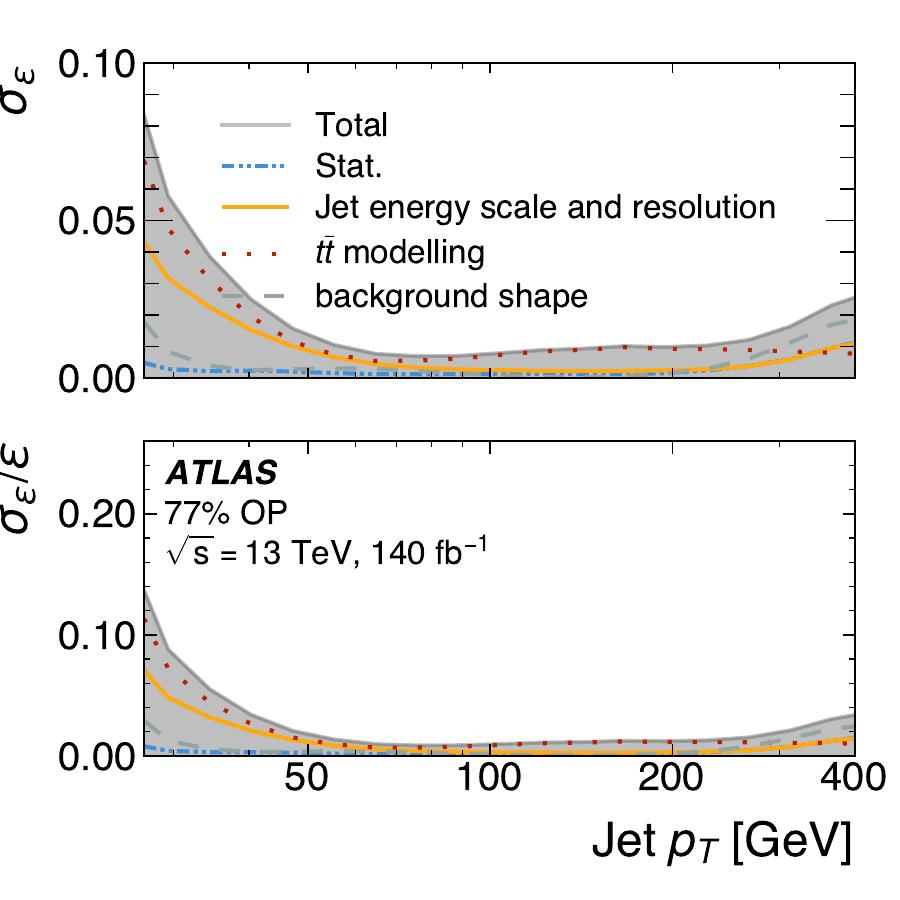}
}
\subfigure[]{
\includegraphics[width=0.4\textwidth]{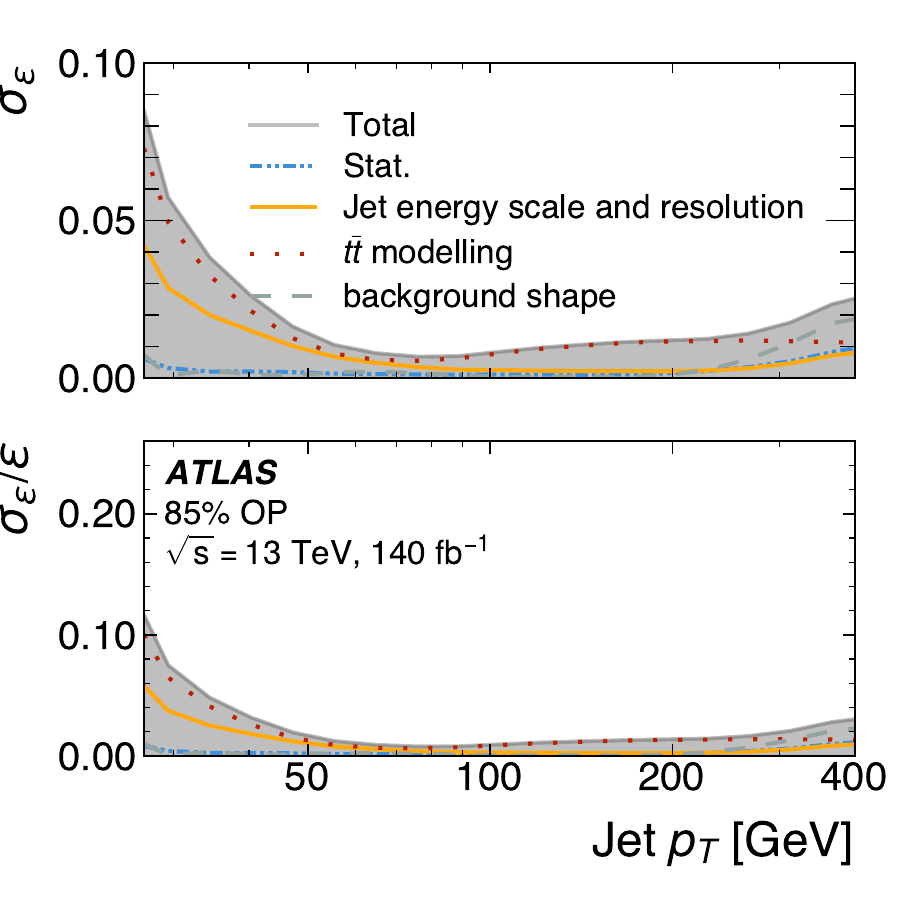}
} \\
\caption{
\label{fig:uncertainties_dl1rb}
Uncertainty breakdown of the calibrated simulation, $\hat T_\#
\psim^\prime$, for \dlonerb \bjet efficiency versus jet
\pt for various \btagging operating points.
The top panel shows the
absolute uncertainties, while the bottom panel shows the relative uncertainties.
The (a) 60\%, (b) 70\%, (c) 77\%, and (d) 85\% efficiency operating
points are shown.
The grey uncertainty band of the calibrated simulation denotes the impact of
all uncertainties described in \cref{sec:uncertainties} on the calibration
maps.
}
\end{figure}

Given the optimal transport maps, efficiencies can be obtained for a given
selection built from the jet flavour probabilities and compared with the
uncalibrated simulation and LHC data:\ if the simulation has been calibrated
adequately, efficiencies in simulation should match those in data.
In this section the four common \dlonerb selections used by the ATLAS
Collaboration during LHC Run~2 (defined as the 85\%, 77\%, 70\%, and 60\%
efficiency operating points derived in SM \ttbar events with at least one prompt
final-state lepton) are considered~\cite{FTAG-2019-07}.
These operating points amount to requiring that the jet \dlonerb score is above
0.67, 2.20, 3.25, or 4.57, respectively.
The \bjet efficiencies in the uncalibrated and calibrated
simulations are compared in \cref{fig:Floweffs_dlonerb} as a
function of the jet transverse momentum for the same operating points, highlighting the impact of the
calibration in correcting the simulation for these selections.
The absolute and relative uncertainties on the b-tagging efficiency vs the jet
\pt for these operating points are shown in
Figure~\ref{fig:uncertainties_dl1rb}.

\begin{figure}
\centering
\subfigure[]{
\includegraphics[width=0.4\textwidth]{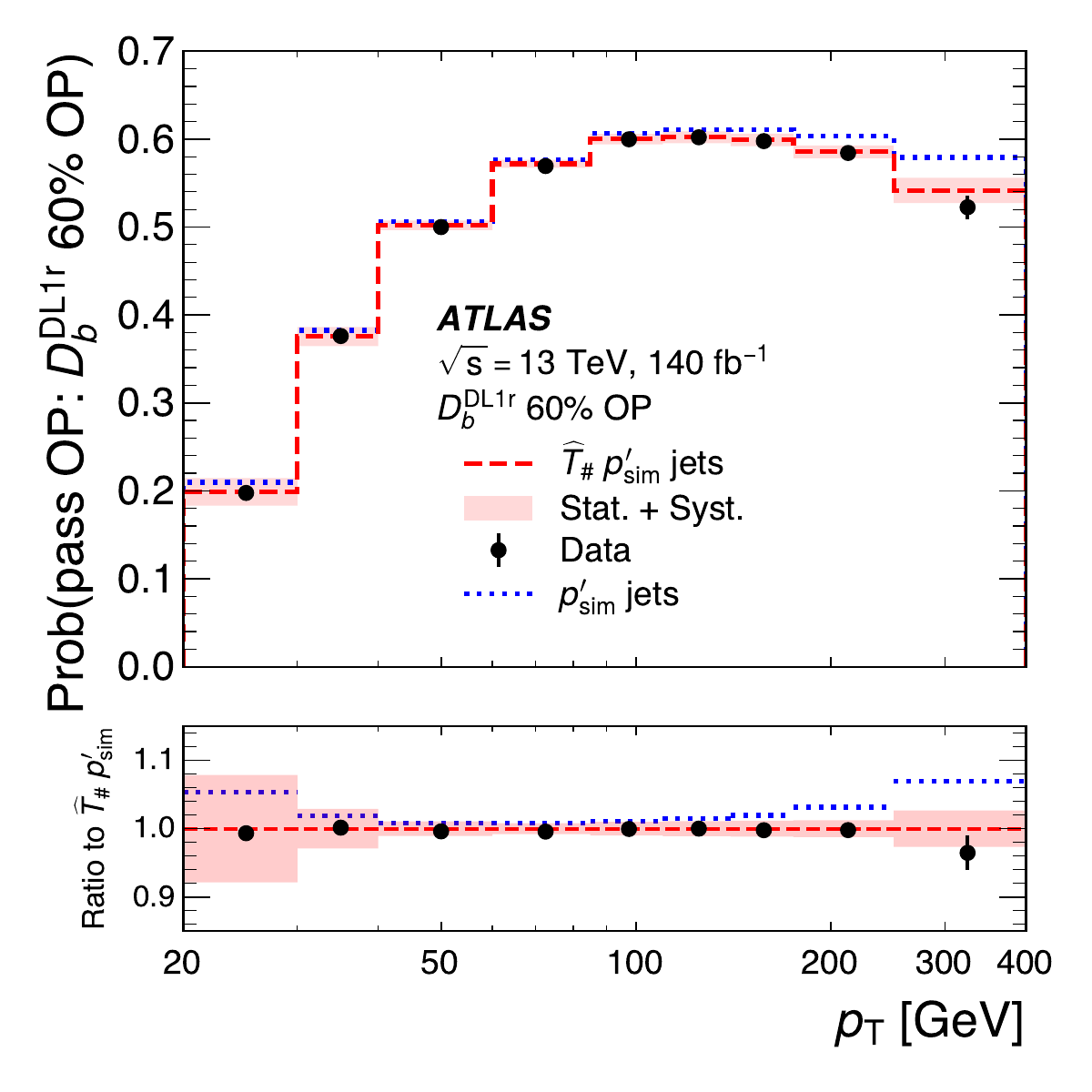}
}
\subfigure[]{
\includegraphics[width=0.4\textwidth]{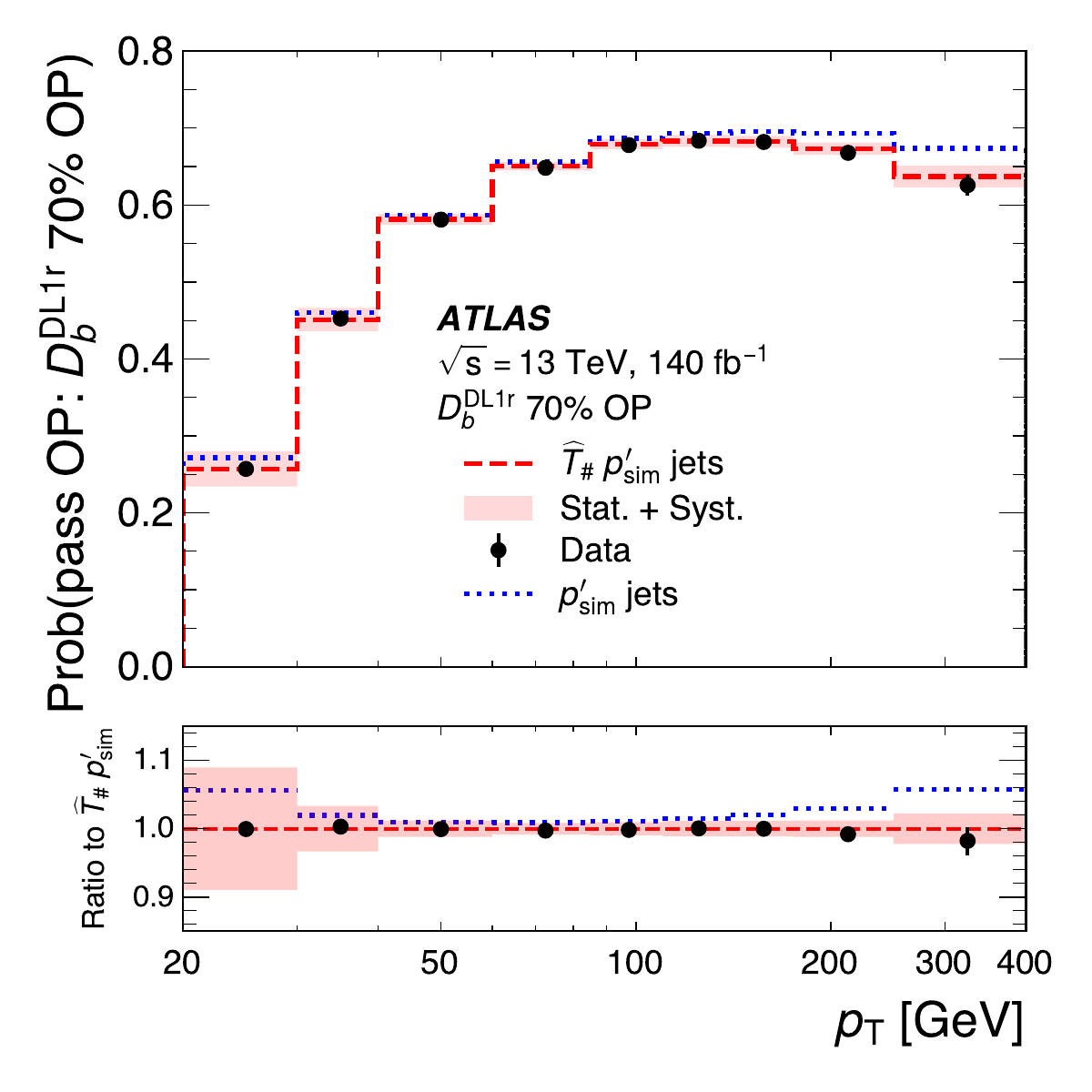}
} \\
\subfigure[]{
\includegraphics[width=0.4\textwidth]{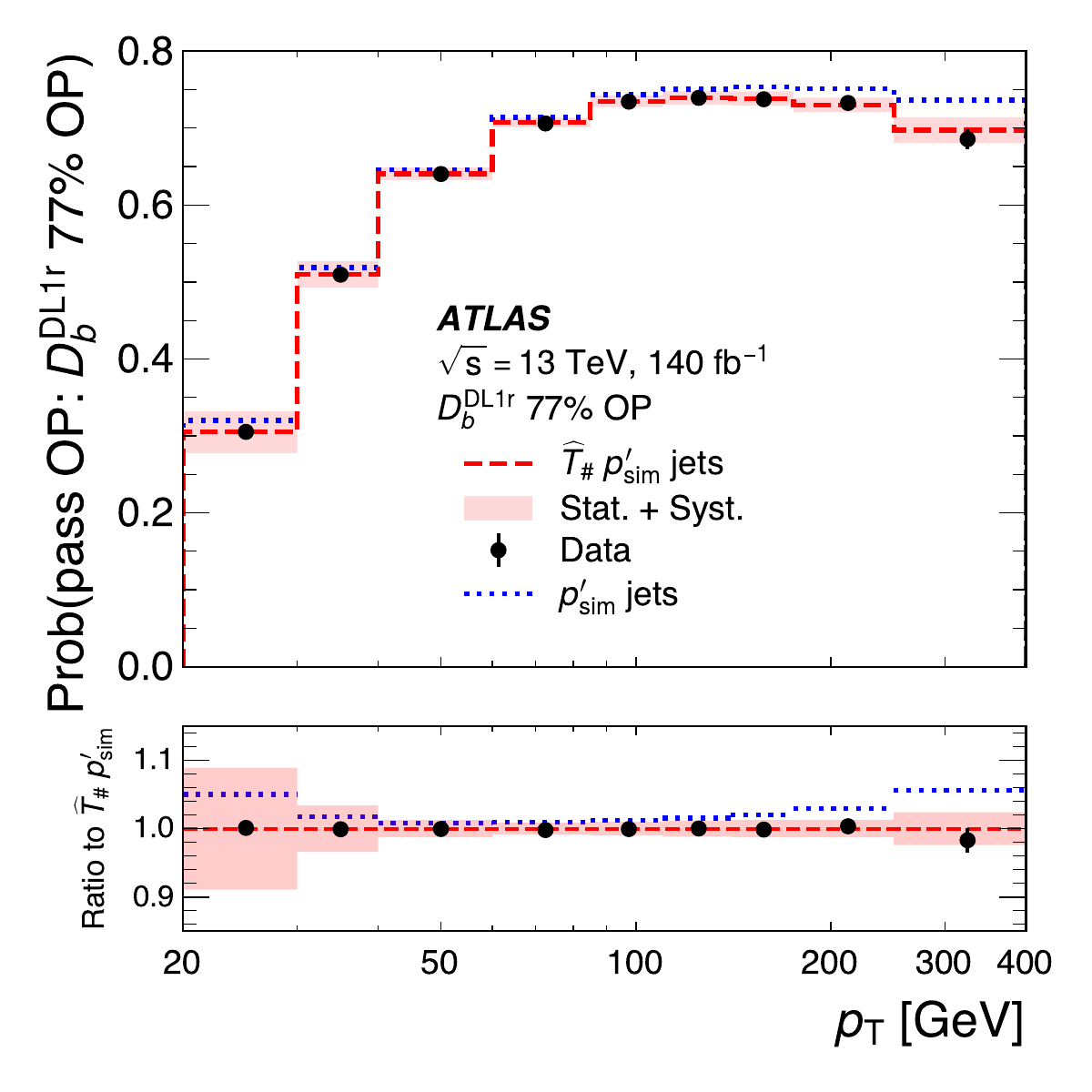}
}
\subfigure[]{
\includegraphics[width=0.4\textwidth]{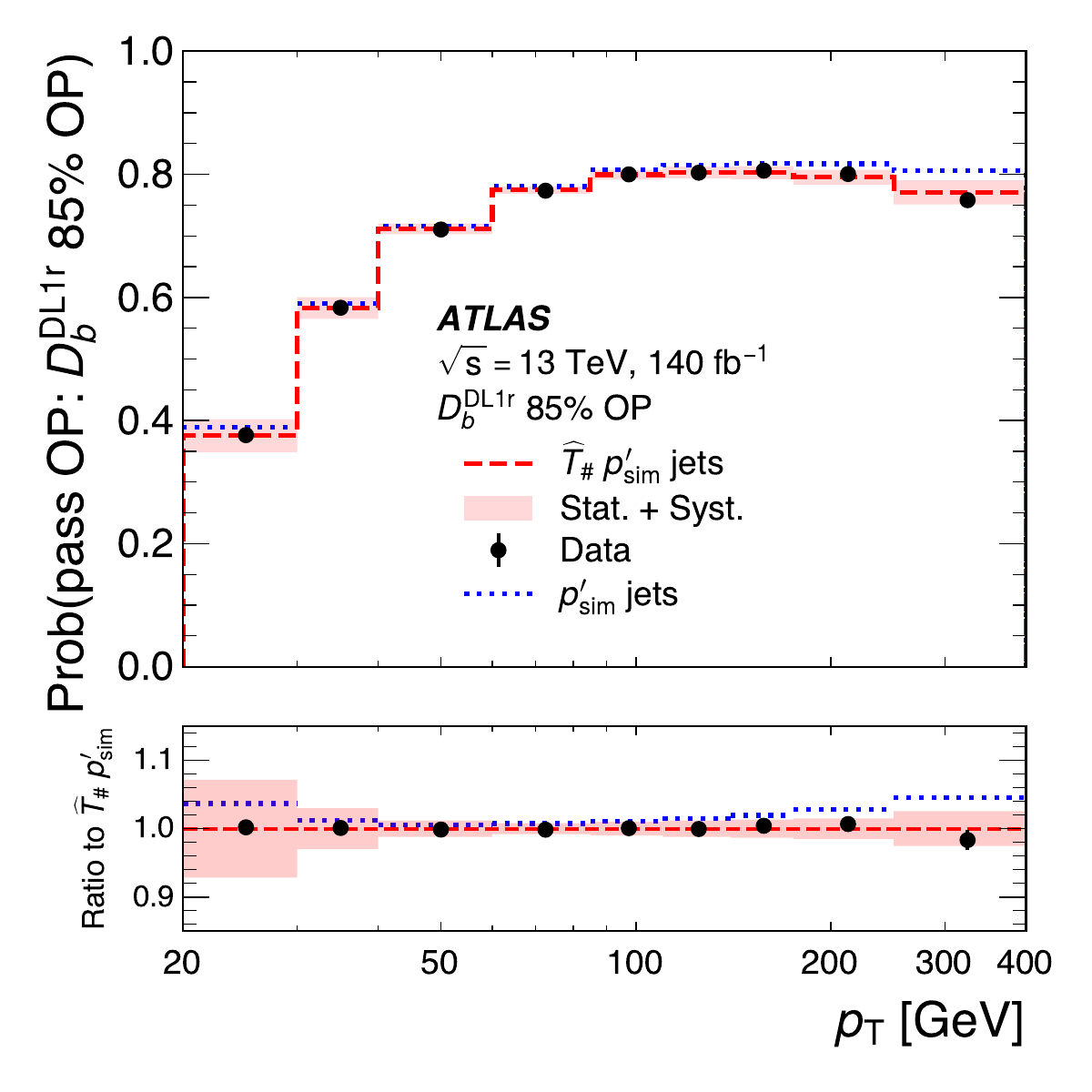}
}
\caption{
\label{fig:dataeffs}
Comparison of the probability of a selected jet to pass various \dlonerb
operating points, between data, the nominal simulated prediction,
$\psim^\prime$, and the calibrated simulation, $\hat T_\# \psim^\prime$, as
a function of the jet \pt.
The \dlonerb 60\%, (b) 70\%, (c) 77\%, and (d) 85\% efficiency operating
points are shown.
The uncertainty band of the calibrated simulation shows the impact of all
uncertainties described in \cref{sec:uncertainties} on the calibration maps.
Non-\bjets in simulation were corrected using transport maps derived in
\cref{sec:nonb_calibration} and are included in curves for the nominal and
calibrated simulation.
}
\end{figure}

After applying the calibration, the probability that simulated jets satisfy some
selection criterion should match the probability observed in data.
The \btagging selection probabilities in simulation and data are compared as
a function of the jet \pt in \cref{fig:dataeffs}; the corrected simulation matches
the observed data well within the calibration uncertainties.

\begin{figure}
\centering
\subfigure[]{
\includegraphics[width=0.4\textwidth]{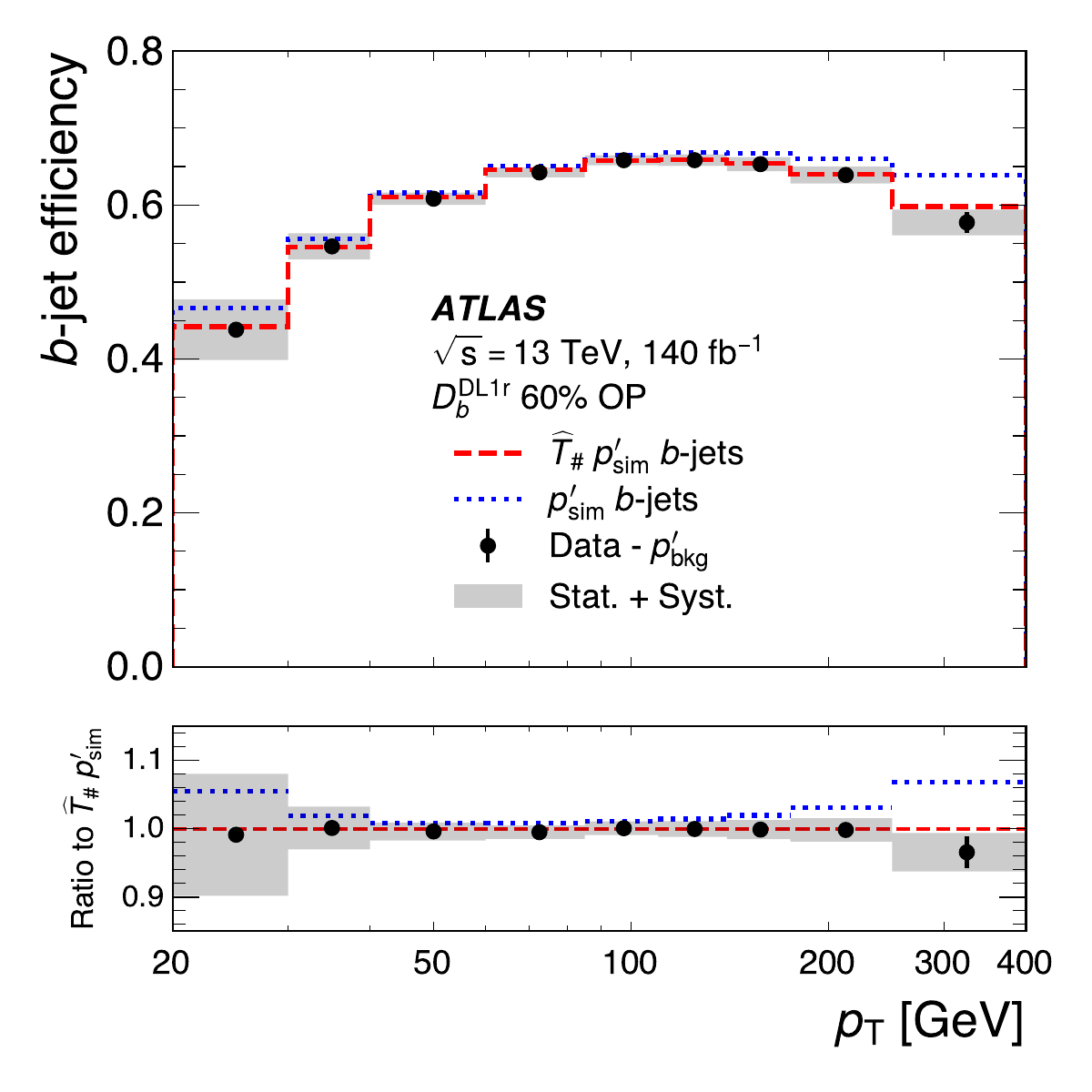}
}
\subfigure[]{
\includegraphics[width=0.4\textwidth]{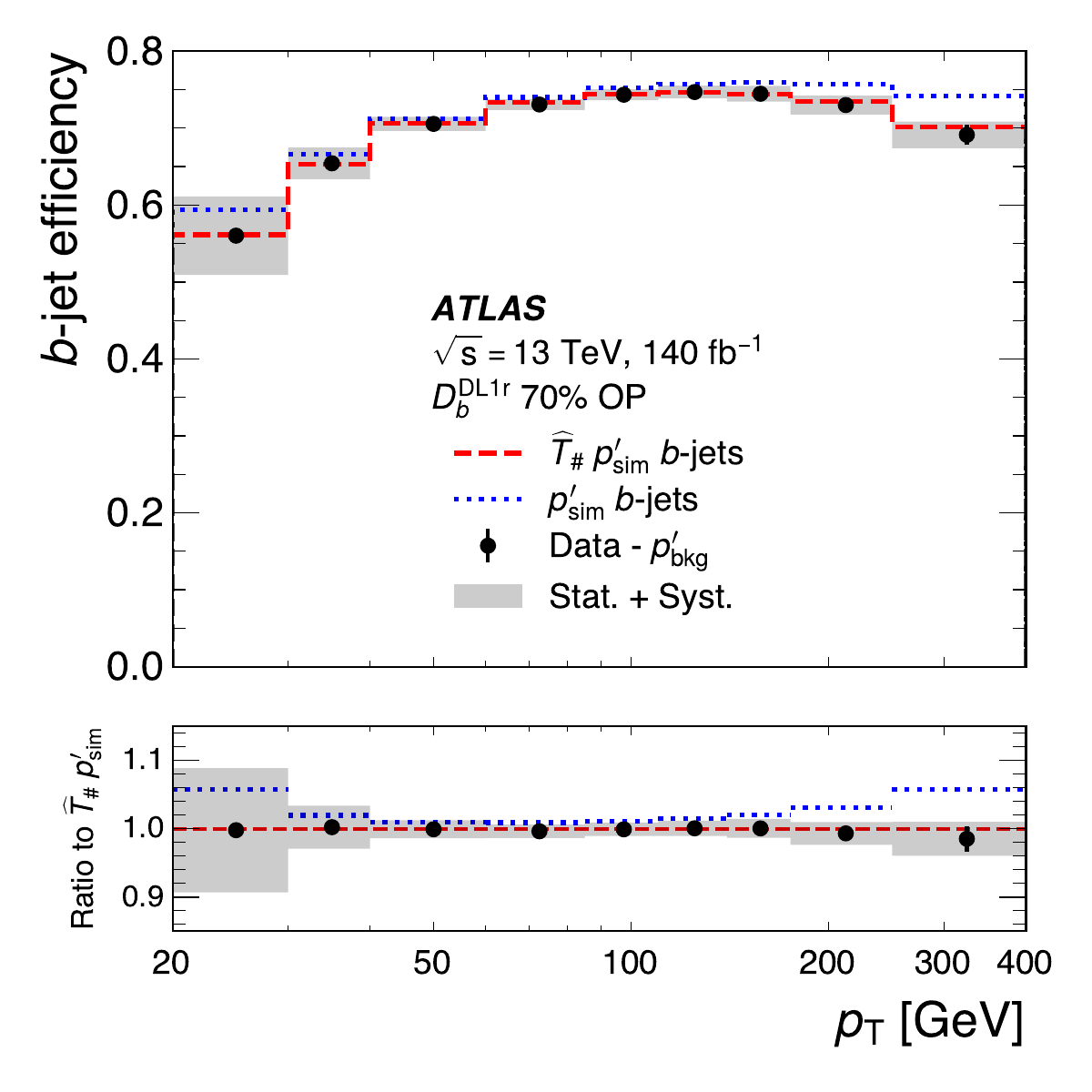}
} \\
\subfigure[]{
\includegraphics[width=0.4\textwidth]{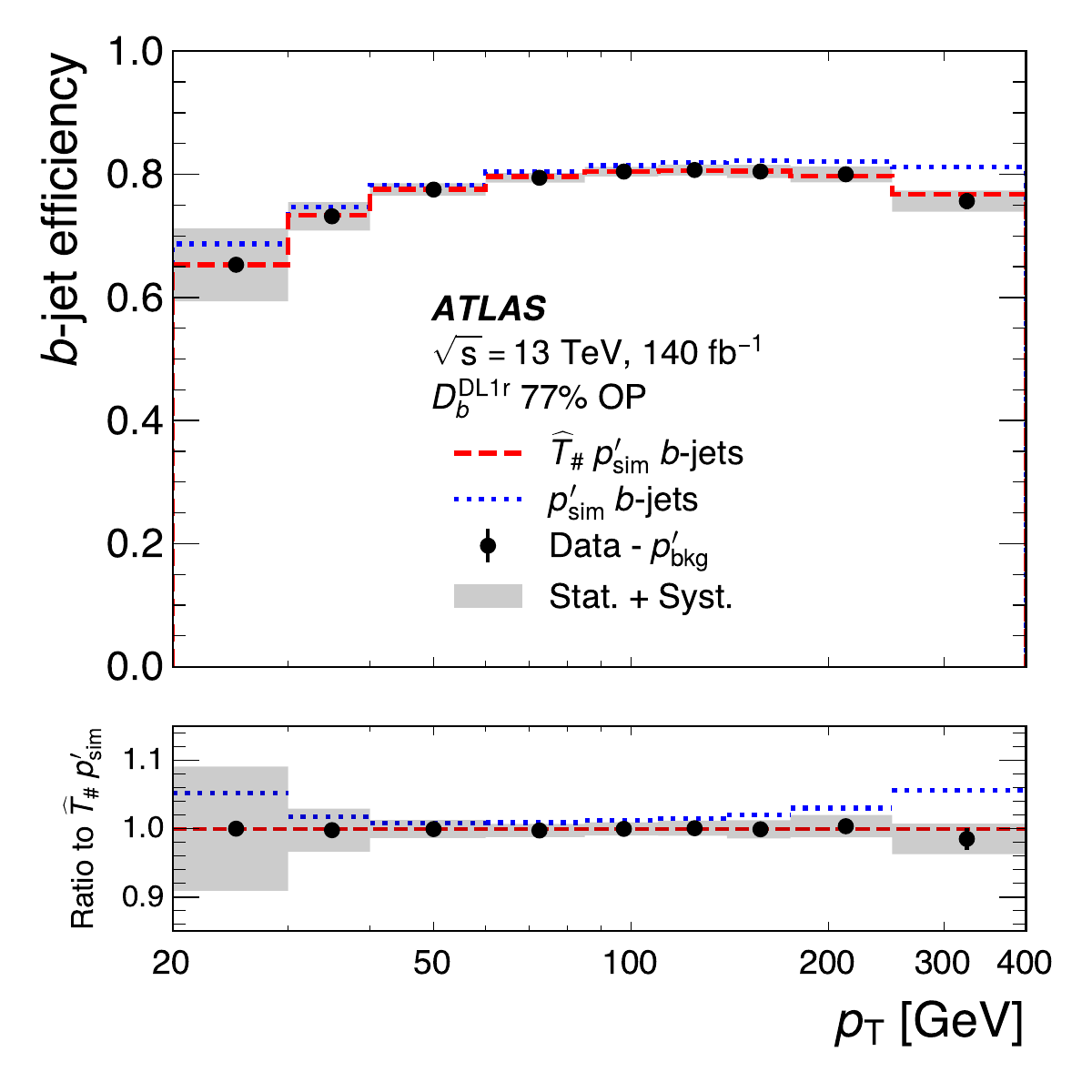}
}
\subfigure[]{
\includegraphics[width=0.4\textwidth]{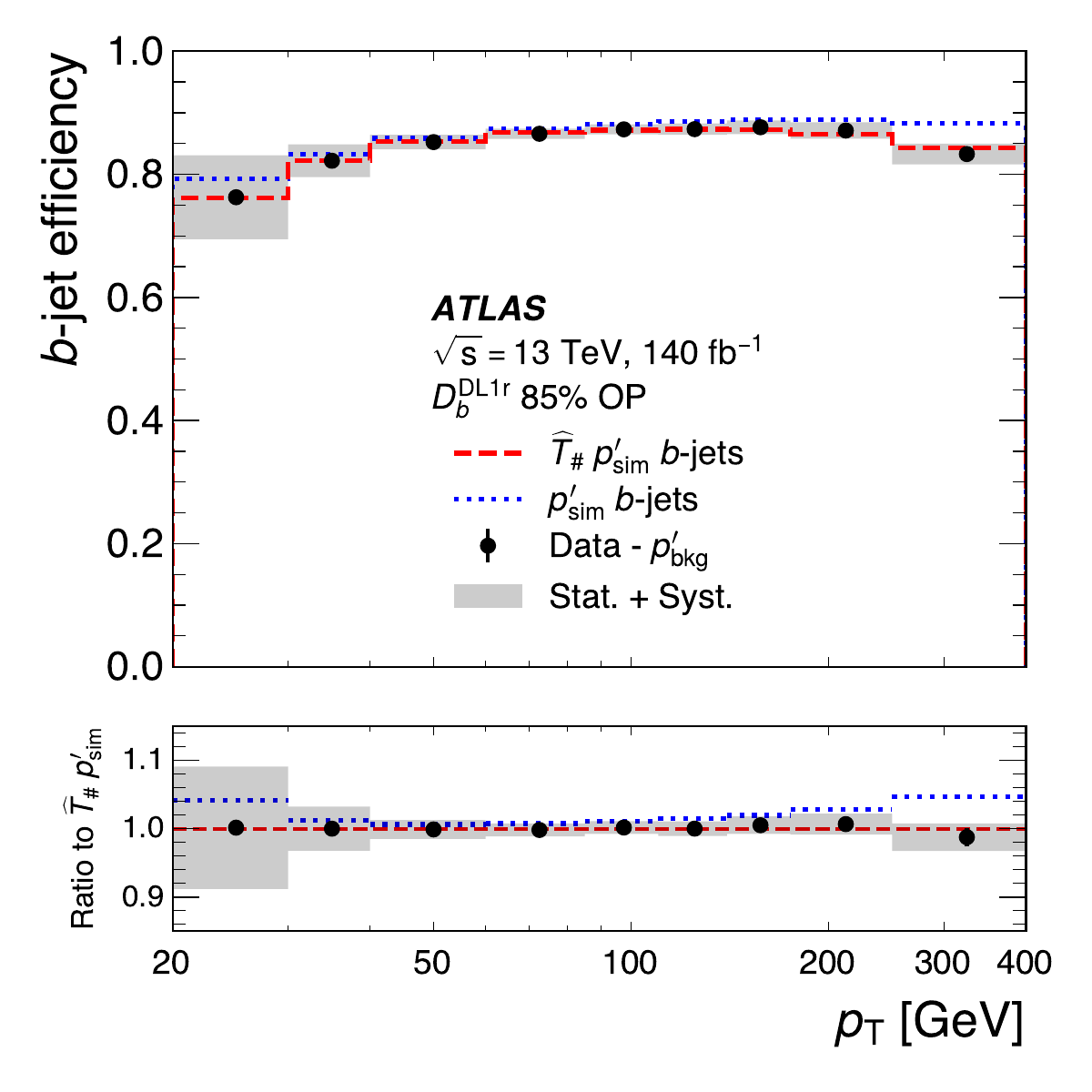}
}
\caption{
\label{fig:sf_closure}
Comparison of \bjet efficiencies at different \dlonerb operating points,
between the nominal simulation, $\psim^\prime$, the calibrated simulation,
$\hat T_\# \psim^\prime$, and background-subtracted data
($\text{Data}-p^\prime_\mathrm{bkg}$), versus jet \pt.
The \dlonerb efficiency operating points shown are (a) 60\%, (b) 70\%, (c)
77\%, and (d) 85\%.
The uncertainty band of $\text{Data}-\pbkg^\prime$ is calculated by applying
each variation of $(1-f_{\mathrm{sig, variation}})\cdot p_{\mathrm{bkg,
variation}}^\prime$, where the variations used are described in
\cref{sec:uncertainties}.
Non-\bjets in simulation were corrected using the transport maps derived in
\cref{sec:nonb_calibration}.
}
\end{figure}

Similarly, the measured \bjet efficiency, $\varepsilon_\mathrm{sig,data}$, of a
selection can be defined as the fraction of jets passing the selection after
accounting for the background prediction:
\begin{equation*}
\varepsilon_\mathrm{sig, data}
= \frac{N^\mathrm{pass}_\mathrm{sig,data}}{N_\mathrm{sig,data}},
\end{equation*}
where $N_\mathrm{sig, data}$ is the number of \bjets in data, and
$N^\mathrm{pass}_\mathrm{sig, data}$ is the number of those that pass some
operating point selection.
By dividing the numerator and denominator by the total number of jets in data,
this efficiency can be approximated by taking \fsig and the probability of a
background jet to satisfy the operating point, $\varepsilon_\mathrm{bkg,sim}$,
from simulation after the correction described in
Section~\ref{sec:nonb_calibration},
\begin{equation*}
\varepsilon_\mathrm{sig}
= \frac{f^\mathrm{pass}_\mathrm{data} - (1 - \fsig) \varepsilon_\mathrm{bkg}}{\fsig},
\end{equation*}
where $f^\mathrm{pass}_\mathrm{data}$ is the fraction of jets in data that pass
the operating point selection.
In order to validate the methodology used to derive the calibration maps,
Figure~\ref{fig:sf_closure} compares the measured efficiency in the \ttbar
phase space with the predicted efficiencies before and after calibrations are applied.
Good agreement between the measured efficiencies and the calibrated
efficiencies is observed across the full \pt range.

\begin{figure}
\centering
\subfigure[]{\includegraphics[width=0.45\textwidth]{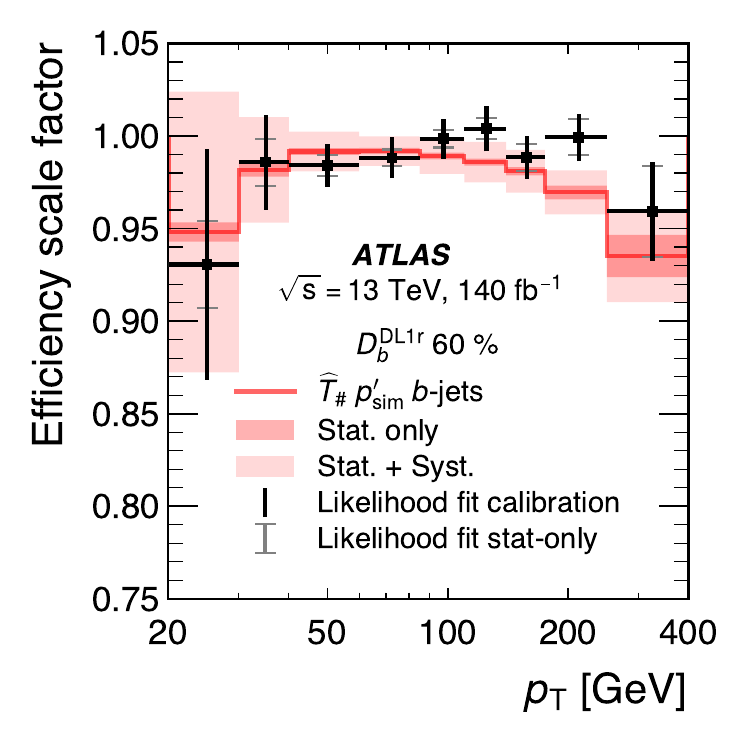}}
\subfigure[]{\includegraphics[width=0.45\textwidth]{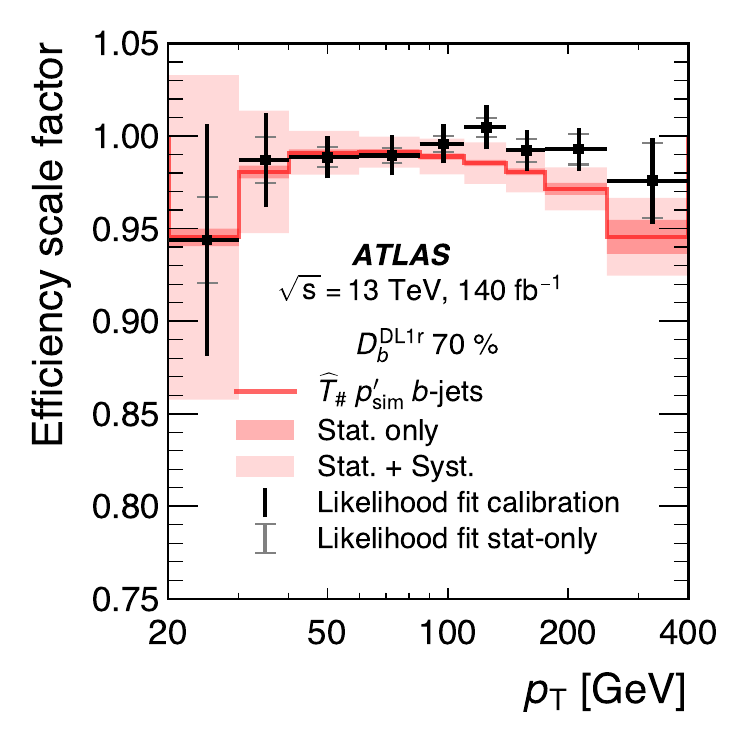}} \\
\subfigure[]{\includegraphics[width=0.45\textwidth]{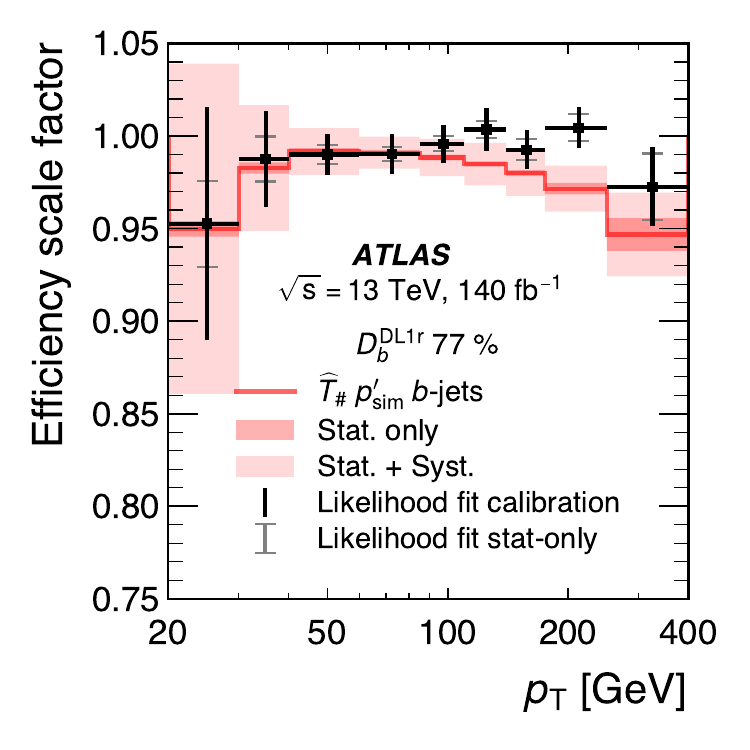}}
\subfigure[]{\includegraphics[width=0.45\textwidth]{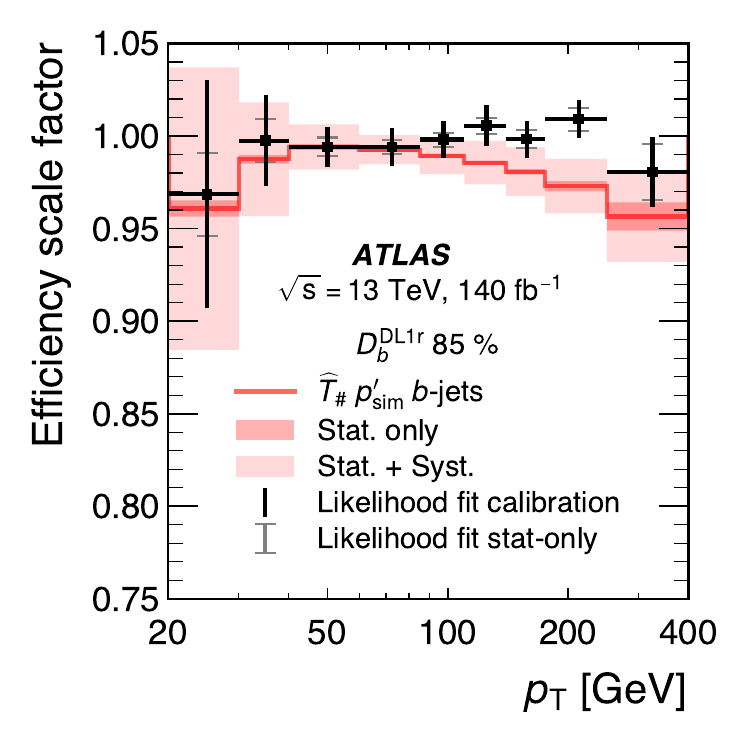}}
\caption{
\label{fig:SFs}
Comparison between \dlonerb efficiency scale factors derived using the
calibration transport maps and a likelihood-fit-based binned scale-factor
approach for the (a) 60\%, (b) 70\%, (c) 77\%, and (d) 85\% efficiency
operating points.
The pale red uncertainty band denotes the total impact of all uncertainties
described in Section~\ref{sec:uncertainties} on the transport-based scale
factors.
It should be noted that the optimal transport calibration comprises a single, general
correction that is universal in the sense that all operating points ought to
be calibrated after its application; the likelihood-fit-based calibrations
were derived individually for particular operating points.
The \enquote{likelihood fit calibration} curve follows the method for measuring
the \btagging efficiency described in Ref.~\cite{FTAG-2018-01}, in which
the jet flavour fractions are extracted jointly with the \btagging
efficiency.
The grey serifs attached to the error bars of the likelihood-based
calibration scale factors show their statistical uncertainties.
}
\end{figure}

Finally, \cref{fig:SFs} shows a comparison of binned and neural estimates
of the \btagging efficiency scale factors versus \pt, where the binned scale
factors were derived by applying the likelihood-fit approach in
Ref.~\cite{FTAG-2018-01} to each operating point individually.
The two calibrations have similar precision, but two important differences
should be noted.
First, the optimal transport calibrations are inherently continuous both versus
the flavour-tagging discriminant and versus \pt since they are approximated as
neural networks.
This requirement reduces the statistical uncertainty compared to the binned
calibration, which has no continuity imposed.
Second, the likelihood-fit approach for measuring efficiencies in \pt bins
includes \emph{in situ} constraints on the jet flavour fractions and therefore
has smaller uncertainties for backgrounds in the \ttbar phase space.
In particular, this mitigates the dominant uncertainties from the modelling of
\ttbar production; the transport-based calibration does not currently benefit
from this \emph{in situ} constraint.

\subsection{Closure tests for charm-tagging selections}

With continuous calibration maps derived for the full tagging probability space,
the efficiency of more sophisticated selections based on a jet's flavour
probability scores should also be properly calibrated.
To illustrate this, the \bjet efficiencies of selections used in the recent
ATLAS search for $H \to c\bar c$ decays via $VH$
production~\cite{HIGG-2020-20} are compared between simulation,
simulation corrected by the transport calibration, and observed data.
Only jets that do not satisfy the \dloner 70\% \btagging efficiency operating point are
considered as potential charm-jet candidates in the \VHcc analysis.
The remaining jets are categorized by their \dlonerc score:\ those with
$\dlonerc > 1.76$ are labelled \ctight; the remaining jets are labelled \cloose if
they satisfy $\dlonerc > 0.6735$ and \untagged otherwise.
In a \ttbar sample with at least one leptonic $W$-boson decay, about 25\% (20\%)
of \cjets pass the \ctight (\cloose) requirement, and about 50\% are \untagged;
for \bjets these fraction are 10\% for \ctight, 5\% for \cloose, and 15\%
for untagged.

\begin{figure}
\centering
\subfigure[]{
\includegraphics[width=0.31\textwidth]{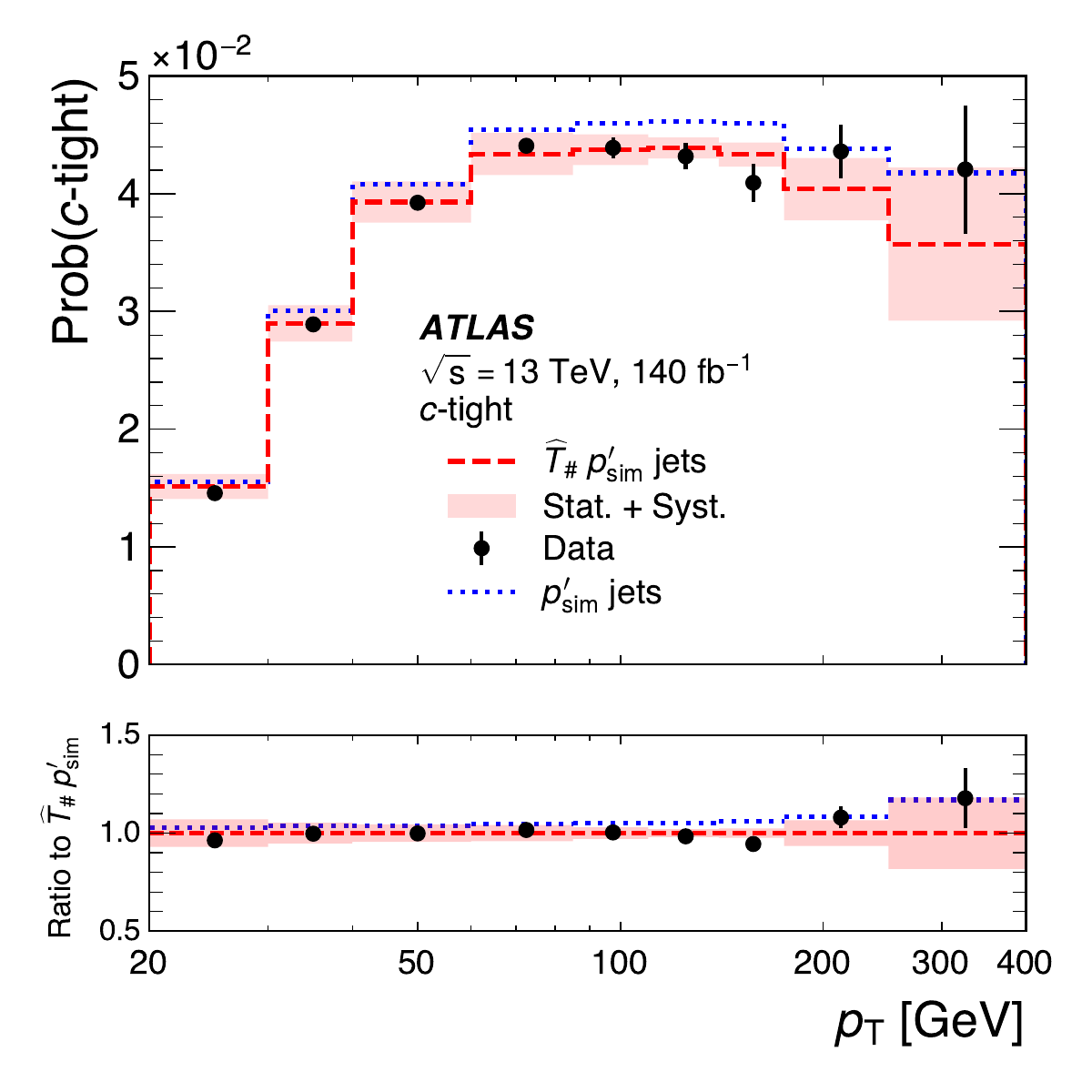}
}
\subfigure[]{
\includegraphics[width=0.31\textwidth]{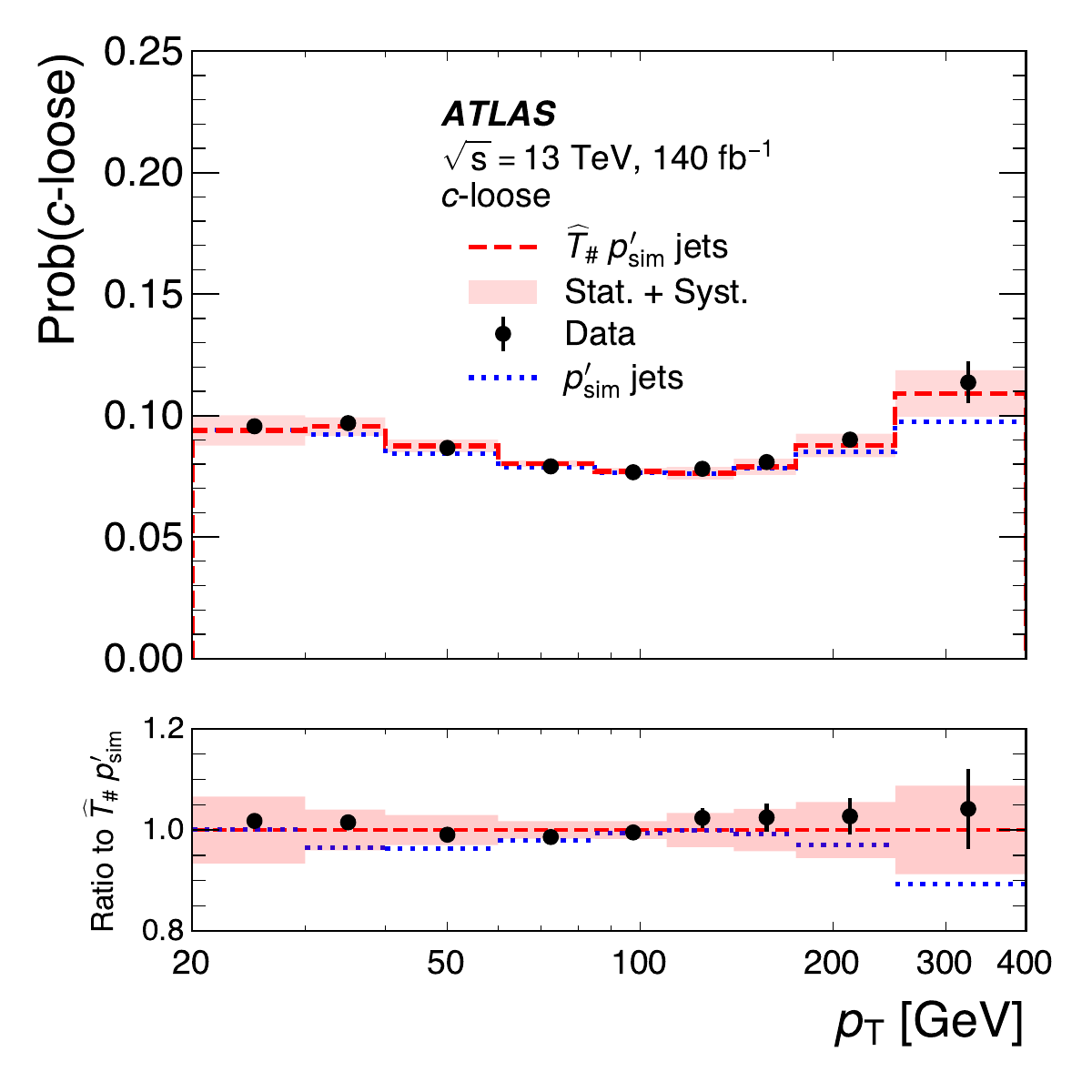}
}
\subfigure[]{
\includegraphics[width=0.31\textwidth]{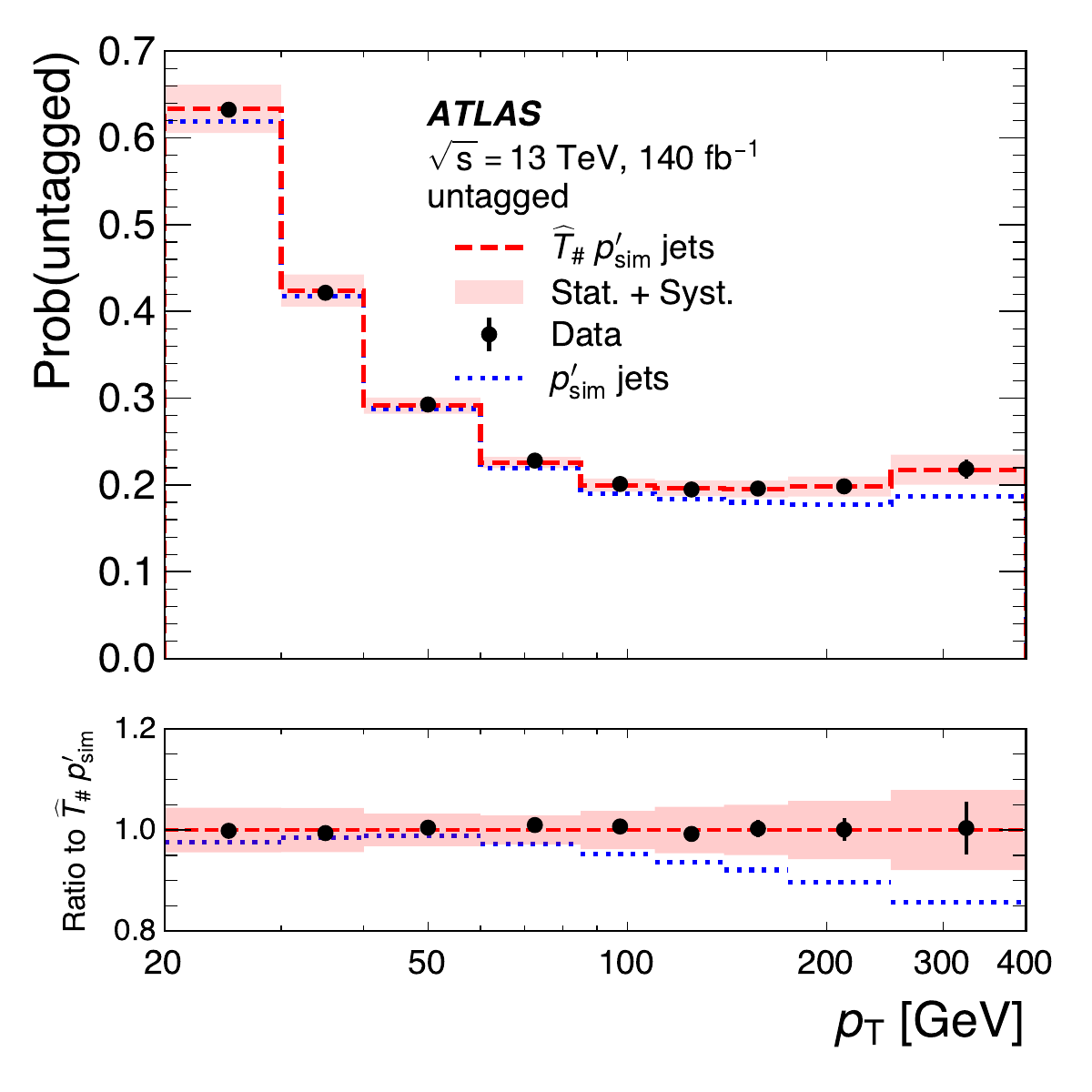}
}
\caption{
\label{fig:VHcc_dataeffs}
Comparison of the probability that a selected jet passes various \VHcc
$c$-tagging selections, between data (\enquote{Data}), the nominal simulated
prediction, $\psim^\prime$, and the calibrated simulation, $\hat T_\#
\psim^\prime$, as a function of the jet \pt in the \ttbar phase space.
The (a) \ctight, (b) \cloose, and (c) \untagged $c$-tagging categories are
shown.
The uncertainty band of the corrected simulation denotes the impact of
all uncertainties described in \cref{sec:uncertainties} on the calibration
maps.
Error bars denote statistical uncertainties.
All efficiencies are shown for the \pt-corrected density
$\psim^\prime(\qvec,\pt) = \psim(\qvec | \pt) \pdata(\pt)$.
Non-\bjets in simulation were corrected using transport maps derived in
\cref{sec:nonb_calibration} and are included in curves for the nominal and
calibrated simulation.
}
\end{figure}

\begin{figure}
\centering
\subfigure[]{
\includegraphics[width=0.31\textwidth]{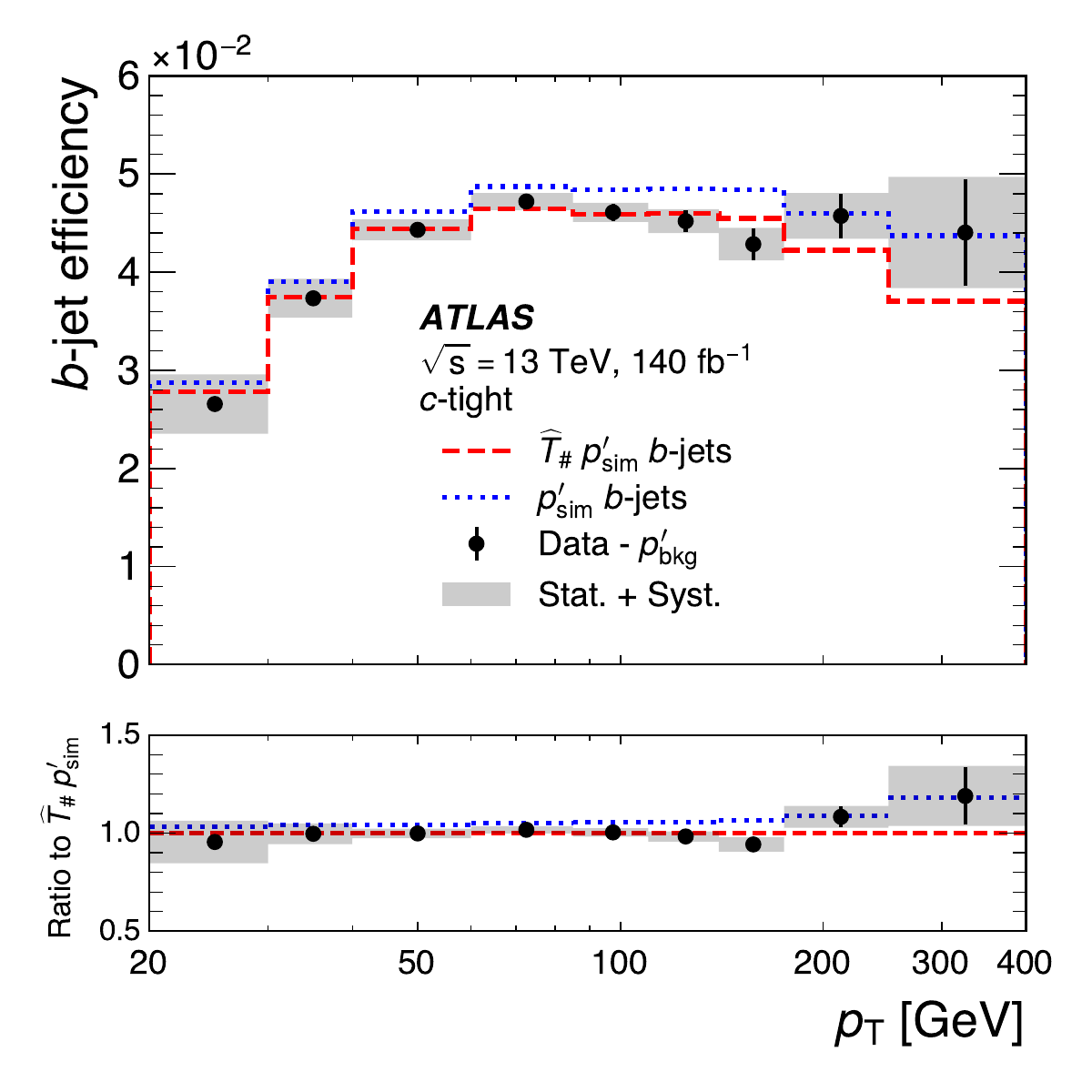}
}
\subfigure[]{
\includegraphics[width=0.31\textwidth]{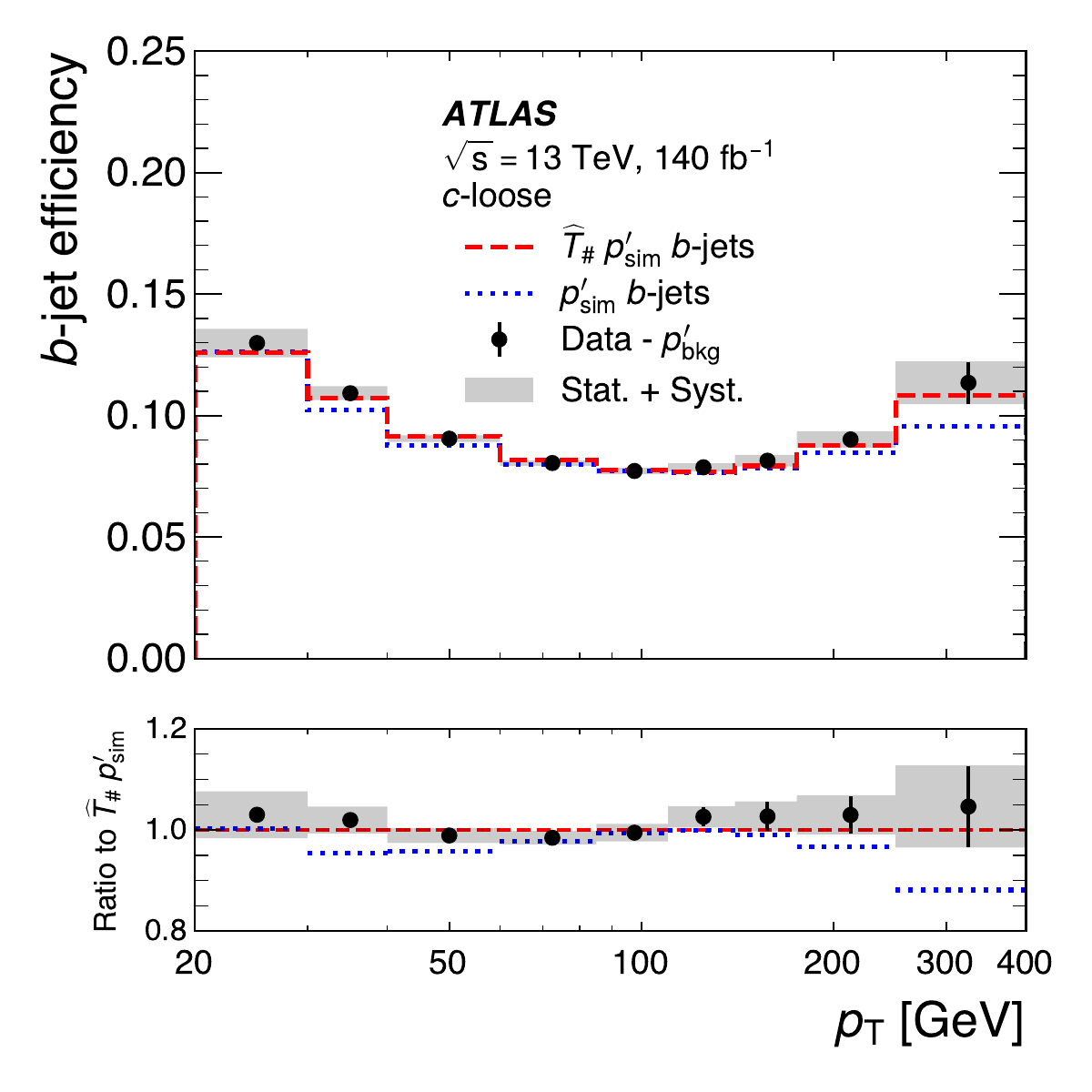}
}
\subfigure[]{
\includegraphics[width=0.31\textwidth]{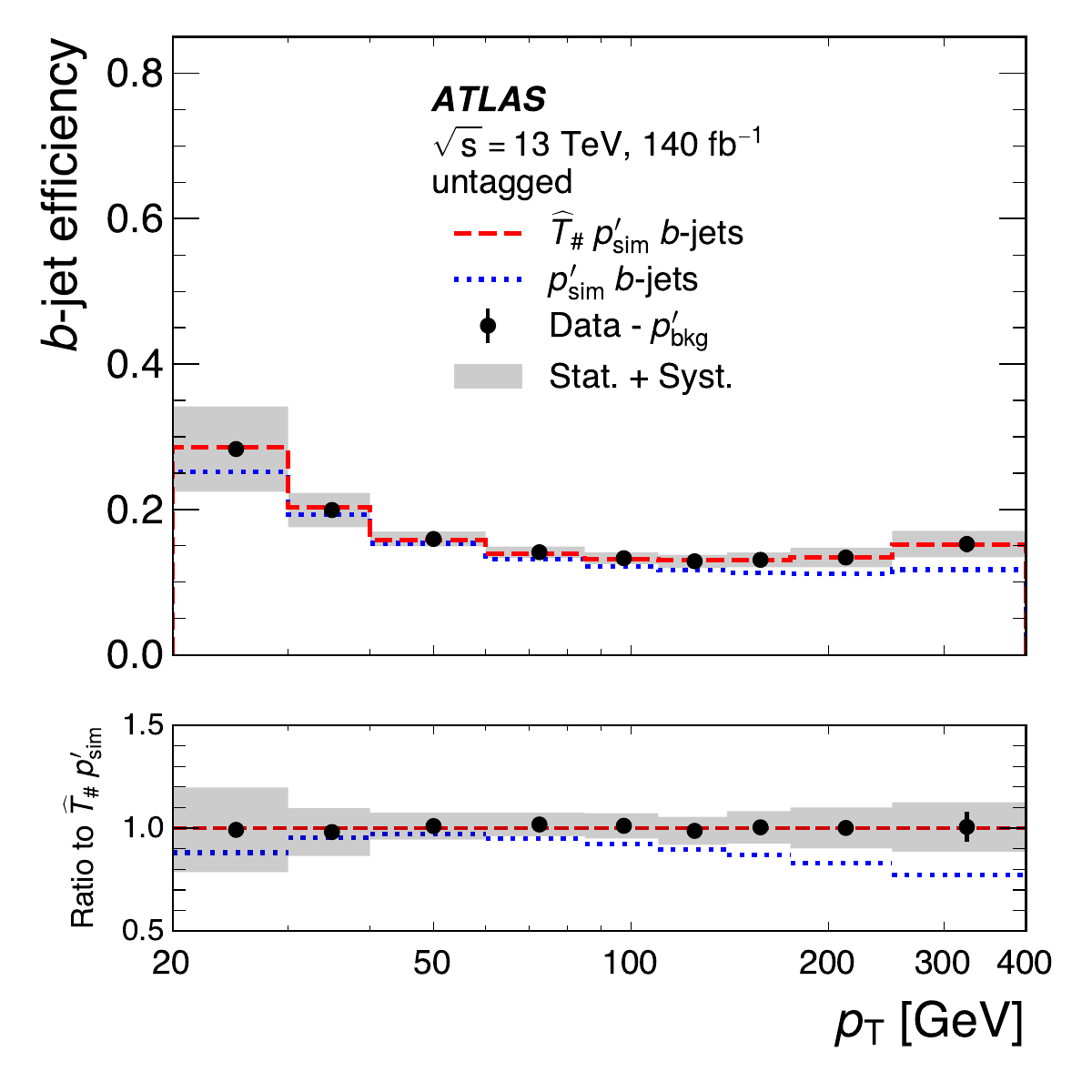}
}
\caption{
\label{fig:VHcc_flow_sf_closure}
Comparison of \bjet efficiencies for the \VHcc $c$-tagging categories,
between the \pt-corrected simulation, $\psim^\prime$, calibrated simulation,
$\hat T_\# \psim^\prime$, and background-subtracted data
(data$-\pbkg^\prime$), versus jet \pt in the \ttbar phase space.
The \bjets efficiency in data is calculated by subtracting
the simulated non-\bjet contribution from data ($\text{Data}-(1-f_{\text{sig}})\cdot \pbkg^\prime$).
The (a) \ctight, (b) \cloose, and (c) \untagged $c$-tagging categories are
shown.
The uncertainty band of $\text{Data}-\pbkg^\prime$ is calculated by subtracting
each $(1-f_{\mathrm{sig, variation}})\cdot p_{\mathrm{bkg, variation}}^\prime$ from the data.
The variations used are described in \cref{sec:uncertainties}.
Non-\bjets in simulation were corrected using transport maps derived in
\cref{sec:nonb_calibration}.
}
\end{figure}

\begin{figure}
\centering
\subfigure[]{
\includegraphics[width=0.31\textwidth]{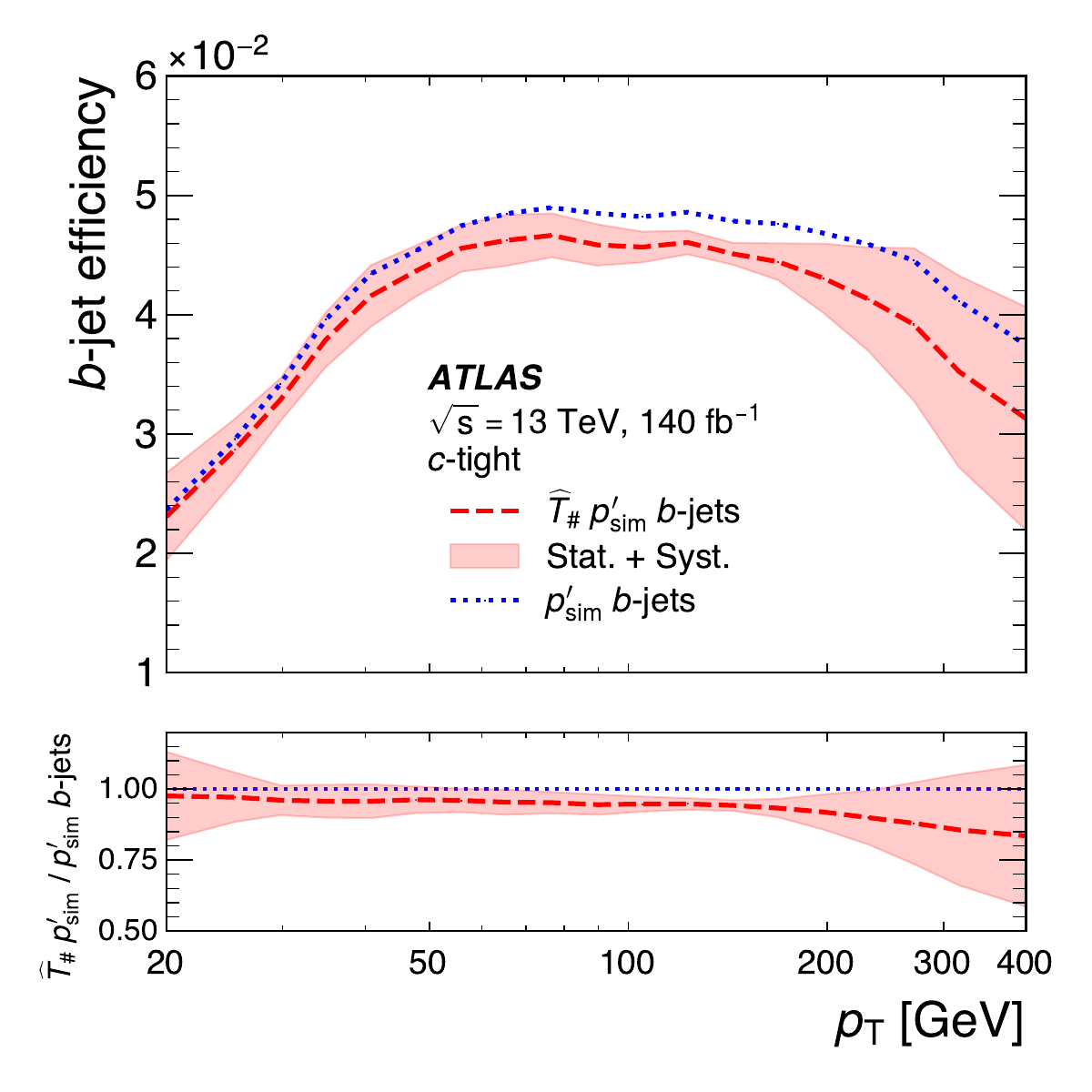}
}
\subfigure[]{
\includegraphics[width=0.31\textwidth]{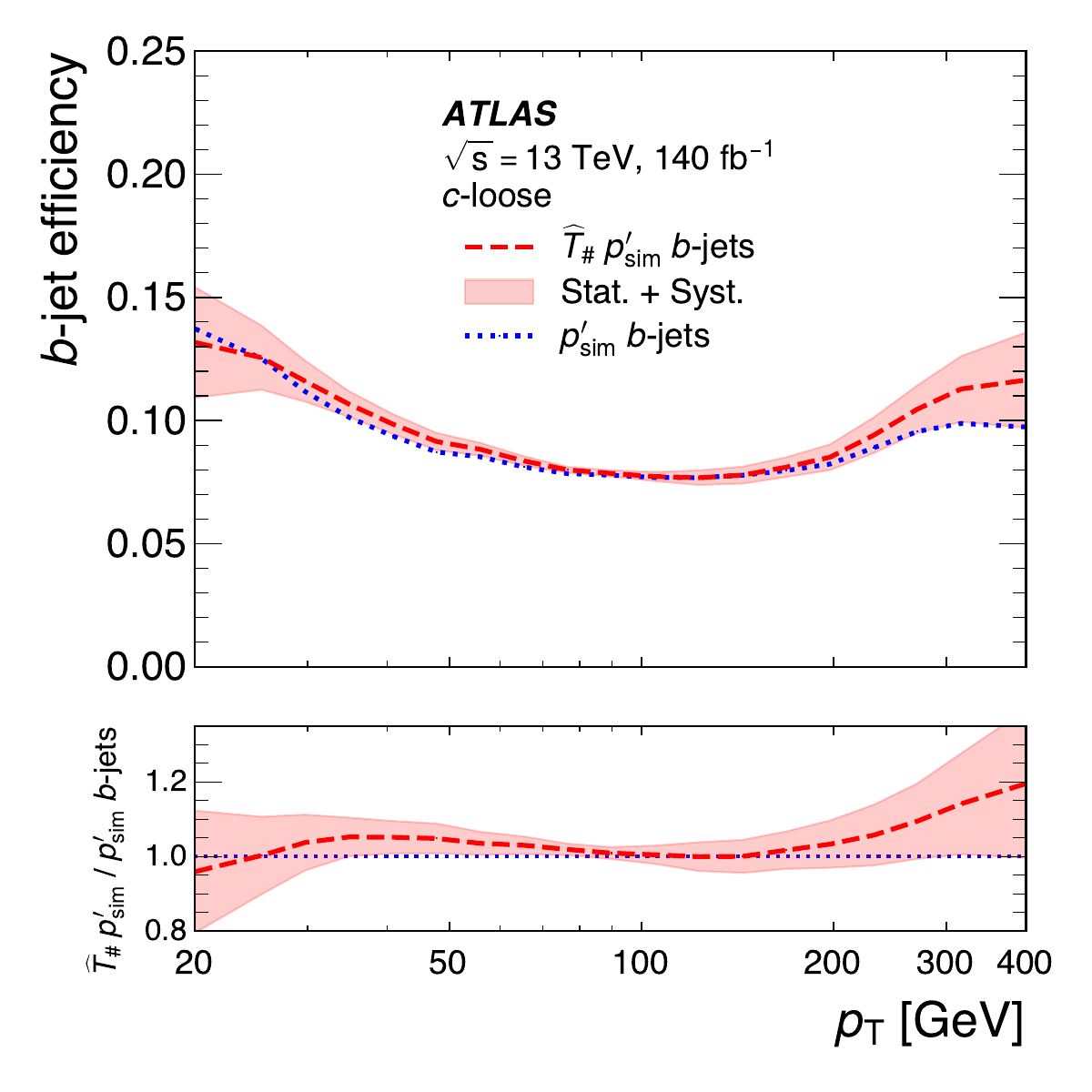}
}
\subfigure[]{
\includegraphics[width=0.31\textwidth]{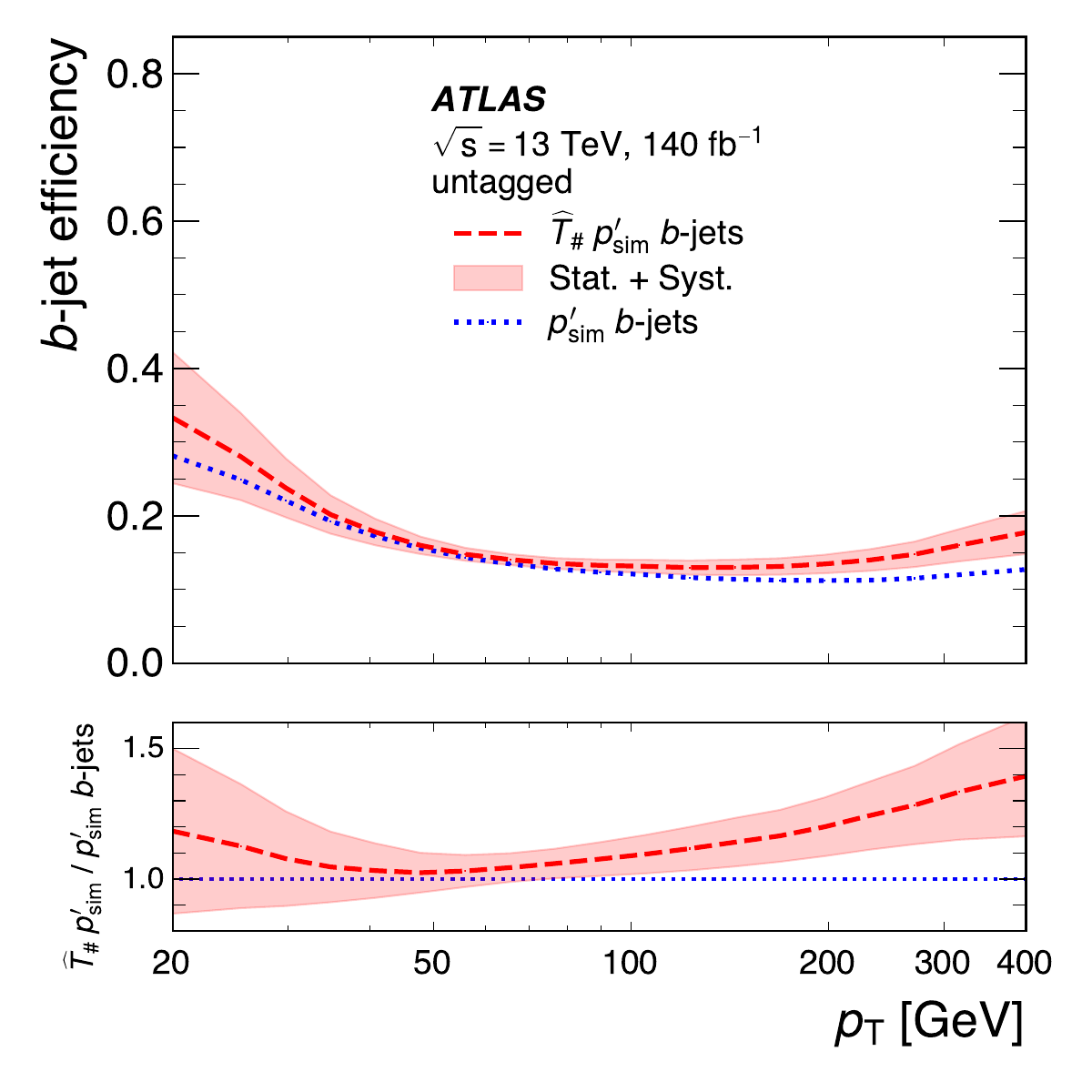}
}
\caption{
\label{fig:VHcc_Floweffs}
Comparison of \bjet efficiencies for the \VHcc $c$-tagging categories, between
the nominal simulation, $\psim^\prime$, and calibrated simulation, $\hat
T_\# \psim^\prime$, in the \ttbar phase space.
The (a) \ctight, (b) \cloose, and (c) \untagged $c$-tagging categories are
shown.
The uncertainty band of the corrected simulation denotes the impact of
all uncertainties described in \cref{sec:uncertainties} on the calibration
maps.
Error bars show statistical uncertainties.
All efficiencies are shown for the \pt-corrected density
$\psim^\prime(\qvec,\pt) = \psim(\qvec | \pt) \pdata(\pt)$.
}
\end{figure}

The measured probability that a jet in the \ttbar phase space satisfies the various
\VHcc charm-tagging categories as a function of jet \pt is shown in
\cref{fig:VHcc_dataeffs}.
Good agreement with the data is observed after calibration.
Using the background-subtraction procedure introduced in
Section~\ref{subsec:dlonerb-closure}, the \bjet efficiencies in data are
similarly measured for the charm-tagging working points and shown as a function
of jet \pt in \cref{fig:VHcc_flow_sf_closure}.
\cref{fig:VHcc_Floweffs} shows a comparison between the efficiencies obtained
from the uncalibrated and calibrated simulations as a function of the jet \pt in the
three $c$-tagging categories.

\begin{figure}
\centering

\subfigure[]{
\includegraphics[width=0.31\textwidth]{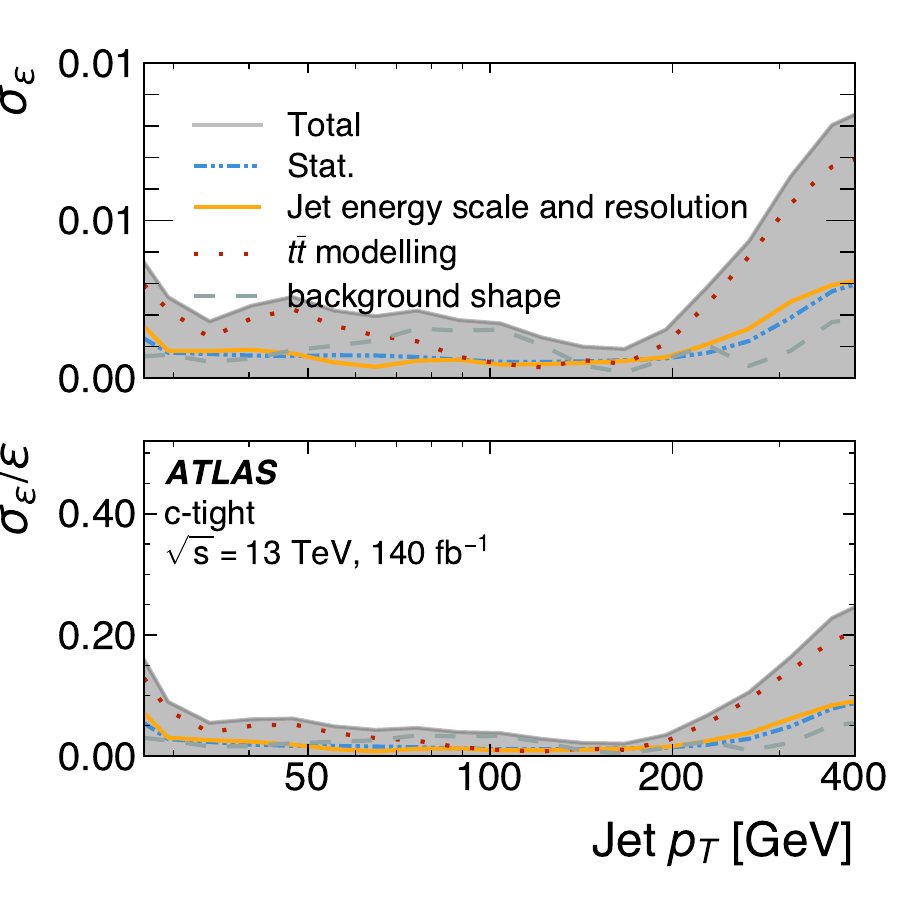}
}
\subfigure[]{
\includegraphics[width=0.31\textwidth]{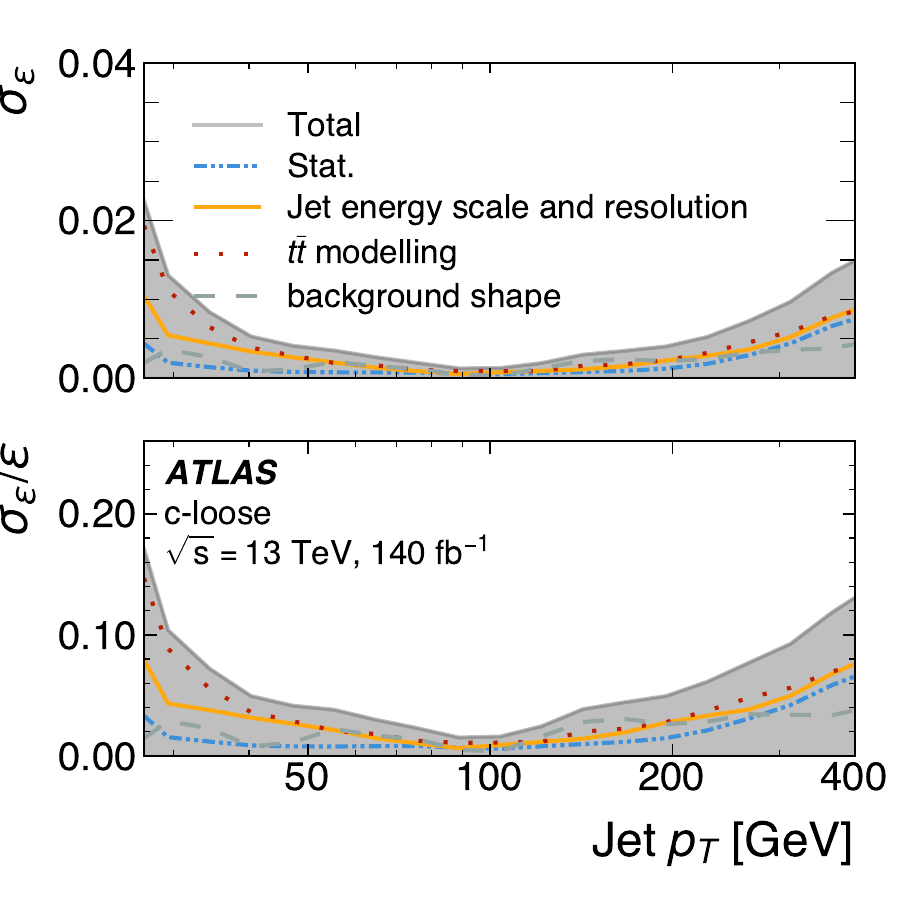}
}
\subfigure[]{
\includegraphics[width=0.31\textwidth]{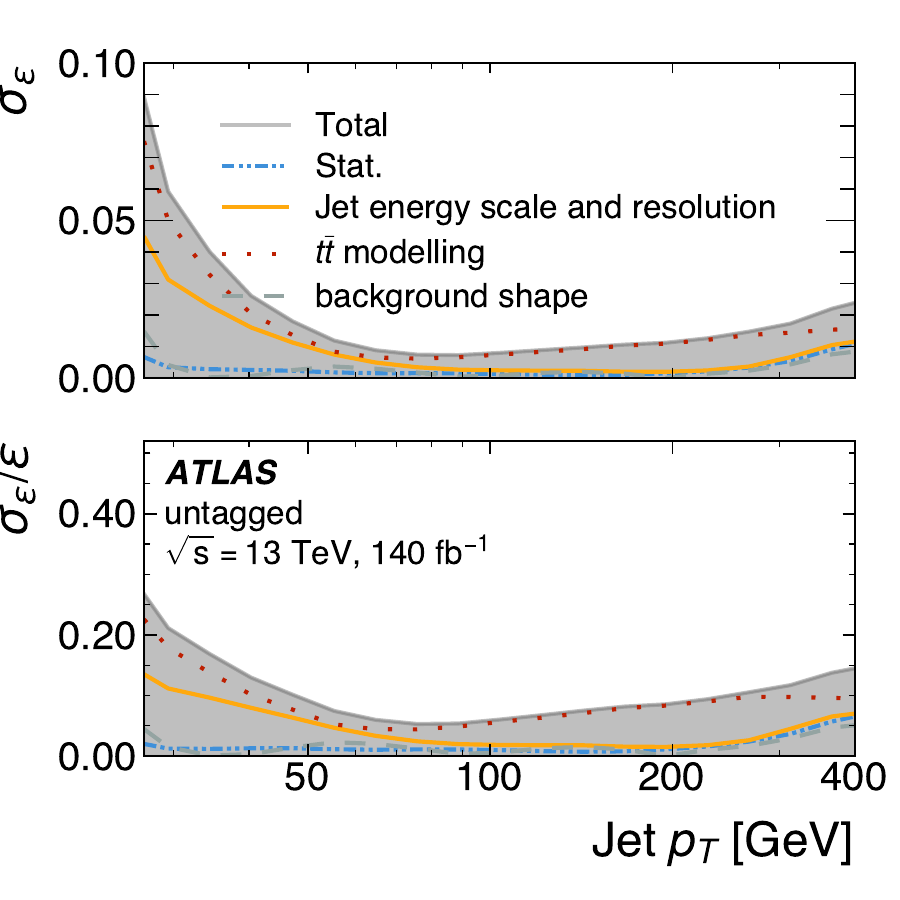}
}
\caption{
Uncertainty breakdown of the calibrated simulation, $\hat T_\#
\psim^\prime$, for corrected \bjet efficiencies versus jet
\pt for various \ctagging operating points.
The top panel shows the absolute uncertainties, while the bottom panel shows
the relative uncertainties.
The (a) \ctight, (b) \cloose, and (c) \untagged operating
points are shown.
The grey uncertainty band of the calibrated simulation denotes the impact of
all uncertainties described in \cref{sec:uncertainties} on the calibration
maps.
}
\label{fig:VHcc_continuous_uncertainties}
\end{figure}

Finally, \cref{fig:VHcc_continuous_uncertainties} shows the absolute and
relative uncertainties as a function of jet \pt in the three \ctagging
categories.

% End of text imported from the .//results.tex input file

%
%
\FloatBarrier

\clearpage
% The next lines are included from the .//conclusion.tex input file
\section{Conclusion}
\label{sec:conclusion}

Flavour tagging is a key component of a broad range of analyses of LHC data
collected by the ATLAS detector.
Recent advances in flavour-tagging performance, driven in part by the uptake of
state-of-the-art machine-learning techniques, have improved the sensitivity to
possible new phenomena across a variety of final states.

In this study, a new strategy for deriving unbinned calibrations for
high-dimensional simulations is developed.
Neural solvers are used to find optimal transport maps to calibrate the
multi-class flavour-tagging discriminators used to analyse ATLAS data.
As opposed to traditional strategies, which involve deriving per-jet weights
from data-to-simulation efficiency ratios, the optimal-transportation-based
calibration presented here alters the flavour probabilities for individual jets
such that the performance in simulation matches that in data.
Moreover, the derived transportation maps are constructed such that the
simulation is minimally altered while achieving closure with the performance
in data.

This calibration strategy is applied in a sample of \bjets obtained from \ttbar
events in Run~2 LHC $pp$ collision data with $\sqrt{s} = 13$~TeV collected by
the ATLAS detector.
The outcome is a set of continuous corrections that yield a calibrated
simulation in the three-dimensional space of jet flavour probabilities
conditional on the jet \pt.
After applying the calibration transport map to the simulation, the performance
observed in data is reproduced for a wide range of \btagging operating points,
with efficiencies ranging from 60\% to 85\%.
The corrected simulation also accurately reflects the performance achieved in
data with more sophisticated selections built from the jet classification scores,
such as those recently developed in the ATLAS search for $H \to c\bar c$ decays.

The presented calibration provides a robust set of three-dimensional corrections
to the ATLAS simulation for the jet flavour-tagging algorithms used to analyse
LHC collision data.
This enables a more effective exploitation of jet flavour information in
ATLAS analyses, including those which benefit from simultaneous bottom- and
charm-jet identification such as joint measurements of $H\to bb$ and $H\to
cc$~\cite{HIGG-2020-20}; that require very strong \cjet rejection, such as
searches for rare top-quark processes~cite{TOPQ-2018-06}; or which would gain
from tagging operating points that smoothing vary with event kinematics, such as
searches for high-mass resonances~\cite{EXOT-2016-33}.
This development also opens several promising avenues for future research,
including calibrations derived for finer-grained features, conditioned on
more relevant kinematic information, or integrated into the training of
inference
networks~\cite{wehenkel2024addressingmisspecificationsimulationbasedinference}.
These developments lay the groundwork for a more general and flexible framework
for reliable inference based on imperfect simulation.

% End of text imported from the .//conclusion.tex input file
\clearpage

\section*{Acknowledgements}
%

% The next lines are included from the .//acknowledgements/Acknowledgements.tex input file
%
%

%
%

We thank CERN for the very successful operation of the LHC and its injectors, as well as the support staff at
CERN and at our institutions worldwide without whom ATLAS could not be operated efficiently.

The crucial computing support from all WLCG partners is acknowledged gratefully, in particular from CERN, the ATLAS Tier-1 facilities at TRIUMF/SFU (Canada), NDGF (Denmark, Norway, Sweden), CC-IN2P3 (France), KIT/GridKA (Germany), INFN-CNAF (Italy), NL-T1 (Netherlands), PIC (Spain), RAL (UK) and BNL (USA), the Tier-2 facilities worldwide and large non-WLCG resource providers. Major contributors of computing resources are listed in Ref.~\cite{ATL-SOFT-PUB-2025-001}.

We gratefully acknowledge the support of ANPCyT, Argentina; YerPhI, Armenia; ARC, Australia; BMWFW and FWF, Austria; ANAS, Azerbaijan; CNPq and FAPESP, Brazil; NSERC, NRC and CFI, Canada; CERN; ANID, Chile; CAS, MOST and NSFC, China; Minciencias, Colombia; MEYS CR, Czech Republic; DNRF and DNSRC, Denmark; IN2P3-CNRS and CEA-DRF/IRFU, France; SRNSFG, Georgia; BMFTR, HGF and MPG, Germany; GSRI, Greece; RGC and Hong Kong SAR, China; ISF and Benoziyo Center, Israel; INFN, Italy; MEXT and JSPS, Japan; CNRST, Morocco; NWO, Netherlands; RCN, Norway; MNiSW, Poland; FCT, Portugal; MNE/IFA, Romania; MSTDI, Serbia; MSSR, Slovakia; ARIS and MVZI, Slovenia; DSI/NRF, South Africa; MICIU/AEI, Spain; SRC and Wallenberg Foundation, Sweden; SERI, SNSF and Cantons of Bern and Geneva, Switzerland; NSTC, Taipei; TENMAK, T\"urkiye; STFC/UKRI, United Kingdom; DOE and NSF, United States of America.

Individual groups and members have received support from BCKDF, CANARIE, CRC and DRAC, Canada; CERN-CZ, FORTE and PRIMUS, Czech Republic; COST, ERC, ERDF, Horizon 2020, ICSC-NextGenerationEU and Marie Sk{\l}odowska-Curie Actions, European Union; Investissements d'Avenir Labex, Investissements d'Avenir Idex and ANR, France; DFG and AvH Foundation, Germany; Herakleitos, Thales and Aristeia programmes co-financed by EU-ESF and the Greek NSRF, Greece; BSF-NSF and MINERVA, Israel; NCN and NAWA, Poland; La Caixa Banking Foundation, CERCA Programme Generalitat de Catalunya and PROMETEO and GenT Programmes Generalitat Valenciana, Spain; G\"{o}ran Gustafssons Stiftelse, Sweden; The Royal Society and Leverhulme Trust, United Kingdom.

In addition, individual members wish to acknowledge support from Armenia: Yerevan Physics Institute (FAPERJ); CERN: European Organization for Nuclear Research (CERN PJAS); Chile: Agencia Nacional de Investigaci\'on y Desarrollo (FONDECYT 1230812, FONDECYT 1240864); China: Chinese Ministry of Science and Technology (MOST-2023YFA1605700, MOST-2023YFA1609300), National Natural Science Foundation of China (NSFC - 12175119, NSFC 12275265, NSFC-12075060); Czech Republic: Czech Science Foundation (GACR - 24-11373S), Ministry of Education Youth and Sports (ERC-CZ-LL2327, FORTE CZ.02.01.01/00/22\_008/0004632), PRIMUS Research Programme (PRIMUS/21/SCI/017); EU: H2020 European Research Council (ERC - 101002463); European Union: European Research Council (ERC - 948254, ERC 101089007), Horizon 2020 Framework Programme (MUCCA - CHIST-ERA-19-XAI-00), European Union, Future Artificial Intelligence Research (FAIR-NextGenerationEU PE00000013), Italian Center for High Performance Computing, Big Data and Quantum Computing (ICSC, NextGenerationEU); France: Agence Nationale de la Recherche (ANR-20-CE31-0013, ANR-21-CE31-0013, ANR-21-CE31-0022, ANR-22-EDIR-0002); Germany: Baden-Württemberg Stiftung (BW Stiftung-Postdoc Eliteprogramme), Deutsche Forschungsgemeinschaft (DFG - 469666862, DFG - CR 312/5-2); Italy: Istituto Nazionale di Fisica Nucleare (ICSC, NextGenerationEU), Ministero dell'Università e della Ricerca (NextGenEU 153D23001490006 M4C2.1.1, NextGenEU I53D23000820006 M4C2.1.1, NextGenEU I53D23001490006 M4C2.1.1); Japan: Japan Society for the Promotion of Science (JSPS KAKENHI JP22H01227, JSPS KAKENHI JP22H04944, JSPS KAKENHI JP22KK0227, JSPS KAKENHI JP23KK0245); Netherlands: Netherlands Organisation for Scientific Research (NWO Veni 2020 - VI.Veni.202.179); Norway: Research Council of Norway (RCN-314472); Poland: Ministry of Science and Higher Education (IDUB AGH, POB8, D4 no 9722), Polish National Agency for Academic Exchange (PPN/PPO/2020/1/00002/U/00001), Polish National Science Centre (NCN 2021/42/E/ST2/00350, NCN OPUS 2023/51/B/ST2/02507, NCN OPUS nr 2022/47/B/ST2/03059, NCN UMO-2019/34/E/ST2/00393, NCN \& H2020 MSCA 945339, UMO-2020/37/B/ST2/01043, UMO-2021/40/C/ST2/00187, UMO-2022/47/O/ST2/00148, UMO-2023/49/B/ST2/04085, UMO-2023/51/B/ST2/00920); Portugal: Foundation for Science and Technology (FCT); Slovenia: Slovenian Research Agency (ARIS grant J1-3010); Spain: Generalitat Valenciana (Artemisa, FEDER, IDIFEDER/2018/048), Ministry of Science and Innovation (MCIN \& NextGenEU PCI2022-135018-2, MICIN \& FEDER PID2021-125273NB, RYC2019-028510-I, RYC2020-030254-I, RYC2021-031273-I, RYC2022-038164-I); Sweden: Carl Trygger Foundation (Carl Trygger Foundation CTS 22:2312), Swedish Research Council (Swedish Research Council 2023-04654, VR 2018-00482, VR 2021-03651, VR 2022-03845, VR 2022-04683, VR 2023-03403), Knut and Alice Wallenberg Foundation (KAW 2018.0458, KAW 2019.0447, KAW 2022.0358); Switzerland: Swiss National Science Foundation (SNSF - PCEFP2\_194658); United Kingdom: Leverhulme Trust (Leverhulme Trust RPG-2020-004), Royal Society (NIF-R1-231091); United States of America: U.S. Department of Energy (ECA DE-AC02-76SF00515), Neubauer Family Foundation.

%
%

% End of text imported from the .//acknowledgements/Acknowledgements.tex input file

%
\clearpage

\printbibliography

@article{Buckley:2011ms,
    author = "Buckley, Andy and others",
    title = "{General-purpose event generators for LHC physics}",
    eprint = "1101.2599",
    archivePrefix = "arXiv",
    primaryClass = "hep-ph",
    reportNumber = "CAVENDISH-HEP-10-21, CERN-PH-TH-2010-298, DCPT-10-202, IPPP-10-101, KA-TP-40-2010, LU-TP-10-28, MAN-HEP-2010-23, SLAC-PUB-14333, HD-THEP-10-24, MCNET-11-01",
    doi = "10.1016/j.physrep.2011.03.005",
    journal = "Phys. Rept.",
    volume = "504",
    pages = "145--233",
    year = "2011"
}

@book{monge1781mémoire,
  title={M{\'e}moire sur la th{\'e}orie des d{\'e}blais et des remblais},
  author={Monge, G.},
  year={1781},
  publisher={Imprimerie royale}
}

@book{villani2008optimal,
  title={Optimal Transport: Old and New},
  author={Villani, C.},
  isbn={9783540710509},
  lccn={2008932183},
  series={Grundlehren der mathematischen Wissenschaften},
  doi={10.1007/978-3-540-71050-9},
  year={2008},
  publisher={Springer Berlin Heidelberg},
  url="https://doi.org/10.1007/978-3-540-71050-9"
}

@inproceedings{amos2017input,
  title={Input convex neural networks},
  author={Amos, Brandon and Xu, Lei and Kolter, J Zico},
  booktitle={International Conference on Machine Learning},
  pages={146--155},
  year={2017},
  organization={PMLR},
  doi = {10.48550/arXiv.1609.07152},
  eprint = "1609.07152",
  archivePrefix = "arXiv",
  primaryClass = "cs.LG",
}

@article{Pollard:2021fqv,
    author = "Pollard, Chris and Windischhofer, Philipp",
    title = "{Transport away your problems: Calibrating stochastic simulations with optimal transport}",
    eprint = "2107.08648",
    archivePrefix = "arXiv",
    primaryClass = "physics.data-an",
    doi = "10.1016/j.nima.2021.166119",
    journal = "Nucl. Instrum. Meth. A",
    volume = "1027",
    pages = "166119",
    year = "2022"
}

@article{bunne2022supervised,
  title={Supervised training of conditional monge maps},
  author={Bunne, Charlotte and Krause, Andreas and Cuturi, Marco},
  journal={NeurIPS},
  volume={35},
  pages={6859--6872},
  year={2023},
  eprint = "2206.14262",
  archivePrefix = "arXiv",
  doi = {10.48550/arXiv.2206.14262},
}

@article{DELPHI:2011aa,
      author         = "{DELPHI Collaboration}",
      title          = "{A study of the $b$-quark fragmentation function with the
                        DELPHI detector at LEP~I and an averaged distribution
                        obtained at the $Z$ pole}",
      collaboration  = "DELPHI",
      journal        = "Eur. Phys. J. C",
      volume         = "71",
      year           = "2011",
      pages          = "1557",
      doi            = "10.1140/epjc/s10052-011-1557-x",
      eprint         = "1102.4748",
      archivePrefix  = "arXiv",
      primaryClass   = "hep-ex",
      reportNumber   = "CERN-PH-EP-2010-057",
      SLACcitation   = "%%CITATION = ARXIV:1102.4748;%%"
}

@article{Abbiendi:2002vt,
      author         = "{OPAL Collaboration}",
      title          = "{Inclusive analysis of the b quark fragmentation function
                        in Z decays at LEP}",
      collaboration  = "OPAL",
      journal        = "Eur. Phys. J. C",
      volume         = "29",
      year           = "2003",
      pages          = "463-478",
      doi            = "10.1140/epjc/s2003-01229-x",
      eprint         = "hep-ex/0210031",
      archivePrefix  = "arXiv",
      reportNumber   = "CERN-EP-2002-051",
      SLACcitation   = "%%CITATION = HEP-EX/0210031;%%"
}

@article{Heister:2001jg,
      author         = "{ALEPH Collaboration}",
      title          = "{Study of the fragmentation of $b$ quarks into $B$ mesons at
                        the $Z$ peak}",
      collaboration  = "ALEPH",
      journal        = "Phys. Lett. B",
      volume         = "512",
      year           = "2001",
      pages          = "30-48",
      doi            = "10.1016/S0370-2693(01)00690-6",
      eprint         = "hep-ex/0106051",
      archivePrefix  = "arXiv",
      reportNumber   = "CERN-EP-2001-039",
      SLACcitation   = "%%CITATION = HEP-EX/0106051;%%"
}

@article{Abe:2002iq,
      author         = "{SLD Collaboration}",
      title          = "{Measurement of the $b$-quark fragmentation function in $Z^0$ decays}",
      collaboration  = "SLD",
      journal        = "Phys. Rev. D",
      volume         = "65",
      year           = "2002",
      pages          = "092006",
      doi            = "10.1103/PhysRevD.65.092006",

      addendum       = "Erratum: \href{https://doi.org/10.1103/PhysRevD.66.079905}{Phys. Rev. D \textbf{66} (2002) 79905}",
      %note           = "[Erratum: Phys. Rev. D \textbf{66} (2002) 079905]",
      eprint         = "hep-ex/0202031",
      archivePrefix  = "arXiv",
      reportNumber   = "SLAC-PUB-9087",
      SLACcitation   = "%%CITATION = HEP-EX/0202031;%%"
}

@InProceedings{pmlr-v119-makkuva20a,
  title = 	 {Optimal transport mapping via input convex neural networks},
  author =       {Makkuva, Ashok and Taghvaei, Amirhossein and Oh, Sewoong and Lee, Jason},
  booktitle = 	 {Proceedings of the 37th International Conference on Machine Learning},
  pages = 	 {6672--6681},
  year = 	 {2020},
  editor = 	 {III, Hal Daumé and Singh, Aarti},
  volume = 	 {119},
  archivePrefix = "arXiv",
  eprint = "1908.10962",
  primaryClass = "cs.LG",
  series = 	 {Proceedings of Machine Learning Research},
  month = 	 {13--18 Jul},
  publisher =    {PMLR}
}

@ARTICLE{9089305,
  author={Kobyzev, Ivan and Prince, Simon J.D. and Brubaker, Marcus A.},
  journal={IEEE PAMI}, 
  title={Normalizing Flows: An Introduction and Review of Current Methods}, 
  year={2021},
  volume={43},
  number={11},
  pages={3964-3979},
  doi={10.1109/TPAMI.2020.2992934},
  eprint = "1908.09257",
  archivePrefix="arXiv",
  primaryClass = "stat.ML",
  }

@article{Algren:2023qnb,
    author = "Algren, Malte and Golling, Tobias and Guth, Manuel and Pollard, Chris and Raine, John Andrew",
    title = "{Flow Away your Differences: Conditional Normalizing Flows as an Improvement to Reweighting}",
    eprint = "2304.14963",
    archivePrefix = "arXiv",
    primaryClass = "hep-ph",
    month = "4",
    year = "2023"
}

@article{Cranmer:2015bka,
    author = "Cranmer, Kyle and Pavez, Juan and Louppe, Gilles",
    title = "{Approximating Likelihood Ratios with Calibrated Discriminative  Classifiers}",
    eprint = "1506.02169",
    archivePrefix = "arXiv",
    primaryClass = "stat.AP",
    month = "6",
    year = "2016"
}

@book{efron1994introduction,
  title={An Introduction to the Bootstrap},
  author={Efron, B. and Tibshirani, R.J.},
  isbn={9780412042317},
  lccn={93004489},
  series={Chapman \& Hall/CRC Monographs on Statistics \& Applied Probability},
  url={https://doi.org/10.1201/9780429246593},
  year={1994},
  publisher={Taylor \& Francis}
}

@article{cpflows,
  author       = {Chin{-}Wei Huang and
                  Ricky T. Q. Chen and
                  Christos Tsirigotis and
                  Aaron C. Courville},
  title        = {Convex Potential Flows: Universal Probability Distributions with Optimal
                  Transport and Convex Optimization},
  journal      = {ICLR},
  year         = {2021},
  eprinttype    = {arXiv},
  eprint       = {2012.05942},
}

@article{Algren_2024,
   title={Decorrelation using optimal transport},
   volume={84},
   DOI={10.1140/epjc/s10052-024-12868-6},
   number={6},
   journal={Eur. Phys. J. C},
   publisher={Springer Science and Business Media LLC},
   author={Algren, Malte and Raine, John Andrew and Golling, Tobias},
   year={2024},
   eprint = "2307.05187",
  archivePrefix = "arXiv",
  primaryClass = "hep-ph",
   month=jun }

@article{ADAMW,
  author       = {Ilya Loshchilov and
                  Frank Hutter},
  title        = {Fixing Weight Decay Regularization in Adam},
  journal      = {ICLR},
  year         = {2018},
  eprinttype    = {arXiv},
  eprint       = {1711.05101},
  timestamp    = {Mon, 13 Aug 2018 16:48:18 +0200},
  % biburl       = {https://dblp.org/rec/journals/corr/abs-1711-05101.bib},
  bibsource    = {dblp computer science bibliography, https://dblp.org}
}

@article{celu,
  author       = {Jonathan T. Barron},
  title        = {Continuously Differentiable Exponential Linear Units},
  journal      = {CoRR},
  year         = {2017},
  eprinttype    = {arXiv},
  eprint       = {1704.07483},
}

@article{CosineAnnealingLR,
  author       = {Ilya Loshchilov and
                  Frank Hutter},
  title        = {{SGDR:} Stochastic Gradient Descent with Restarts},
  journal      = {ICLR},
  year         = {2017},
  eprinttype    = {arXiv},
  eprint       = {1608.03983},
}

@article{Gambhir:2022dut,
    author = "Gambhir, Rikab and Nachman, Benjamin and Thaler, Jesse",
    title = "{Bias and priors in machine learning calibrations for high energy physics}",
    eprint = "2205.05084",
    archivePrefix = "arXiv",
    primaryClass = "hep-ph",
    reportNumber = "MIT-CTP 5432",
    doi = "10.1103/PhysRevD.106.036011",
    journal = "Phys. Rev. D",
    volume = "106",
    number = "3",
    pages = "036011",
    year = "2022"
}

@Article{         numpy,
 title         = {Array programming with {NumPy}},
 author        = {Charles R. Harris and K. Jarrod Millman and St{\'{e}}fan J.
                 van der Walt and Ralf Gommers and Pauli Virtanen and David
                 Cournapeau and Eric Wieser and Julian Taylor and Sebastian
                 Berg and Nathaniel J. Smith and Robert Kern and Matti Picus
                 and Stephan Hoyer and Marten H. van Kerkwijk and Matthew
                 Brett and Allan Haldane and Jaime Fern{\'{a}}ndez del
                 R{\'{i}}o and Mark Wiebe and Pearu Peterson and Pierre
                 G{\'{e}}rard-Marchant and Kevin Sheppard and Tyler Reddy and
                 Warren Weckesser and Hameer Abbasi and Christoph Gohlke and
                 Travis E. Oliphant},
 year          = {2020},
 month         = sep,
 journal       = {Nature},
 volume        = {585},
 number        = {7825},
 pages         = {357--362},
 doi           = {10.1038/s41586-020-2649-2},
 publisher     = {Springer Science and Business Media {LLC}},
}

@incollection{torch,
  title = {PyTorch: An Imperative Style, High-Performance Deep Learning Library},
  author = {Paszke, Adam and Gross, Sam and Massa, Francisco and Lerer, Adam and Bradbury, James and Chanan, Gregory and Killeen, Trevor and Lin, Zeming and Gimelshein, Natalia and Antiga, Luca and Desmaison, Alban and Kopf, Andreas and Yang, Edward and DeVito, Zachary and Raison, Martin and Tejani, Alykhan and Chilamkurthy, Sasank and Steiner, Benoit and Fang, Lu and Bai, Junjie and Chintala, Soumith},
  booktitle = {Advances in Neural Information Processing Systems 32},
  pages = {8024--8035},
  year = {2019},
  publisher = {Curran Associates, Inc.},
  eprint={1912.01703},
  archivePrefix={arXiv},
  primaryClass={cs.LG},
}

@Article{matplotlib,
  Author    = {Hunter, J. D.},
  Title     = {Matplotlib: A 2D graphics environment},
  Journal   = {Computing in Science \& Engineering},
  Volume    = {9},
  Number    = {3},
  Pages     = {90--95},
  publisher = {IEEE COMPUTER SOC},
  doi       = {10.1109/MCSE.2007.55},
  year      = 2007
}

@software{reback2020pandas,
    author       = {{Pandas Development Team}},
    title        = {Pandas},
    month        = feb,
    year         = 2020,
    publisher    = {Zenodo},
    doi          = {10.5281/zenodo.3509134},
    url          = {https://doi.org/10.5281/zenodo.3509134}
}

@article{seaborn,
    doi = {10.21105/joss.03021},
    year = {2021},
    publisher = {The Open Journal},
    volume = {6},
    number = {60},
    pages = {3021},
    author = {Michael L. Waskom},
    title = {Seaborn: Statistical Data Visualization},
    journal = {Journal of Open Source Software}
 }

@article{scikit-hep,
    author = "Rodrigues, Eduardo and others",
    editor = "Doglioni, C. and Kim, D. and Stewart, G. A. and Silvestris, L. and Jackson, P. and Kamleh, W.",
    title = "{The Scikit HEP Project -- overview and prospects}",
    eprint = "2007.03577",
    archivePrefix = "arXiv",
    primaryClass = "physics.comp-ph",
    doi = "10.1051/epjconf/202024506028",
    journal = "EPJ Web Conf.",
    volume = "245",
    pages = "06028",
    year = "2020"
}

@article{Manole:2022bmi,
    author = "Manole, Tudor and Bryant, Patrick and Alison, John and Kuusela, Mikael and Wasserman, Larry",
    title = "{Background Modeling for Double Higgs Boson Production: Density Ratios and Optimal Transport}",
    eprint = "2208.02807",
    archivePrefix = "arXiv",
    primaryClass = "stat.AP",
    month = "8",
    year = "2024"
}

@article{Craig:2024rlv,
    author = "Craig, Nathaniel and Howard, Jessica N. and Li, Hancheng",
    title = "{Exploring Optimal Transport for Event-Level Anomaly Detection at the Large Hadron Collider}",
    eprint = "2401.15542",
    archivePrefix = "arXiv",
    primaryClass = "hep-ph",
    month = "1",
    year = "2024",
    doi = "10.48550/arXiv.2401.15542"
}

@ARTICLE{2020SciPy-NMeth,
  author  = {Virtanen, Pauli and Gommers, Ralf and Oliphant, Travis E. and
            Haberland, Matt and Reddy, Tyler and Cournapeau, David and
            Burovski, Evgeni and Peterson, Pearu and Weckesser, Warren and
            Bright, Jonathan and {van der Walt}, St{\'e}fan J. and
            Brett, Matthew and Wilson, Joshua and Millman, K. Jarrod and
            Mayorov, Nikolay and Nelson, Andrew R. J. and Jones, Eric and
            Kern, Robert and Larson, Eric and Carey, C J and
            Polat, {\.I}lhan and Feng, Yu and Moore, Eric W. and
            {VanderPlas}, Jake and Laxalde, Denis and Perktold, Josef and
            Cimrman, Robert and Henriksen, Ian and Quintero, E. A. and
            Harris, Charles R. and Archibald, Anne M. and
            Ribeiro, Ant{\^o}nio H. and Pedregosa, Fabian and
            {van Mulbregt}, Paul and {SciPy 1.0 Contributors}},
  title   = {{{SciPy} 1.0: Fundamental Algorithms for Scientific
            Computing in Python}},
  journal = {Nature Methods},
  year    = {2020},
  volume  = {17},
  pages   = {261--272},
  adsurl  = {https://rdcu.be/b08Wh},
  doi     = {10.1038/s41592-019-0686-2},
  eprint={1907.10121},
  archivePrefix={arXiv},
  primaryClass={cs.MS},
}

@misc{durkan2019neuralsplineflows,
      title={Neural Spline Flows}, 
      author={Conor Durkan and Artur Bekasov and Iain Murray and George Papamakarios},
      year={2019},
      eprint={1906.04032},
      archivePrefix={arXiv},
      primaryClass={stat.ML},
}

@misc{wehenkel2024addressingmisspecificationsimulationbasedinference,
      title={Addressing Misspecification in Simulation-based Inference through Data-driven Calibration}, 
      author={Antoine Wehenkel and Juan L. Gamella and Ozan Sener and Jens Behrmann and Guillermo Sapiro and Marco Cuturi and Jörn-Henrik Jacobsen},
      year={2024},
      eprint={2405.08719},
      archivePrefix={arXiv},
      primaryClass={stat.ML},
}

@Article{Evans:2008zzb,
      author         = "Evans, Lyndon and Bryant, Philip",
      title          = "{LHC Machine}",
      journal        = "JINST",
      volume         = "3",
      pages          = "S08001",
      doi            = "10.1088/1748-0221/3/08/S08001",
      year           = "2008",
      SLACcitation   = "%%CITATION = JINST,3,S08001;%%",
}

@Article{Cacciari:2008gp,
     author    = "Cacciari, Matteo and Salam, Gavin P. and Soyez, Gregory",
     title     = "{The anti-\(k_{t}\) jet clustering algorithm}",
     journal   = "JHEP",
     volume    = "04",
     year      = "2008",
     pages     = "063",
     eprint    = "0802.1189",
     archivePrefix = "arXiv",
     primaryClass  =  "hep-ph",
     doi       = "10.1088/1126-6708/2008/04/063",
     SLACcitation  = "%%CITATION = 0802.1189;%%"
}

@Article{Fastjet,
      author         = "Cacciari, Matteo and Salam, Gavin P. and Soyez, Gregory",
      title          = "{FastJet user manual}",
      journal        = "Eur. Phys. J. C",
      volume         = "72",
      year           = "2012",
      pages          = "1896",
      doi            = "10.1140/epjc/s10052-012-1896-2",
      eprint         = "1111.6097",
      archivePrefix  = "arXiv",
      primaryClass   = "hep-ph",
      reportNumber   = "CERN-PH-TH-2011-297",
      SLACcitation   = "%%CITATION = ARXIV:1111.6097;%%"
}

@Article{Pumplin:2002vw,
     author    = "Pumplin, J. and others",
     title     = "{New Generation of Parton Distributions with Uncertainties from Global QCD Analysis}",
     journal   = "JHEP",
     volume    = "07",
     year      = "2002",
     pages     = "012",
     doi       = "10.1088/1126-6708/2002/07/012",
     eprint    = "hep-ph/0201195",
     archivePrefix = "arXiv",
     SLACcitation  = "%%CITATION = HEP-PH/0201195;%%"
}

@article{Lai:2010vv,
      author         = "Lai, H.-L. and others",
      title          = "{New parton distributions for collider physics}",
      journal        = "Phys. Rev. D",
      volume         = "82",
      pages          = "074024",
      doi            = "10.1103/PhysRevD.82.074024",
      year           = "2010",
      eprint         = "1007.2241",
      archivePrefix  = "arXiv",
      primaryClass   = "hep-ph",
      reportNumber   = "MSUHEP-100707, SMU-HEP-10-10",
      SLACcitation   = "%%CITATION = ARXIV:1007.2241;%%",
}

@article{Harland-Lang:2014zoa,
      author         = "Harland-Lang, L. A. and Martin, A. D. and Motylinski, P.
                        and Thorne, R. S.",
      title          = "{Parton distributions in the LHC era: MMHT 2014 PDFs}",
      journal        = "Eur. Phys. J. C",
      number         = "5",
      volume         = "75",
      pages          = "204",
      doi            = "10.1140/epjc/s10052-015-3397-6",
      year           = "2015",
      eprint         = "1412.3989",
      archivePrefix  = "arXiv",
      primaryClass   = "hep-ph",
      reportNumber   = "LCTS-2014-47, IPPP-14-97, DCPT-14-194",
      SLACcitation   = "%%CITATION = ARXIV:1412.3989;%%",
}

@article{Ball:2012cx,
      author         = "{NNPDF Collaboration} and Ball, Richard D. and others",
      title          = "{Parton distributions with LHC data}",
      journal        = "Nucl. Phys. B",
      volume         = "867",
      year           = "2013",
      pages          = "244",
      doi            = "10.1016/j.nuclphysb.2012.10.003",
      eprint         = "1207.1303",
      archivePrefix  = "arXiv",
      primaryClass   = "hep-ph",
      reportNumber   = "EDINBURGH-2012-08, IFUM-FT-997, FR-PHENO-2012-014,
                        RWTH-TTK-12-25, CERN-PH-TH-2012-037, SFB-CPP-12-47\,
                        --CERN-PH-TH-2012-037",
      SLACcitation   = "%%CITATION = ARXIV:1207.1303;%%"
}

@article{Ball:2014uwa,
      author         = "{NNPDF Collaboration} and Ball, Richard D. and others",
      title          = "{Parton distributions for the LHC run II}",
      journal        = "JHEP",
      volume         = "04",
      year           = "2015",
      pages          = "040",
      doi            = "10.1007/JHEP04(2015)040",
      eprint         = "1410.8849",
      archivePrefix  = "arXiv",
      primaryClass   = "hep-ph",
      reportNumber   = "EDINBURGH-2014-15, IFUM-1034-FT, CERN-PH-TH-2013-253,
                        OUTP-14-11P, CAVENDISH-HEP-14-11",
      SLACcitation   = "%%CITATION = ARXIV:1410.8849;%%"
}

@article{Skands:2014pea,
      author         = "Skands, Peter and Carrazza, Stefano and Rojo, Juan",
      title          = "{Tuning PYTHIA 8.1: the Monash 2013 Tune}",
      journal        = "Eur. Phys. J. C",
      volume         = "74",
      year           = "2014",
      number         = "8",
      pages          = "3024",
      doi            = "10.1140/epjc/s10052-014-3024-y",
      eprint         = "1404.5630",
      archivePrefix  = "arXiv",
      primaryClass   = "hep-ph",
      reportNumber   = "CERN-PH-TH-2014-069, MCNET-14-08, OUTP-14-05P",
      SLACcitation   = "%%CITATION = ARXIV:1404.5630;%%"
}

@Article{Sjostrand:2007gs,
	author    = "Sj{\"o}strand, T. and Mrenna, S. and Skands, P.",
	title     = "{A brief introduction to PYTHIA 8.1}",
	journal   = "Comput. Phys. Commun.",
	volume    = "178",
	year      = "2008",
	pages     = "852-867",
	eprint    = "0710.3820",
	archivePrefix = "arXiv",
	primaryClass  =  "hep-ph",
	doi       = "10.1016/j.cpc.2008.01.036",
	SLACcitation  = "%%CITATION = 0710.3820;%%"
}

@article{Sjostrand:2014zea,
      author         = "Sj{\"o}strand, Torbj{\"o}rn and Ask, Stefan and Christiansen,
                        Jesper R. and Corke, Richard and Desai, Nishita and Ilten,
                        Philip and Mrenna, Stephen and Prestel, Stefan and
                        Rasmussen, Christine O. and Skands, Peter Z.",
      title          = "{An introduction to PYTHIA 8.2}",
      journal        = "Comput. Phys. Commun.",
      volume         = "191",
      year           = "2015",
      pages          = "159",
      doi            = "10.1016/j.cpc.2015.01.024",
      eprint         = "1410.3012",
      archivePrefix  = "arXiv",
      primaryClass   = "hep-ph",
      reportNumber   = "LU-TP-14-36, MCNET-14-22, CERN-PH-TH-2014-190,
                        FERMILAB-PUB-14-316-CD, DESY-14-178, SLAC-PUB-16122,
                        --FERMILAB-PUB-14-316-CD",
      SLACcitation   = "%%CITATION = ARXIV:1410.3012;%%"
}

@Article{Lange:2001uf,
      author         = "Lange, D. J.",
      title          = "{The EvtGen particle decay simulation package}",
      booktitle      = "{Proceedings, 7th International Conference on B physics
                        at hadron machines (BEAUTY 2000)}",
      journal        = "Nucl. Instrum. Meth. A",
      volume         = "462",
      year           = "2001",
      pages          = "152",
      doi            = "10.1016/S0168-9002(01)00089-4",
      SLACcitation   = "%%CITATION = NUIMA,A462,152;%%"
}

@Article{Frixione:2008yi,
     author    = "Frixione, Stefano and Laenen, Eric and Motylinski, Patrick and White, Chris and Webber, Bryan R.",
     title     = "{Single-top hadroproduction in association with a \(W\) boson}",
     journal   = "JHEP",
     volume    = "07",
     year      = "2008",
     pages     = "029",
     eprint    = "0805.3067",
     archivePrefix = "arXiv",
     primaryClass  =  "hep-ph",
     doi       = "10.1088/1126-6708/2008/07/029"
}

@Article{Bahr:2008pv,
      author         = "B{\"a}hr, M. and others",
      title          = "{Herwig++ physics and manual}",
      journal        = "Eur. Phys. J. C",
      volume         = "58",
      year           = "2008",
      pages          = "639",
      doi            = "10.1140/epjc/s10052-008-0798-9",
      eprint         = "0803.0883",
      archivePrefix  = "arXiv",
      primaryClass   = "hep-ph",
      reportNumber   = "CERN-PH-TH-2008-038, CAVENDISH-HEP-08-03, KA-TP-05-2008,
                        DCPT-08-22, IPPP-08-11, CP3-08-05",
      SLACcitation   = "%%CITATION = ARXIV:0803.0883;%%"
}

@Article{Bellm:2015jjp,
      author         = "Bellm, Johannes and others",
      title          = "{Herwig 7.0/Herwig++ 3.0 release note}",
      journal        = "Eur. Phys. J. C",
      volume         = "76",
      year           = "2016",
      number         = "4",
      pages          = "196",
      doi            = "10.1140/epjc/s10052-016-4018-8",
      eprint         = "1512.01178",
      archivePrefix  = "arXiv",
      primaryClass   = "hep-ph",
      reportNumber   = "CERN-PH-TH-2015-289, MAN-HEP-2015-15, IFJPAN-IV-2015-13,
                        HERWIG-2015-01, KA-TP-18-2015, DCPT-15-142, MCNET-15-28,
                        IPPP-15-71, --HERWIG-2015-01",
      SLACcitation   = "%%CITATION = ARXIV:1512.01178;%%"
}

@Article{Nason:2004rx,
      author         = "Nason, Paolo",
      title          = "{A new method for combining NLO QCD with shower Monte Carlo algorithms}",
      journal        = "JHEP",
      volume         = "11",
      pages          = "040",
      doi            = "10.1088/1126-6708/2004/11/040",
      year           = "2004",
      eprint         = "hep-ph/0409146",
      archivePrefix  = "arXiv",
}

@Article{Frixione:2007vw,
      author         = "Frixione, Stefano and Nason, Paolo and Oleari, Carlo",
      title          = "{Matching NLO QCD computations with parton shower
                        simulations: the POWHEG method}",
      journal        = "JHEP",
      volume         = "11",
      pages          = "070",
      doi            = "10.1088/1126-6708/2007/11/070",
      year           = "2007",
      eprint         = "0709.2092",
      archivePrefix  = "arXiv",
      primaryClass   = "hep-ph",
}

@Article{Alioli:2010xd,
      author         = "Alioli, Simone and Nason, Paolo and Oleari, Carlo and Re,
                        Emanuele",
      title          = "{A general framework for implementing NLO calculations in
                        shower Monte Carlo programs: the POWHEG BOX}",
      journal        = "JHEP",
      volume         = "06",
      pages          = "043",
      doi            = "10.1007/JHEP06(2010)043",
      year           = "2010",
      eprint         = "1002.2581",
      archivePrefix  = "arXiv",
      primaryClass   = "hep-ph",
}

@article{Nason:2013ydw,
      author         = "Nason, Paolo and Zanderighi, Giulia",
      title          = "{\(W^+ W^-\), \(W Z\) and \(Z Z\) production in the
                        POWHEG-BOX-V2}",
      journal        = "Eur. Phys. J. C",
      volume         = "74",
      year           = "2014",
      number         = "1",
      pages          = "2702",
      doi            = "10.1140/epjc/s10052-013-2702-5",
      eprint         = "1311.1365",
      archivePrefix  = "arXiv",
      primaryClass   = "hep-ph",
      SLACcitation   = "%%CITATION = ARXIV:1311.1365;%%"
}

@article{Anastasiou:2003ds,
      author         = "Anastasiou, Charalampos and Dixon, Lance and Melnikov,
                        Kirill and Petriello, Frank",
      title          = "{High-precision QCD at hadron colliders: Electroweak
                        gauge boson rapidity distributions at next-to-next-to leading order}",
      journal        = "Phys. Rev. D",
      volume         = "69",
      year           = "2004",
      pages          = "094008",
      doi            = "10.1103/PhysRevD.69.094008",
      eprint         = "hep-ph/0312266",
      archivePrefix  = "arXiv",
      reportNumber   = "SLAC-PUB-10288, UH-511-1042-03",
      SLACcitation   = "%%CITATION = HEP-PH/0312266;%%"
}

@article{Frixione:2007nw,
      author         = "Frixione, Stefano and Ridolfi, Giovanni and Nason, Paolo",
      title          = "{A positive-weight next-to-leading-order Monte Carlo for
                        heavy flavour hadroproduction}",
      journal        = "JHEP",
      volume         = "09",
      pages          = "126",
      doi            = "10.1088/1126-6708/2007/09/126",
      year           = "2007",
      eprint         = "0707.3088",
      archivePrefix  = "arXiv",
      primaryClass   = "hep-ph",
      SLACcitation   = "%%CITATION = ARXIV:0707.3088;%%",
}

@Article{Gleisberg:2008ta,
      author         = "Gleisberg, T. and H{\"o}che, S. and Krauss, F. and Sch{\"o}nherr, M. and Schumann, S. and Siegert, F. and Winter J.",
      title          = "{Event generation with SHERPA 1.1}",
      journal        = "JHEP",
      volume         = "02",
      pages          = "007",
      doi            = "10.1088/1126-6708/2009/02/007",
      year           = "2009",
      eprint         = "0811.4622",
      archivePrefix  = "arXiv",
      primaryClass   = "hep-ph",
}

@Article{Hoeche:2009rj,
      author         = "H{\"o}che, Stefan and Krauss, Frank and Schumann, Steffen and Siegert, Frank",
      title          = "{QCD matrix elements and truncated showers}",
      journal        = "JHEP",
      volume         = "05",
      pages          = "053",
      doi            = "10.1088/1126-6708/2009/05/053",
      year           = "2009",
      eprint         = "0903.1219",
      archivePrefix  = "arXiv",
      primaryClass   = "hep-ph",
}

@Article{Gleisberg:2008fv,
      author         = "Gleisberg, Tanju and H{\"o}che, Stefan",
      title          = "{Comix, a new matrix element generator}",
      journal        = "JHEP",
      volume         = "12",
      pages          = "039",
      doi            = "10.1088/1126-6708/2008/12/039",
      year           = "2008",
      eprint         = "0808.3674",
      archivePrefix  = "arXiv",
      primaryClass   = "hep-ph",
}

@Article{Schumann:2007mg,
      author         = "Schumann, Steffen and Krauss, Frank",
      title          = "{A parton shower algorithm based on Catani--Seymour dipole factorisation}",
      journal        = "JHEP",
      volume         = "03",
      pages          = "038",
      doi            = "10.1088/1126-6708/2008/03/038",
      year           = "2008",
      eprint         = "0709.1027",
      archivePrefix  = "arXiv",
      primaryClass   = "hep-ph",
}

@article{Hoeche:2012yf,
      author         = "H{\"o}che, Stefan and Krauss, Frank and Sch{\"o}nherr, Marek and
                        Siegert, Frank",
      title          = "{QCD matrix elements + parton showers. The NLO case}",
      journal        = "JHEP",
      volume         = "04",
      year           = "2013",
      pages          = "027",
      doi            = "10.1007/JHEP04(2013)027",
      eprint         = "1207.5030",
      archivePrefix  = "arXiv",
      primaryClass   = "hep-ph",
      reportNumber   = "SLAC-PUB-15191, IPPP-12-52, DCPT-12-104, LPN12-081,
                        FR-PHENO-2012-017, MCNET-12-09, --FR-PHENO-2012-017",
      SLACcitation   = "%%CITATION = ARXIV:1207.5030;%%"
}

@article{Bothmann:2019yzt,
      author = "Bothmann, Enrico and others",
      title = "{Event generation with Sherpa 2.2}",
      journal = "SciPost Phys.",
      volume = "7",
      year = "2019",
      number = "3",
      pages = "034",
      doi = "10.21468/SciPostPhys.7.3.034",
      reportNumber = "FERMILAB-PUB-19-218-T, SLAC-PUB-17433, IPPP/19/42, MCNET-19-11",
      eprint = "1905.09127",
      archivePrefix = "arXiv",
      primaryClass = "hep-ph",
}

@article{Hoeche:2011fd,
      author         = "H{\"o}che, Stefan and Krauss, Frank and Sch{\"o}nherr, Marek and
                        Siegert, Frank",
      title          = "{A critical appraisal of NLO+PS matching methods}",
      journal        = "JHEP",
      volume         = "09",
      year           = "2012",
      pages          = "049",
      doi            = "10.1007/JHEP09(2012)049",
      eprint         = "1111.1220",
      archivePrefix  = "arXiv",
      primaryClass   = "hep-ph",
      reportNumber   = "SLAC-PUB-14661, IPPP-11-67, DCPT-11-134, LPN11-58,
                        FR-PHENO-2011-019, MCNET-11-24",
      SLACcitation   = "%%CITATION = ARXIV:1111.1220;%%"
}

@article{Catani:2001cc,
      author         = "Catani, S. and Krauss, F. and Webber, B. R. and Kuhn, R.",
      title          = "{QCD Matrix Elements + Parton Showers}",
      journal        = "JHEP",
      volume         = "11",
      year           = "2001",
      pages          = "063",
      doi            = "10.1088/1126-6708/2001/11/063",
      eprint         = "hep-ph/0109231",
      archivePrefix  = "arXiv",
      reportNumber   = "CERN-TH-2000-367, CAVENDISH-HEP-00-03",
      SLACcitation   = "%%CITATION = HEP-PH/0109231;%%"
}

@article{Cascioli:2011va,
      author         = "Cascioli, Fabio and Maierh{\"o}fer, Philipp and Pozzorini, Stefano",
      title          = "{Scattering Amplitudes with Open Loops}",
      journal        = "Phys. Rev. Lett.",
      volume         = "108",
      year           = "2012",
      pages          = "111601",
      doi            = "10.1103/PhysRevLett.108.111601",
      eprint         = "1111.5206",
      archivePrefix  = "arXiv",
      primaryClass   = "hep-ph",
      reportNumber   = "ZU-TH-23-11, LPN11-66",
      SLACcitation   = "%%CITATION = ARXIV:1111.5206;%%"
}

@article{Buccioni:2019sur,
    author = "Buccioni, Federico and Lang, Jean-Nicolas and Lindert, Jonas M. and Maierh{\"o}fer, Philipp and Pozzorini, Stefano and Zhang, Hantian and Zoller, Max F.",
    title = "{OpenLoops 2}",
    eprint = "1907.13071",
    archivePrefix = "arXiv",
    primaryClass = "hep-ph",
    reportNumber = "IPPP/19/62, FR-PHENO-2019-12, PSI-PR-19-15, ZU-TH 37/19",
    doi = "10.1140/epjc/s10052-019-7306-2",
    journal = "Eur. Phys. J. C",
    volume = "79",
    number = "10",
    pages = "866",
    year = "2019"
}

@article{Denner:2016kdg,
      author         = "Denner, Ansgar and Dittmaier, Stefan and Hofer, Lars",
      title          = "{\textsc{Collier}: A fortran-based complex one-loop library in
                        extended regularizations}",
      journal        = "Comput. Phys. Commun.",
      volume         = "212",
      year           = "2017",
      pages          = "220-238",
      doi            = "10.1016/j.cpc.2016.10.013",
      eprint         = "1604.06792",
      archivePrefix  = "arXiv",
      primaryClass   = "hep-ph",
      reportNumber   = "FR-PHENO-2016-003, ICCUB-16-016",
      SLACcitation   = "%%CITATION = ARXIV:1604.06792;%%"
}

@article{Agostinelli:2002hh,
      author         = "Agostinelli, S. and others",
      title          = "{\textsc{Geant4} -- a simulation toolkit}",
      journal        = "Nucl. Instrum. Meth. A",
      volume         = "506",
      year           = "2003",
      pages          = "250",
      doi            = "10.1016/S0168-9002(03)01368-8",
      reportNumber   = "SLAC-PUB-9350, FERMILAB-PUB-03-339",
      SLACcitation   = "%%CITATION = NUIMA,A506,250;%%"
}

@Article{PERF-2007-01,
    author         = "{ATLAS Collaboration}",
    title          = "{The ATLAS Experiment at the CERN Large Hadron Collider}",
    journal        = "JINST",
    volume         = "3",
    year           = "2008",
    pages          = "S08003",
    doi            = "10.1088/1748-0221/3/08/S08003",
    primaryClass   = "hep-ex",
}

@Article{SOFT-2010-01,
    author         = "{ATLAS Collaboration}",
    title          = "{The ATLAS Simulation Infrastructure}",
    journal        = "Eur. Phys. J. C",
    volume         = "70",
    year           = "2010",
    pages          = "823",
    doi            = "10.1140/epjc/s10052-010-1429-9",
    eprint         = "1005.4568",
    archivePrefix  = "arXiv",
    primaryClass   = "physics.ins-det",
}

@Article{STDM-2012-23,
    author         = "{ATLAS Collaboration}",
    title          = "{Measurement of the \(Z/\gamma^*\) boson transverse momentum distribution in \(pp\) collisions at \(\sqrt{s} = 7\,\text{TeV}\) with the ATLAS detector}",
    journal        = "JHEP",
    volume         = "09",
    year           = "2014",
    pages          = "145",
    doi            = "10.1007/JHEP09(2014)145",
    reportNumber   = "CERN-PH-EP-2014-075",
    eprint         = "1406.3660",
    archivePrefix  = "arXiv",
    primaryClass   = "hep-ex",
}

@Article{PERF-2015-01,
    author         = "{ATLAS Collaboration}",
    title          = "{Reconstruction of primary vertices at the ATLAS experiment in Run~1 proton--proton collisions at the LHC}",
    journal        = "Eur. Phys. J. C",
    volume         = "77",
    year           = "2017",
    pages          = "332",
    doi            = "10.1140/epjc/s10052-017-4887-5",
    reportNumber   = "CERN-EP-2016-150",
    eprint         = "1611.10235",
    archivePrefix  = "arXiv",
    primaryClass   = "physics.ins-det",
}

@Article{PERF-2015-08,
    author         = "{ATLAS Collaboration}",
    title          = "{Performance of the ATLAS track reconstruction algorithms in dense environments in LHC Run~2}",
    journal        = "Eur. Phys. J. C",
    volume         = "77",
    year           = "2017",
    pages          = "673",
    doi            = "10.1140/epjc/s10052-017-5225-7",
    reportNumber   = "CERN-EP-2017-045",
    eprint         = "1704.07983",
    archivePrefix  = "arXiv",
    primaryClass   = "hep-ex",
}

@Article{PERF-2015-09,
    author         = "{ATLAS Collaboration}",
    title          = "{Jet reconstruction and performance using particle flow with the ATLAS Detector}",
    journal        = "Eur. Phys. J. C",
    volume         = "77",
    year           = "2017",
    pages          = "466",
    doi            = "10.1140/epjc/s10052-017-5031-2",
    reportNumber   = "CERN-EP-2017-024",
    eprint         = "1703.10485",
    archivePrefix  = "arXiv",
    primaryClass   = "hep-ex",
}

@Article{EXOT-2016-33,
    author         = "{ATLAS Collaboration}",
    title          = "{Search for resonances in the mass distribution of jet pairs with one or two jets identified as \(b\)-jets in proton--proton collisions at \(\sqrt{s} = 13\,\text{TeV}\) with the ATLAS detector}",
    journal        = "Phys. Rev. D",
    volume         = "98",
    year           = "2018",
    pages          = "032016",
    doi            = "10.1103/PhysRevD.98.032016",
    reportNumber   = "CERN-EP-2018-075",
    eprint         = "1805.09299",
    archivePrefix  = "arXiv",
    primaryClass   = "hep-ex",
}

@Article{TRIG-2016-01,
    author         = "{ATLAS Collaboration}",
    title          = "{Performance of the ATLAS trigger system in 2015}",
    journal        = "Eur. Phys. J. C",
    volume         = "77",
    year           = "2017",
    pages          = "317",
    doi            = "10.1140/epjc/s10052-017-4852-3",
    reportNumber   = "CERN-EP-2016-241",
    eprint         = "1611.09661",
    archivePrefix  = "arXiv",
    primaryClass   = "hep-ex",
}

@Article{TOPQ-2017-03,
    author         = "{ATLAS Collaboration}",
    title          = "{Measurement of the top quark mass in the \(t\bar{t} \to\) lepton+jets channel from \(\sqrt{s} = 8\,\text{TeV}\) ATLAS data and combination with previous results}",
    journal        = "Eur. Phys. J. C",
    volume         = "79",
    year           = "2019",
    pages          = "290",
    doi            = "10.1140/epjc/s10052-019-6757-9",
    reportNumber   = "CERN-EP-2018-238",
    eprint         = "1810.01772",
    archivePrefix  = "arXiv",
    primaryClass   = "hep-ex",
}

@Article{TOPQ-2017-19,
    author         = "{ATLAS Collaboration}",
    title          = "{Measurements of jet observables sensitive to \(b\)-quark fragmentation in \(t\bar{t}\) events at the LHC with the ATLAS detector}",
    journal        = "Phys. Rev. D",
    volume         = "106",
    year           = "2022",
    pages          = "032008",
    doi            = "10.1103/PhysRevD.106.032008",
    reportNumber   = "CERN-EP-2021-116",
    eprint         = "2202.13901",
    archivePrefix  = "arXiv",
    primaryClass   = "hep-ex",
}

@Article{DAPR-2018-01,
    author         = "{ATLAS Collaboration}",
    title          = "{ATLAS data quality operations and performance for 2015--2018 data-taking}",
    journal        = "JINST",
    volume         = "15",
    year           = "2020",
    pages          = "P04003",
    doi            = "10.1088/1748-0221/15/04/P04003",
    reportNumber   = "CERN-EP-2019-207",
    eprint         = "1911.04632",
    archivePrefix  = "arXiv",
    primaryClass   = "physics.ins-det",
}

@Article{EGAM-2018-01,
    author         = "{ATLAS Collaboration}",
    title          = "{Electron and photon performance measurements with the ATLAS detector using the 2015--2017 LHC proton--proton collision data}",
    journal        = "JINST",
    volume         = "14",
    year           = "2019",
    pages          = "P12006",
    doi            = "10.1088/1748-0221/14/12/P12006",
    reportNumber   = "CERN-EP-2019-145",
    eprint         = "1908.00005",
    archivePrefix  = "arXiv",
    primaryClass   = "hep-ex",
}

@Article{EXOT-2018-48,
    author         = "{ATLAS Collaboration}",
    title          = "{Search for  \(t\bar{t}\) resonances in fully hadronic final states in \(pp\) collisions at \(\sqrt{s} = 13\,\text{TeV}\) with the ATLAS detector}",
    journal        = "JHEP",
    volume         = "10",
    year           = "2020",
    pages          = "061",
    doi            = "10.1007/JHEP10(2020)061",
    reportNumber   = "CERN-EP-2020-055",
    eprint         = "2005.05138",
    archivePrefix  = "arXiv",
    primaryClass   = "hep-ex",
}

@Article{FTAG-2018-01,
    author         = "{ATLAS Collaboration}",
    title          = "{ATLAS \(b\)-jet identification performance and efficiency measurement with \(t\bar{t}\) events in \(pp\) collisions at \(\sqrt{s} = 13\,\text{TeV}\)}",
    journal        = "Eur. Phys. J. C",
    volume         = "79",
    year           = "2019",
    pages          = "970",
    doi            = "10.1140/epjc/s10052-019-7450-8",
    reportNumber   = "CERN-EP-2019-132",
    eprint         = "1907.05120",
    archivePrefix  = "arXiv",
    primaryClass   = "hep-ex",
}

@Article{HDBS-2018-34,
    author         = "{ATLAS Collaboration}",
    title          = "{Search for Higgs boson pair production in the two bottom quarks plus two photons final state in \(pp\) collisions at \(\sqrt{s} = 13\,\text{TeV}\) with the ATLAS detector}",
    journal        = "Phys. Rev. D",
    volume         = "106",
    year           = "2022",
    pages          = "052001",
    doi            = "10.1103/PhysRevD.106.052001",
    reportNumber   = "CERN-EP-2021-180",
    eprint         = "2112.11876",
    archivePrefix  = "arXiv",
    primaryClass   = "hep-ex",
}

@Article{HDBS-2018-40,
    author         = "{ATLAS Collaboration}",
    title          = "{Search for resonant and non-resonant Higgs boson pair production in the \(b\bar{b}\tau^+\tau^-\) decay channel using \(13\,\text{TeV}\) \(pp\) collision data from the ATLAS detector}",
    journal        = "JHEP",
    volume         = "07",
    year           = "2023",
    pages          = "040",
    doi            = "10.1007/JHEP07(2023)040",
    reportNumber   = "CERN-EP-2022-109",
    eprint         = "2209.10910",
    archivePrefix  = "arXiv",
    primaryClass   = "hep-ex",
}

@Article{HIGG-2018-49,
    author         = "{ATLAS Collaboration}",
    title          = "{Measurements of differential cross sections of Higgs boson production through gluon fusion in the \(H \to WW^{*} \to e\nu\mu\nu\) final state at \(\sqrt{s} = 13\,\text{TeV}\) with the ATLAS detector}",
    journal        = "Eur. Phys. J. C",
    volume         = "83",
    year           = "2023",
    pages          = "774",
    doi            = "10.1140/epjc/s10052-023-11873-5",
    reportNumber   = "CERN-EP-2022-228",
    eprint         = "2301.06822",
    archivePrefix  = "arXiv",
    primaryClass   = "hep-ex",
}

@Article{HIGG-2018-51,
    author         = "{ATLAS Collaboration}",
    title          = "{Measurements of \(WH\) and \(ZH\) production in the \(H \to b\bar{b}\) decay channel in \(pp\) collisions at \(13\,\text{TeV}\) with the ATLAS detector}",
    journal        = "Eur. Phys. J. C",
    volume         = "81",
    year           = "2021",
    pages          = "178",
    doi            = "10.1140/epjc/s10052-020-08677-2",
    reportNumber   = "CERN-EP-2020-087",
    eprint         = "2007.02873",
    archivePrefix  = "arXiv",
    primaryClass   = "hep-ex",
}

@Article{JETM-2018-05,
    author         = "{ATLAS Collaboration}",
    title          = "{Jet energy scale and resolution measured in proton--proton collisions at \(\sqrt{s} = 13\,\text{TeV}\) with the ATLAS detector}",
    journal        = "Eur. Phys. J. C",
    volume         = "81",
    year           = "2021",
    pages          = "689",
    doi            = "10.1140/epjc/s10052-021-09402-3",
    reportNumber   = "CERN-EP-2020-083",
    eprint         = "2007.02645",
    archivePrefix  = "arXiv",
    primaryClass   = "hep-ex",
}

@Article{MUON-2018-03,
    author         = "{ATLAS Collaboration}",
    title          = "{Muon reconstruction and identification efficiency in ATLAS using the full Run~2 \(pp\) collision data set at \(\sqrt{s} = 13\,\text{TeV}\)}",
    journal        = "Eur. Phys. J. C",
    volume         = "81",
    year           = "2021",
    pages          = "578",
    doi            = "10.1140/epjc/s10052-021-09233-2",
    reportNumber   = "CERN-EP-2020-199",
    eprint         = "2012.00578",
    archivePrefix  = "arXiv",
    primaryClass   = "hep-ex",
}

@Article{PIX-2018-001,
    author         = "Abbott, B. and others",
    title          = "{Production and integration of the ATLAS Insertable B-Layer}",
    journal        = "JINST",
    volume         = "13",
    year           = "2018",
    pages          = "T05008",
    doi            = "10.1088/1748-0221/13/05/T05008",
    eprint         = "1803.00844",
    archivePrefix  = "arXiv",
    primaryClass   = "physics.ins-det",
}

@Article{STDM-2018-52,
    author         = "{ATLAS Collaboration}",
    title          = "{Measurement of \(b\)-quark fragmentation properties in jets using the decay \(B^{\pm} \to J/\psi K^{\pm}\) in \(pp\) collisions at \(\sqrt{s} = 13\,\text{TeV}\) with the ATLAS detector}",
    journal        = "JHEP",
    volume         = "12",
    year           = "2021",
    pages          = "131",
    doi            = "10.1007/JHEP12(2021)131",
    reportNumber   = "CERN-EP-2021-123",
    eprint         = "2108.11650",
    archivePrefix  = "arXiv",
    primaryClass   = "hep-ex",
}

@Article{TOPQ-2018-06,
    author         = "{ATLAS Collaboration}",
    title          = "{Search for flavour-changing neutral-current interactions of a top quark and a gluon in \(pp\) collisions at \(\sqrt{s} = 13\,\text{TeV}\) with the ATLAS detector}",
    journal        = "Eur. Phys. J. C",
    volume         = "82",
    year           = "2022",
    pages          = "334",
    doi            = "10.1140/epjc/s10052-022-10182-7",
    reportNumber   = "CERN-EP-2021-234",
    eprint         = "2112.01302",
    archivePrefix  = "arXiv",
    primaryClass   = "hep-ex",
}

@Article{TOPQ-2018-26,
    author         = "{ATLAS Collaboration}",
    title          = "{Inclusive and differential cross-sections for dilepton \(t\bar{t}\) production measured in \(\sqrt{s} = 13\,\text{TeV}\) \(pp\) collisions with the ATLAS detector}",
    journal        = "JHEP",
    volume         = "07",
    year           = "2023",
    pages          = "141",
    doi            = "10.1007/JHEP07(2023)141",
    reportNumber   = "CERN-EP-2023-016",
    eprint         = "2303.15340",
    archivePrefix  = "arXiv",
    primaryClass   = "hep-ex",
}

@Article{TRIG-2018-01,
    author         = "{ATLAS Collaboration}",
    title          = "{Performance of the ATLAS muon triggers in Run~2}",
    journal        = "JINST",
    volume         = "15",
    year           = "2020",
    pages          = "P09015",
    doi            = "10.1088/1748-0221/15/09/p09015",
    reportNumber   = "CERN-EP-2020-031",
    eprint         = "2004.13447",
    archivePrefix  = "arXiv",
    primaryClass   = "physics.ins-det",
}

@Article{TRIG-2018-05,
    author         = "{ATLAS Collaboration}",
    title          = "{Performance of electron and photon triggers in ATLAS during LHC Run~2}",
    journal        = "Eur. Phys. J. C",
    volume         = "80",
    year           = "2020",
    pages          = "47",
    doi            = "10.1140/epjc/s10052-019-7500-2",
    reportNumber   = "CERN-EP-2019-169",
    eprint         = "1909.00761",
    archivePrefix  = "arXiv",
    primaryClass   = "hep-ex",
}

@Article{FTAG-2019-02,
    author         = "{ATLAS Collaboration}",
    title          = "{Calibration of the light-flavour jet mistagging efficiency of the \(b\)-tagging algorithms with \(Z\)+jets events using \(139\,\text{fb}^{-1}\) of ATLAS proton--proton collision data at \(\sqrt{s} = 13\,\text{TeV}\)}",
    journal        = "Eur. Phys. J. C",
    volume         = "83",
    year           = "2023",
    pages          = "728",
    doi            = "10.1140/epjc/s10052-023-11736-z",
    reportNumber   = "CERN-EP-2022-211",
    eprint         = "2301.06319",
    archivePrefix  = "arXiv",
    primaryClass   = "hep-ex",
}

@Article{FTAG-2019-07,
    author         = "{ATLAS Collaboration}",
    title          = "{ATLAS flavour-tagging algorithms for the LHC Run~2 \(pp\) collision dataset}",
    journal        = "Eur. Phys. J. C",
    volume         = "83",
    year           = "2023",
    pages          = "681",
    doi            = "10.1140/epjc/s10052-023-11699-1",
    reportNumber   = "CERN-EP-2022-226",
    eprint         = "2211.16345",
    archivePrefix  = "arXiv",
    primaryClass   = "physics.data-an",
}

@Article{HDBS-2019-29,
    author         = "{ATLAS Collaboration}",
    title          = "{Search for nonresonant pair production of Higgs bosons in the \(b\bar{b}b\bar{b}\) final state in \(pp\) collisions at \(\sqrt{s} = 13\,\text{TeV}\) with the ATLAS detector}",
    journal        = "Phys. Rev. D",
    volume         = "108",
    year           = "2023",
    pages          = "052003",
    doi            = "10.1103/PhysRevD.108.052003",
    reportNumber   = "CERN-EP-2022-235",
    eprint         = "2301.03212",
    archivePrefix  = "arXiv",
    primaryClass   = "hep-ex",
}

@Article{IDTR-2019-05,
    author         = "{ATLAS Collaboration}",
    title          = "{Alignment of the ATLAS Inner Detector in Run 2}",
    journal        = "Eur. Phys. J. C",
    volume         = "80",
    year           = "2020",
    pages          = "1194",
    doi            = "10.1140/epjc/s10052-020-08700-6",
    reportNumber   = "CERN-EP-2020-108",
    eprint         = "2007.07624",
    archivePrefix  = "arXiv",
    primaryClass   = "hep-ex",
}

@Article{FTAG-2020-08,
    author         = "{ATLAS Collaboration}",
    title          = "{Measurement of the \(c\)-jet mistagging efficiency in \(t\bar{t}\) events using \(pp\) collision data at \(\sqrt{s} = 13\,\text{TeV}\) collected with the ATLAS detector}",
    journal        = "Eur. Phys. J. C",
    volume         = "82",
    year           = "2022",
    pages          = "95",
    doi            = "10.1140/epjc/s10052-021-09843-w",
    reportNumber   = "CERN-EP-2021-166",
    eprint         = "2109.10627",
    archivePrefix  = "arXiv",
    primaryClass   = "hep-ex",
}

@Article{HIGG-2020-03,
    author         = "{ATLAS Collaboration}",
    title          = "{Probing the \(CP\) nature of the top-Higgs Yukawa coupling in \(t\bar{t}H\) and \(tH\) events with \(H \to b\bar{b}\) decays using the ATLAS detector at the LHC}",
    journal        = "Phys. Lett. B",
    volume         = "849",
    year           = "2024",
    pages          = "138469",
    doi            = "10.1016/j.physletb.2024.138469",
    reportNumber   = "CERN-EP-2022-208",
    eprint         = "2303.05974",
    archivePrefix  = "arXiv",
    primaryClass   = "hep-ex",
}

@Article{HIGG-2020-20,
    author         = "{ATLAS Collaboration}",
    title          = "{Measurements of \(WH\) and \(ZH\) production with Higgs boson decays into bottom quarks and direct constraints on the charm Yukawa coupling in \(13\,\text{TeV}\) \(pp\) collisions with the ATLAS detector}",
    year           = "2024",
    reportNumber   = "CERN-EP-2024-237",
    eprint         = "2410.19611",
    archivePrefix  = "arXiv",
    primaryClass   = "hep-ex",
}

@Article{HIGG-2020-25,
    author         = "{ATLAS Collaboration}",
    title          = "{Integrated and differential fiducial cross-section measurements for the vector boson fusion production of the Higgs boson in the \(H \to WW^{*} \to e\nu\mu\nu\) decay channel at \(13\,\text{TeV}\) with the ATLAS detector}",
    journal        = "Phys. Rev. D",
    volume         = "108",
    year           = "2023",
    pages          = "072003",
    doi            = "10.1103/PhysRevD.108.072003",
    reportNumber   = "CERN-EP-2023-025",
    eprint         = "2304.03053",
    archivePrefix  = "arXiv",
    primaryClass   = "hep-ex",
}

@Article{STDM-2020-20,
    author         = "{ATLAS Collaboration}",
    title          = "{Measurements of multijet event isotropies using optimal transport with the ATLAS detector}",
    journal        = "JHEP",
    volume         = "10",
    year           = "2023",
    pages          = "060",
    doi            = "10.1007/JHEP10(2023)060",
    reportNumber   = "CERN-EP-2023-079",
    eprint         = "2305.16930",
    archivePrefix  = "arXiv",
    primaryClass   = "hep-ex",
}

@Article{DAPR-2021-01,
    author         = "{ATLAS Collaboration}",
    title          = "{Luminosity determination in \(pp\) collisions at \(\sqrt{s} = 13\,\text{TeV}\) using the ATLAS detector at the LHC}",
    journal        = "Eur. Phys. J. C",
    volume         = "83",
    year           = "2023",
    pages          = "982",
    doi            = "10.1140/epjc/s10052-023-11747-w",
    reportNumber   = "CERN-EP-2022-281",
    eprint         = "2212.09379",
    archivePrefix  = "arXiv",
    primaryClass   = "hep-ex",
}

@Article{EGAM-2021-01,
    author         = "{ATLAS Collaboration}",
    title          = "{Electron and photon efficiencies in LHC Run~2 with the ATLAS experiment}",
    journal        = "JHEP",
    volume         = "05",
    year           = "2024",
    pages          = "162",
    doi            = "10.1007/JHEP05(2024)162",
    reportNumber   = "CERN-EP-2023-182",
    eprint         = "2308.13362",
    archivePrefix  = "arXiv",
    primaryClass   = "hep-ex",
}

@Article{HIGG-2021-12,
    author         = "{ATLAS Collaboration}",
    title          = "{Direct constraint on the Higgs-charm coupling from a search for Higgs boson decays into charm quarks with the ATLAS detector}",
    journal        = "Eur. Phys. J. C",
    volume         = "82",
    year           = "2022",
    pages          = "717",
    doi            = "10.1140/epjc/s10052-022-10588-3",
    reportNumber   = "CERN-EP-2021-251",
    eprint         = "2201.11428",
    archivePrefix  = "arXiv",
    primaryClass   = "hep-ex",
}

@Article{HIGG-2021-23,
    author         = "{ATLAS Collaboration}",
    title          = "{A detailed map of Higgs boson interactions by the ATLAS experiment ten years after the discovery}",
    journal        = "Nature",
    volume         = "607",
    year           = "2022",
    pages          = "52--59",
    doi            = "10.1038/s41586-022-04893-w",
    reportNumber   = "CERN-EP-2022-057",
    eprint         = "2207.00092",
    archivePrefix  = "arXiv",
    primaryClass   = "hep-ex",
    related        = "HIGG-2021-23-err",
    relatedstring  = "Erratum:",
}

@Article{HDBS-2022-03,
    author         = "{ATLAS Collaboration}",
    title          = "{Constraints on the Higgs boson self-coupling from single- and double-Higgs production with the ATLAS detector using \(pp\) collisions at \(\sqrt{s} = 13\,\text{TeV}\)}",
    journal        = "Phys. Lett. B",
    volume         = "843",
    year           = "2023",
    pages          = "137745",
    doi            = "10.1016/j.physletb.2023.137745",
    reportNumber   = "CERN-EP-2022-149",
    eprint         = "2211.01216",
    archivePrefix  = "arXiv",
    primaryClass   = "hep-ex",
}

@Article{MUON-2022-01,
    author         = "{ATLAS Collaboration}",
    title          = "{Studies of the muon momentum calibration and performance of the ATLAS detector with \(pp\) collisions at \(\sqrt{s} = 13\,\text{TeV}\)}",
    journal        = "Eur. Phys. J. C",
    volume         = "83",
    year           = "2023",
    pages          = "686",
    doi            = "10.1140/epjc/s10052-023-11584-x",
    reportNumber   = "CERN-EP-2022-274",
    eprint         = "2212.07338",
    archivePrefix  = "arXiv",
    primaryClass   = "hep-ex",
}

@Article{SOFT-2022-02,
    author         = "{ATLAS Collaboration}",
    title          = "{Software and computing for Run~3 of the ATLAS experiment at the LHC}",
    journal        = "Eur. Phys. J. C",
    volume         = "85",
    year           = "2025",
    pages          = "234",
    doi            = "10.1140/epjc/s10052-024-13701-w",
    reportNumber   = "CERN-EP-2024-100",
    eprint         = "2404.06335",
    archivePrefix  = "arXiv",
    primaryClass   = "hep-ex",
}

@Booklet{ATL-SOFT-PUB-2025-001,
    author         = "{ATLAS Collaboration}",
    title          = "{ATLAS Computing Acknowledgements}",
    howpublished   = "{ATL-SOFT-PUB-2025-001}",
    url            = "https://cds.cern.ch/record/2922210",
    year           = "2025",
}

@Report{ATLAS-TDR-19,
    author         = "{ATLAS Collaboration}",
    title          = "{ATLAS Insertable B-Layer: Technical Design Report}",
    type           = "ATLAS-TDR-19; CERN-LHCC-2010-013",
    year           = "2010",
    url            = "https://cds.cern.ch/record/1291633",
    related        = "ATLAS-TDR-19-addm",
    relatedstring  = "Addendum:",
}

@Article{LUCID2,
    author         = "Avoni, G. and others",
    title          = "{The new LUCID-2 detector for luminosity measurement and monitoring in ATLAS}",
    journal        = "JINST",
    volume         = "13",
    year           = "2018",
    pages          = "P07017",
    doi            = "10.1088/1748-0221/13/07/P07017",
}

@Booklet{ATLAS-CONF-2014-018,
    author         = "{ATLAS Collaboration}",
    title          = "{Tagging and suppression of pileup jets with the ATLAS detector}",
    howpublished   = "{ATLAS-CONF-2014-018}",
    url            = "https://cds.cern.ch/record/1700870",
    year           = "2014",
}

@Booklet{ATLAS-CONF-2018-006,
    author         = "{ATLAS Collaboration}",
    title          = "{Calibration of light-flavour \(b\)-jet mistagging rates using ATLAS proton--proton collision data at \(\sqrt{s} = 13~\text{TeV}\)}",
    howpublished   = "{ATLAS-CONF-2018-006}",
    url            = "https://cds.cern.ch/record/2314418",
    year           = "2018",
}

@Booklet{ATL-PHYS-PUB-2014-021,
    author         = "{ATLAS Collaboration}",
    title          = "{ATLAS Pythia~8 tunes to \(7~\text{TeV}\) data}",
    howpublished   = "{ATL-PHYS-PUB-2014-021}",
    url            = "https://cds.cern.ch/record/1966419",
    year           = "2014",
}

@Booklet{ATL-PHYS-PUB-2016-017,
    author         = "{ATLAS Collaboration}",
    title          = "{The Pythia~8 A3 tune description of ATLAS minimum bias and inelastic measurements incorporating the Donnachie--Landshoff diffractive model}",
    howpublished   = "{ATL-PHYS-PUB-2016-017}",
    url            = "https://cds.cern.ch/record/2206965",
    year           = "2016",
}

@Booklet{ATL-PHYS-PUB-2016-020,
    author         = "{ATLAS Collaboration}",
    title          = "{Studies on top-quark Monte Carlo modelling for Top2016}",
    howpublished   = "{ATL-PHYS-PUB-2016-020}",
    url            = "https://cds.cern.ch/record/2216168",
    year           = "2016",
}

@Booklet{ATL-PHYS-PUB-2020-014,
    author         = "{ATLAS Collaboration}",
    title          = "{Deep Sets based Neural Networks for Impact Parameter Flavour Tagging in ATLAS}",
    howpublished   = "{ATL-PHYS-PUB-2020-014}",
    url            = "https://cds.cern.ch/record/2718948",
    year           = "2020",
}

@Booklet{ATL-PHYS-PUB-2022-027,
    author         = "{ATLAS Collaboration}",
    title          = "{Graph Neural Network Jet Flavour Tagging with the ATLAS Detector}",
    howpublished   = "{ATL-PHYS-PUB-2022-027}",
    url            = "https://cds.cern.ch/record/2811135",
    year           = "2022",
}

@Booklet{ATL-PHYS-PUB-2022-035,
    author         = "{ATLAS Collaboration}",
    title          = "{Reweighting heavy-flavor production fractions to reduce flavor modelling uncertainties for ATLAS}",
    howpublished   = "{ATL-PHYS-PUB-2022-035}",
    url            = "https://cds.cern.ch/record/2816367",
    year           = "2022",
}
\clearpage
% ATLAS Collaboration author list
% Reference date of FTAG-2023-06 is 2024-08-09
% Author list last updated on date 04-AUG-25
% Data extracted on 04-Aug-2025 for paper reference FTAG-2023-06
% at 1:02pm
 
\begin{flushleft}
\hypersetup{urlcolor=black}
{\Large The ATLAS Collaboration}

\bigskip

\AtlasOrcid[0000-0002-6665-4934]{G.~Aad}$^\textrm{\scriptsize 105}$,
\AtlasOrcid[0000-0001-7616-1554]{E.~Aakvaag}$^\textrm{\scriptsize 17}$,
\AtlasOrcid[0000-0002-5888-2734]{B.~Abbott}$^\textrm{\scriptsize 124}$,
\AtlasOrcid[0000-0002-0287-5869]{S.~Abdelhameed}$^\textrm{\scriptsize 120a}$,
\AtlasOrcid[0000-0002-1002-1652]{K.~Abeling}$^\textrm{\scriptsize 57}$,
\AtlasOrcid[0000-0001-5763-2760]{N.J.~Abicht}$^\textrm{\scriptsize 51}$,
\AtlasOrcid[0000-0002-8496-9294]{S.H.~Abidi}$^\textrm{\scriptsize 30}$,
\AtlasOrcid[0009-0003-6578-220X]{M.~Aboelela}$^\textrm{\scriptsize 46}$,
\AtlasOrcid[0000-0002-9987-2292]{A.~Aboulhorma}$^\textrm{\scriptsize 36e}$,
\AtlasOrcid[0000-0001-5329-6640]{H.~Abramowicz}$^\textrm{\scriptsize 156}$
% ...
\bigskip
\\

% ...
$^{17}$Department for Physics and Technology, University of Bergen, Bergen; Norway.\\
% ...
$^{30}$Physics Department, Brookhaven National Laboratory, Upton NY; United States of America.\\
% ...
$^{36}$$^{(a)}$Facult\'e des Sciences Ain Chock, Universit\'e Hassan II de Casablanca;$^{(b)}$Facult\'{e} des Sciences, Universit\'{e} Ibn-Tofail, K\'{e}nitra;$^{(c)}$Facult\'e des Sciences Semlalia, Universit\'e Cadi Ayyad, LPHEA-Marrakech;$^{(d)}$LPMR, Facult\'e des Sciences, Universit\'e Mohamed Premier, Oujda;$^{(e)}$Facult\'e des sciences, Universit\'e Mohammed V, Rabat;$^{(f)}$Institute of Applied Physics, Mohammed VI Polytechnic University, Ben Guerir; Morocco.\\
% ...
$^{46}$Physics Department, Southern Methodist University, Dallas TX; United States of America.\\
% ...
$^{51}$Fakult\"{a}t Physik , Technische Universit{\"a}t Dortmund, Dortmund; Germany.\\
% ...
$^{57}$II. Physikalisches Institut, Georg-August-Universit\"{a}t G\"ottingen, G\"ottingen; Germany.\\
% ...
$^{105}$CPPM, Aix-Marseille Universit\'e, CNRS/IN2P3, Marseille; France.\\
% ...
$^{120}$$^{(a)}$New York University Abu Dhabi, Abu Dhabi;$^{(b)}$United Arab Emirates University, Al Ain; United Arab Emirates.\\
% ...
$^{124}$Homer L. Dodge Department of Physics and Astronomy, University of Oklahoma, Norman OK; United States of America.\\
% ...
$^{156}$Raymond and Beverly Sackler School of Physics and Astronomy, Tel Aviv University, Tel Aviv; Israel.\\
% ...
$^{a}$ Also Affiliated with an institute covered by a cooperation agreement with CERN.\\
$^{b}$ Also at An-Najah National University, Nablus; Palestine.\\
$^{c}$ Also at Borough of Manhattan Community College, City University of New York, New York NY; United States of America.\\
$^{d}$ Also at Center for High Energy Physics, Peking University; China.\\
$^{e}$ Also at Center for Interdisciplinary Research and Innovation (CIRI-AUTH), Thessaloniki; Greece.\\
$^{f}$ Also at CERN, Geneva; Switzerland.\\
% ...

\end{flushleft}

% Created with Glance <Atlas.Glance@cern.ch>

%

%
%

%

%
%
%
%

\end{document}